\documentclass{amsart}
\usepackage[margin=0.9in]{geometry}
\usepackage{amssymb}
\usepackage{amsmath, amsthm, amsopn, amsfonts}
\usepackage[dvips]{graphicx}
\usepackage{graphpap}
\usepackage{amscd}
\usepackage{fancyhdr}
\usepackage{lastpage}
\usepackage{float}
\usepackage{caption}
\usepackage{color}
\usepackage{tikz} 
\usetikzlibrary{decorations.markings}
\usepackage[utf8]{inputenc}
\usepackage[english]{babel}
\usepackage{cite}
\usepackage{graphicx}
\usepackage{booktabs,caption,fixltx2e}
\usepackage[flushleft]{threeparttable}
\usepackage{footmisc}
\newcommand{\lm}{\textnormal{\tiny\textit{L}}\hspace{-5pt}}
\newcommand{\lex}{\textnormal{\tiny\textit{LE}}\hspace{-5pt}}
\newcommand{\bir}{\textnormal{\tiny\textit{BR}}\hspace{-5pt}}
\newcommand{\mig}{\textnormal{\tiny\textit{MR}}\hspace{-5pt}}
\usepackage{enumitem}
\usepackage{thmtools, thm-restate}
\usepackage{hyperref}
\usepackage{tabularx}

\DeclareMathOperator*{\essinf}{ess\,inf}
\DeclareMathOperator*{\esssup}{ess\,sup}
	
\newcommand{\high}{\textnormal{\tiny \textit{high}}\hspace{-5pt}}
\newcommand{\low}{\textnormal{\tiny\textit{low}}\hspace{-5pt}}

\newcommand{\itn}{\textnormal{\tiny\text{ITN}}\hspace{-5pt}}
\usepackage[numbers]{natbib}

\newtheorem{theorem}{Theorem}[section]
\newtheorem{lemma}{Lemma}

\newtheorem{definition}{Definition} 
\newtheorem{Remark}{Remark}[section]
\begin{document}
%################
\selectlanguage{english}
\title
[Asymptomatic Malaria Model]
{\bf AN EPIDEMIOLOGICAL MODEL OF MALARIA ACCOUNTING FOR ASYMPTOMATIC CARRIERS\footnote{* jgutierr@uga.edu}}
\author{Jacob B. Aguilar} 
\address{Department of Mathematics, University of Georgia, Athens, GA 30602}
\author {Juan B. Gutierrez *}
\address{Department of Mathematics, University of Georgia, Athens, GA 30602}
%\date{}
%################
\begin{abstract}
Asymptomatic individuals in the context of malarial disease refers to subjects who carry a parasite load but do not show clinical symptoms. A correct understanding of the influence of asymptomatic individuals on transmission dynamics will provide a comprehensive description of the complex interplay between the definitive host (female \textit{Anopheles} mosquito), intermediate host (human) and agent (\textit{Plasmodium} parasite). The goal of this article is to conduct a rigorous mathematical analysis of a new compartmentalized malaria model accounting for asymptomatic human hosts for the purpose of calculating the basic reproductive number ($\mathcal{R}_0$), and determining the bifurcations that might occur at the onset of disease free equilibrium. A point of departure of this model from others appearing in literature is that the asymptomatic compartment is decomposed into two mutually disjoint sub-compartments by making use of the naturally acquired immunity (NAI) of the population under consideration. After deriving the model, a qualitative analysis is carried out to classify the stability of the equilibria of the system. Our results show that the dynamical system is locally asymptotically stable provided that $\mathcal{R}_0<1$. However this stability is not global, owning to the occurrence of a sub-critical bifurcation in which additional non-trivial sub-threshold equilibrium solutions appear in response to a specified parameter being perturbed. To ensure that the model does not undergo a backward bifurcation, we demand that an auxiliary parameter denoted $\Lambda<1$ in addition to the threshold constraint $\mathcal{R}_0<1$. The authors hope that this qualitative analysis will fill in the gaps of what is currently known about asymptomatic malaria and aid in designing strategies that assist the further development of malaria control and eradication efforts. 
\end{abstract}  
%################
\maketitle
%################
\vspace{10pt}
\begin{center}
\textit{Dedicated to my mother Jacqueline Donna}
\end{center}
\vspace{10pt}
%################
\section{Introduction}\label{sec:Introduction}
Malaria is one of the most lethal and complex parasitic diseases in the world \cite{WHO}.  Throughout human history, malaria has burdened most regions of our planet and has had a profound impact in human history and evolution; e.g. it has been credited for contributing to the decline of the Roman Empire \citep[p.~14]{ROME}. %During the early nineties in a site located at Lugnano in Teverina (approximately 70 miles up the Tiber valley in Rome), archaeologists unearthed a cemetery \citep[p.~319]{sallares2004spread}. Dating back to c.450 AD and primarily consisting of infants, this cemetery stands to be one of the largest of its kind ever found in Roman Italy. There is evidence to suggest that the most likely cause of the infant deaths was a malaria epidemic, \citep[p.~35]{soren1995late}.
In 2015, the World Health Organization (WHO) reported the occurrence of approximately 214 million new cases of malaria (range: 149–303 million), which resulted in 438 thousand disease-induced deaths (range: 236–635 thousand) \cite{WHO}. It was estimated that 70\% of these fatalities were experienced by children under the age of five.

The life-cycle of the \textit{Plasmodium} parasite can be broken down into two separate subcycles: the asexual cycle, occurring in humans (intermediate host) and the sexual cycle in mosquitoes (definitive host), in which maturity is reached. %The sequences of developmental stages undergone are dependent on the species of \textit{Plasmodium} and vertebrate host. 
The \textbf{sexual cycle} begins when a susceptible mosquito feeds on the blood of an infectious human, ingesting sexual forms of the \textit{Plasmodium} parasite previously developed in the human body, known as \textit{gametocytes}. While in the midgut lumen of the mosquito, these \textit{gametocytes} fuse to form diploid \textit{zygotes}, which grow into enlongated \textit{ookinetes}. The motile \textit{ookinetes} burrow into the outer membrane of the mosquito midgut and form ellipsoid shaped \textit{oocysts}. Eventually, the \textit{oocysts} rupture releasing thousands of haploid forms called \textit{sporozoites} \cite{rosenberg1991number}. These \textit{sporozoites} accumulate in the salivary glands of the mosquito, causing it to become infectious. 

The \textbf{asexual cycle} begins when an infectious mosquito bites the host and injects saliva with anti-coagulant agents that keep the wound open thus allowing a blood meal, and simultaneously injecting \textit{sporozoites} into the skin \cite{CAD}. The \textit{sporozoites} travel through the blood vascular system to the liver where they invade the cells of the liver, known as \textit{hepatocytes}. Inside the human, a \textit{Plasmodium} infection goes through two cycles: a initial liver (hepatic, or exo-erythrocytic) stage lasting a few days, followed by a blood (erythrocytic) stage that lasts until the hosts clears naturally the infections, receives treatment, or dies. The \textbf{hepatic stage} begins in the \textit{hepatocytes}, where a proportion of the \textit{sporozoites} undergo a process called \textit{pre-erythrocytic or hepatic schizogony}, in which they multiply asexually to produce thousands of haploid daughter cells, known as \textit{merozoites}. During this process, \textit{schizonts} are formed, causing the \textit{hepatocytes} to rupture. This allows the \textit{merozoites} to enter the bloodstream. The \textbf{erythrocytic stage} begins when free-floating \textit{merozoites} invade \textit{erythrocytes} in a matter of minutes. Inside the \textit{erythrocyte}, parasites enter the ring stage in which some mutate into an enlarged ring-shaped form called \textit{trophozoites} that mature into \textit{schizonts}, causing the cell to burst and releasing more \textit{merozoites} into the bloodstream. In the case of \textit{P. falciparum}, this process of invasion and rupture of RBCs occurs synchronously every 48 hours. At this point in the subcycle, a small portion of the \textit{merozoites}, for reasons currently unknown, morph into \textit{gametocytes}, or sexual forms; however, no sexual reproduction occurs inside the host \cite{PARA}. 

The infected human host is now ready to infect new susceptible mosquitos, thus completing the \textit{Plasmodium }life cycle. It is worth mentioning that the blood stage parasites are responsible for most clinical symptoms associated with the disease \cite{PARA}. The periodic rupturing of the RBCs results in the release of various debri and waste products, which in turn activate the immune system and cause symptoms such as chills, fatigue, pain, and fever. The average duration for the infection of an RBC is dependent on the \textit{Plasmodium} species. \textit{P. falciparum} has the interesting pathological effect of sequestration, which occurs when infected RBCs containing mature forms of the parasite, i.e. \textit{trophozoites} and  \textit{schizonts}, adhere to the walls of small diameter blood vessels, e.g. the endothelium of capillaries and venules \cite{PFAL}. As a result of sequestration, the microcirculation is reduced and in some cases inflammatory processes take place. One of the common complications of this sequestration is cerebral malaria, which might cause patients to sustain brain injury resulting in long-term neuro-cognitive impairment \cite{CEREB}. 

A human host is called asymptomatic when it is a carrier for the \textit{Plasmodium} parasite, but displays no clinical symptoms. Asymptomatic carriers contribute to \textit{gametocyte} circulation by providing a hidden reservoir for the parasite to take refuge. As pointed out in \cite[2012]{LPJJ}, asymptomatic infections often go undetected, resulting in a major source of \textit{gametocytes} for local mosquito vectors. Accordingly, asymptomatic carriers contribute to the persistence of malaria transmission within their localized populations \cite{BTG}. Fequent exposure to the \textit{Plasmodium} parasites leads to naturally acquired immunity to the symptoms of the disease, but not necessarily to the parasite and as a result it creates asymptomatic carriers in a given population \cite{SH}.  Asymptomatic malaria infections have been reported in various high and intermediate transmission areas, such as Kenya and Nigeria \cite{BTG,ECN}. Recently, asymptomatic infections have been reported in relatively low endemic areas, such as Colombia and the Amazonian region of Brazil \cite{COLA,ASYBRAZ}. There is much evidence that asymptomatic malaria infections play a fundamental role in malaria transmission, cf. \cite{LSA}. Disease transmission dynamics are greatly affected by the amount of asymptomatic carriers in a given population over a specified time interval. Indeed, a positive correlation between high transmission and high asymptomatic prevalence has been reported in Nigeria, Senegal, Gabon and the Amazonian regions of Brazil \cite{DMKKK,ECN,AFR,ALCO}.

In accordance with \cite[1980]{Bruce}, 
%
%ARE THE DEFINITIONS THAT FOLLOW STILL USED IN RECENT LITERATURE????
%
we define malaria immunity as the state of resistance to the infection brought about by all processes which are involved in destroying the plasmodia or limiting their multiplication. \textit{Natural innate immunity} is an intrinsic property of the host. This type of immunity is characterized by an immediate inhibitory response to the introduction of the parasite which is independent of any previous infection. Their are two types of acquired immunity, namely: \textit{active acquired immunity} and \textit{passive acquired immunity}. \textit{Active acquired immunity} is defined as an enhancement of the hosts defense mechanism due to previous contact with the pathogen. \textit{Passive acquired immunity} is characterized by either the mother to child transfer of protective antibodies in the pre or post natal developmental periods or by the injection of such antibodies. In this work we are focused on \textit{Active acquired immunity}, more specifically a type of immunity that is acquired through means of exposure.        

Humans experience various kinds of \textit{Active acquired immunity} which provide different kinds of protection.  Making use of the definitions in \cite[2009]{AcqIM}, we define protection to be objective evidence of a lower risk of clinical disease, indicated by the absence of fever, i.e. the oral temperature does not exceed the threshold $37^\circ$C \cite{Clark}, with parasitemia. \textit{Anti-disease immunity} is conferred protection against clinical disease; which affects the overall risk and extent of morbidity associated with a given parasite density. \textit{Anti-parasite immunity} is conferred protection against parasitemia; which affects the parasite density. \textit{Premunition} 
%
%   DEFINITION OF PREMUNITION ?!?!?!?!
%
provides protection against new infections by maintaining a generally asymptomatic parasitemia, \cite{Koch1,Koch2,Sergent}. In this article, we make use of a kind of \textit{premunition} called naturally acquired immunity (NAI). In holoendemic regions across sub-Saharan Africa most people are continuously infected by \textit{P. falciparum} while the majority of infected adults rarely experience observable disease. As reported in \cite[2009]{AcqIM}, this valid protection against infection is NAI corresponding to \textit{P. falciparum}.

Mathematical models of malaria transmission have been studied by various authors, for a survey we refer the reader to \cite[2011]{MSS}. Asymptomatic malaria has also been previously modeled and studied. Of recent, an asymptomatic malaria model was introduced in \cite[2007]{FLIPE}. This model depends on a state-invariant control parameter $\phi$, which stands for the proportion of human infections that develop disease. After letting $1/h$ denote the mean latent period in humans, the progression rates from the exposed to the symptomatic and asymptomatic classes were defined to be the products $h\phi$ and $h(1-\phi)$, respectively. Additionally, asymptomatic humans were included in the recovered compartment.  

In this article we depart from the previous models of asymptomatic malaria by creating explicitly different compartments for symptomatic ($Y$) and asymptomatic ($A$) subjects, which in addition to susceptibles ($S$), exposed ($E$), and recovered ($R$) yields the acronym $SEY \hspace{-3pt}AR$. Another important point of departure with respect to previous models of asymptomatic malaria is that in the $SEY \hspace{-3pt}AR$ model the progression rates (\ref{SEYAR_DS}) to symtomatic and asymptomatic are nonlinear functions of the time-dependent exposed proportion. Therefore, the $SEY \hspace{-3pt}AR$ model does not fall into a sub-class of such models currently appearing in literature. Moreover, we do not include the asymptomatic humans in the recovered compartment. This allows an effective isolation of the effect that asymptomatic carriers have on the disease transmission dynamics. The derivation of the $SEY \hspace{-3pt}AR$ model (\ref{SEYAR_DS}) hinges on a specific decomposition of the infected human compartment into two mutually disjoint sub-compartments accounting for asymptomatic and symptomatic carriers. This decomposition is accomplish by making use of a nonlinear exposure dependent NAI function, which is the solution to the initial-value problem (\ref{NAI_IVP}) derived in Section \ref{sec:Methods}. Although there is much in the literature, the proper inclusion of asymptomatic carriers into the epidemiological modeling of malaria warrants a formal mathematical understanding. 

This manuscript provides a new malarial model accounting for asymptomatic human hosts in terms of the NAI of the population under consideration and is organized as follows: Section \ref{sec:Introduction} presents a summary of the state of the art, Section \ref{sec:Methods} presents the model formulation, Section \ref{sec:WellPosedness} covers the issue of well-posedness of the initial value problem and provides an analysis of the total population dynamics, Section \ref{sec:Rnumber} contains a rigorous study of the local asymptotic stability of disease-free equilibrium (DFE) for the model with a mathematical and epidemiological interpretation of the reproductive threshold, Section \ref{sec:RnumberComparison} introduces a mathematical framework to formally address the impact of the asymptomatic class on the reproductive threshold, Section \ref{sec:EE} is focused on nonlinear stability analysis and provides a classification parameter in which its size determines the type of bifurcation undergone by the dynamical system, Section \ref{sec:ParamEst} provides a unified derivation of many static quantities widely used in malaria epidemiology, Section \ref{sec:ControlMeas} incorporates generalized control measures into the dynamical system, Section \ref{sec:SeyarViv} introduces a modification of the model including relapse rates,
Section \ref{sec:SenAna} is focused on a sensitivity analysis of the reproductive threshold arising from the model, Section \ref{sec:NumRes} consists of numerical results corresponding to the following three high transmission sites: Kaduna in Nigeria, Namawala in Tanzania, and Butelgut in Papua New Guinea, Section \ref{sec:Conclusion} consists of a summary of the results contained in Sections \ref{sec:Methods}-\ref{sec:NumRes} and a discussion regarding future direction and extensions, Section \ref{sec:A1} contains formal proofs of the lemmas and theorems contained in Sections \ref{sec:Methods}-\ref{sec:ControlMeas}, Section \ref{sec:A2} is a summary of the main stability theorems used in the investigation of the local asymptotic stability of the equilibrium solutions studied in Sections \ref{sec:Rnumber} and \ref{sec:EE}, and Section \ref{sec:Pvalues} contains tables of numerical rates corresponding to the high transmission sites studied in Section \ref{sec:NumRes}.  
%################
%################
%Model Formulation
\section{Methods: Model Formulation}\label{sec:Methods}
%----------------------------
The formulation of the $SEY \hspace{-3pt}AR$ model for the spread of malaria in the human and mosquito populations begins with dividing the total host-vector population into two compartments, denoted by $N_H(t)$ and $N_M(t)$, which stand for the total population sizes of the humans and mosquitoes, respectively, at a given time $t$. From this point on, whenever implied by the context of the discussion, the time $t$ dependency is suppressed to avoid a cluttering of the notation. Assuming a homogeneously mixed host population, we further decompose the compartments into the following five epidemiological classes: susceptible human $S$, exposed human $E$, symptomatic human $Y$, asymptomatic human $A$, and recovered human $R$. So that, $N_H=S+E+Y+A+R$. For simplicity of exposition the state variable is identified with its corresponding class. For example, when we are considering a human from class $A$, it is understood that $A$ is a function and not a class, in general. A point of departure from the usual $SEIR$ models, as studied in \cite{ROOP,li1995global,li1999global,d2002stability,smith2001global}, resides in the mutually disjoint partitioning of the infected compartment into two sub-compartments, labeled asymptomatic $A$ and symptomatic $Y$, i.e. $I=Y \cup A$ where $Y \cap A= \varnothing$. For the mosquito population we have the following three classes: susceptible mosquito $M_S$, exposed mosquito $M_E$ and infected mosquito $M_I$. Accordingly, the total mosquito population is given by $N_M=M_S+M_E+M_I$.

A fundamental step in the accurate modeling of any infectious disease resides in the formulation for the force of infection, i.e. the probability per unit time for a susceptible to become infected. As mentioned in \cite{FLIPE}, it is known that the infection rates between human and mosquito populations depend on numerous factors, including: the man-biting rate of the mosquito $\sigma$ (which is the number of bites per mosquito), transmission probabilities (to be later defined), and the number of infectious and susceptible of each species involved. Furthermore, we assume that the average number of mosquito bites suffered by humans depends on the total sizes of their respective populations in the community. As a result, the number of bites per human is $\sigma \frac{N_M}{N_H}$. Therefore, the force of infection from mosquitoes to humans $\lambda_{SE}$ is defined to be the product of the number of bites per human, the transmission probability $\beta_M$ from a mosquito in the class $M_I$ to a human in class $S$ and the probability that a mosquito is infectious $\frac{M_I}{N_M}$, i.e. 
%----------------------------
\begin{equation*}
	\lambda_{SE} =\omega_{M} = \sigma \frac{N_M}{N_H} \ \beta_M \ \frac{M_I}{N_M}=\sigma \beta_M \frac{M_I}{N_H}.
\end{equation*}
%----------------------------
Although asymptomatic carriers do not get clinically ill, they still harbor low levels of \textit{gametocytes} in their bloodstreams and are able to pass the infection onto mosquitoes, \cite{VG}. When a mosquito from class $M_S$ bites a human from class $Y$, the force of infection $\omega_{Y}$ is defined as the product of the number of bites per mosquito $\sigma$, the transmission probability $\beta_Y$ from a human in $Y$ to a mosquito in $M_S$ and the probability that a human is in the symptomatic class $\frac{Y}{N_H}$. When a mosquito from the class $M_S$ bites a human from class $A$, the corresponding force of infection $\omega_{A}$ is the product of the number of bites per mosquito $\sigma$, the transmission probability $\beta_A$ from a human in $A$ to a mosquito in $M_S$ and the probability that a human is in the asymptomatic class $\frac{A}{N_H}$. As pointed out in \cite{laishram2012complexities}, the parasites carried by asymptomatic hosts can be more infectious than those of symptomatic hosts. One could assume that a typical asymptomatic carrier has a higher NAI level than a symptomatic, so that $\beta_A \leq \beta_Y$, or visa versa. However, from a mathematical standpoint this assumption is unnecessary and will not be made. Accordingly, the force of infection from humans to mosquitoes $V_{SE}$ is defined to be the sum of the forces of infection corresponding to the humans in classes $Y$ and $A$, i.e.
\begin{equation*}
\nu_{SE}=\omega_{Y} + \omega_{A} =\sigma \Big ( \beta_Y \frac{Y}{N_H} +\beta_A \frac{A}{N_H}\Big ).
\end{equation*}
Let $\nu_{EI}=\tau$, where $\tau$ is the mean duration of the definite host latent period, i.e. the reciprocal of the \textit{Plasmodium} incubation period corresponding to mosquito species being studied. 
%----------------------------
\begin{Remark}
It is worth pointing out that the progression rate $\nu_{SE}$ is equal to the sum of the human to mosquito infection forces $\omega_Y$ and $\omega_A$, while $\lambda_{SE}$ is equal to the singular force of infection from mosquitoes to humans, denoted as $\omega_M$. The decomposition of the human infected compartment results in two different infection forces corresponding to the asymptomatic and symptomatic carriers. However, no decomposition is applied for the infected mosquito compartment.
\end{Remark}
%----------------------------
At a given time $t \in \mathbb{R}_+$ an individual's experience of malaria is dependent upon the degree of naturally acquired immunity that he or she has gained. Effective anti-parasitic immunity is achieved only after many frequent infections, \cite{JS,GM,EVASS}. This important epidemiological observation, in combination with the discussion regarding naturally acquired immunity (NAI) in Section (\ref{sec:Introduction}), implies that the rate of progression $\lambda_{EA}$ from the exposed class $E$ to asymptomatic class $A$ depends on the proportion of human individuals receiving sufficient protection from the average NAI accumulated in the population with respect to natural exposure. Let $u(t)$ denote the proportion of the human population fully protected by NAI. Since naturally acquired immunity to the \textit{Plasmodium} parasite is acquired and accumulates over time in response to frequent exposure. The rate that this proportion of protected individuals changes depends on the the rate that the human population is being exposed, up to a threshold value. To uncover this exposure dependency, firstly let the lower and upper protected proportion thresholds be given by $u_{_{\low}\hspace{2pt}}$ and $u_{_{\high}\hspace{5pt}}$, respectively. It should be noted that $0<u_{_{\low}\hspace{2pt}}<u_{_{\high}\hspace{3pt}}<1$. An increase or decrease in the protected proportion $u$ corresponds to an increase or decrease in the exposed proportion $\frac{E}{N_H}$. Let $\varepsilon:=\frac{E}{N_H}$, upon assuming that the initial NAI protected proportion is given by the lower threshold $u(0):=u_{_{\low}\hspace{5pt}}$, these epidemiological principles lead to the following initial value problem (IVP) being posed
%----------------------------
\begin{equation}
		\begin{cases}
			\dot{u}= (u_{_{\high}}- u) \dot{\varepsilon},\\
			u(0)=u_{_{\low}\hspace{5pt}}.
		\end{cases} 
	\label{NAI_IVP}
\end{equation}
%----------------------------
By making use of the integrating factor $L=e^{\int_{0}^t \dot{\varepsilon}(s)ds}:=e^{\varepsilon}$, it follows that
%----------------------------
\begin{align*} 
      	\dot{(Lu)} &=  L\dot{\varepsilon} u_{_{\high}\hspace{5pt}},\\ 
	       	   Lu  &= e^{\varepsilon_0}u_{_{\low}} +(L-e^{\varepsilon_0})u_{_{\high}\hspace{5pt}},\\   
	          	u  &= L^{-1}e^{\varepsilon_0}u_{_{\low}} + (1-L^{-1}e^{\varepsilon_0})u_{_{\high}\hspace{5pt}},\\ 
              	u  &= e^{\varepsilon_0-\varepsilon} u_{_{\low}} + (1-e^{\varepsilon_0-\varepsilon})u_{_{\high}\hspace{5pt}}.\\       
\end{align*} 
%----------------------------
Upon rearranging terms and invoking a slight abuse of notation, to emphasize the exposure $\varepsilon$ dependency of $u$, the solution is represented by the following equation 
%----------------------------
\begin{equation}
		u(\varepsilon)=e^{\varepsilon_0-\varepsilon}(u_{_{\low}}-u_{_{\high}\hspace{5pt}}) + u_{_{\high}\hspace{5pt}}.
		\label{Irate}
\end{equation}
%----------------------------
In the above, the symbol $\varepsilon_0:=\varepsilon(0)$ stands for the initial exposed proportion of the human host population. Care should be taken to ensure that the progression rate is mathematically well-defined and epidemiological sensible, i.e. a singularity should not arise and it should be non negative. Provided that $\mathbf{x}$ is such that $\varepsilon \in C^1(\mathbb{R}_+)$, then clearly the progression rate will not experience a singularity. However, assuming that the initial exposed proportion is zero for some time $t_0$, a subtle complication arises for existence times such that the human population is extinct, i.e. $N_H(t_0)=0$. If the state variable $E$ is positive, then clearly $N_H$ is positive and thus non-zero, so $\varepsilon$ is will not undergo a singularity. However, in the case that $E$ is zero for such existence times, then it is possible that $N_H$ may simultaneously be equal to zero, thus forcing a singularity to arise. Therefore, the following piecewise function $\tilde{\varepsilon}$ is defined to remedy this subtle issue and ensure that the proportion of exposed humans is taken to be zero in the case that the human population is extinct
%---------------------------------------------------------------------------------
\begin{equation*}
\tilde{\varepsilon}:=
		\begin{cases}
			\frac{E}{N_H} ,\quad \text{if} \quad N_H \neq 0,\\
			0,\quad \hspace{9pt} \text{if} \quad N_H = 0.\\
		\end{cases} 
	\label{Total_ineqs}
\end{equation*}
%---------------------------------------------------------------------------------
Time dependency is implicit in the above function definition. It is worth emphasizing that the above function holds for any time (intervals or discrete values) such that $N_H=0$. Mathematically speaking, if $\mathbf{x}$ is such that $\varepsilon \notin C^1(\mathbb{R}_+)$, the only complication occurs for $t \in \mathbb{R}_+$ such that $N_H=E=0$. In this case, a singularity arises in the composite function $u=(u \circ \varepsilon)$. In order to account for this, epidemiological unreasonable, minor issue, $\varepsilon$ is extended to its infimum value by the above function definition. From this point on, an abuse of notation is invoked in which an identification of $\varepsilon$ with its continuous extension is made. In other words, for all practical purposes, we take $\tilde{\varepsilon}=\varepsilon$.

Besides the trivial singularity issue covered above, care must be taken to ensure the non-negativity of the nonlinear progression rate $u$. As a result, the solution to the IVP (\ref{NAI_IVP}) implicitly imposes an additional constraint upon the initial data of the $SEY \hspace{-3pt}AR$ model (\ref{SEYAR_DS}). The constraint is listed below in lemma (\ref{thm:LAS}), which is proven in Appendix (\ref{sec:A1}).
%----------------------------
%################
%----------------------------
%################
\begin{theorem}{(Initial Data Constraint for the SEYAR Model).}
Let $\vartheta$ be defined as follows 
\begin{equation}
\vartheta:=\ln \left(\frac{u_{_{\high}\hspace{5pt}}}{u_{_{\high}\hspace{5pt}}-u_{_{\low}\hspace{5pt}}}\right).
\label{IntCon}
\end{equation}
To ensure the non-negativity of $u$, the initial data is required to satisfy the inequality $E_0 \leq \vartheta N_0$.
\label{thm:IntData}
\end{theorem}
%################
%----------------------------
%################
%----------------------------
%----------------------------
Let the eight dimensional vector of functions $\mathbf{x}=\left(S,E,Y,A,R,M_S,M_E,M_I\right)^T$ be such that $\varepsilon \in C^1(\mathbb{R}_+)$ and $\vartheta$ be defined as in Lemma (\ref{thm:IntData}) above. It directly follows that $u$ is trapped in the compact sub-interval $T_u$, defined in Section \ref{sec:Rnumber}. Therefore, the following bounds are established
%----------------------------
\begin{align*}
\essinf\limits_{t \in \mathbb{R}^+} \| u\| &= u_{_{\low}\hspace{5pt}}, \\
\| u\|_\infty &= u_{_{\high}\hspace{5pt}}. \\
\end{align*}
%----------------------------
After placing all the above together, the nonlinear progression rate $u$ is well-defined in a mathematical an epidemiological sense. It is worth mentioning that due to the constructive assumptions mentioned above, the vector of functions $(\varepsilon,u(\varepsilon))$ is trapped in the unit rectangle $(0,1) \times (0,1)$, for all $t \in \mathbb{R}_+$ . 

Define $\lambda_{EA}=\gamma u(\varepsilon)$ where $\gamma$ is the mean duration of the human latent period, i.e. the time that elapses before the presence of a disease is manifested by symptoms. It is a direct consequence that $\lambda_{EY}=\gamma (1-u(\varepsilon))$, so that $\lambda_{EA}+\lambda_{EY}=\gamma$. Since the naturally acquired immune proportion will grow in response to exposure, the rate of progression from the exposed class $E$ to asymptomatic class $A$ should increase, warranting the choice of $\lambda_{EA}$. Furthermore, on time intervals such that the exposure rate increase without bound, i.e. $\dot{\varepsilon} \rightarrow +\infty$, it follows that $u(\varepsilon) \rightarrow \| u\|_\infty:=u_{_{\high}\hspace{5pt}}$. In this case the progression rate $\lambda_{EA}=\gamma u_{_{\high}\hspace{5pt}}$ is maximized. This is consistent with the observation that the average amount of asymptomatic human hosts in a population should increase after frequent exposure over a sufficient time period. When the exposed proportion is equal to zero over a prescribed time interval, it follows that $u=e^{\varepsilon_0}(u_{_{\low}}-u_{_{\high}\hspace{5pt}}) + u_{_{\high}\hspace{5pt}}$ over the interval. This quantity is a sum consisting of the upper threshold and a negative scaled difference of the lower and upper thresholds. If this infimum is achieved, then the progression rate $\lambda_{EA}$ will be minimal. This is due to the fact that if there is little exposure, then there is little NAI developed in the population, so that the rate of progression from $E$ to $Y$ will be maximal, i.e.  $\lambda_{EY}=\gamma (1-\essinf\limits_{t \in \mathbb{R}^+} \| u\| )=\gamma (1-u_{_{\low}\hspace{5pt}})$. Furthermore, the progression rate from $E$ to $Y$ should decrease to the smaller threshold value $\lambda_{EY}=\gamma (1-\| u\|_\infty)=\gamma (1-u_{_{\high}\hspace{5pt}})$, as the exposure rate increases. Listed below is the flow diagram for the $SEY \hspace{-3pt}AR$ model (\ref{SEYAR_DS}). 
%----------------------------

%SEYAR DIAGRAM
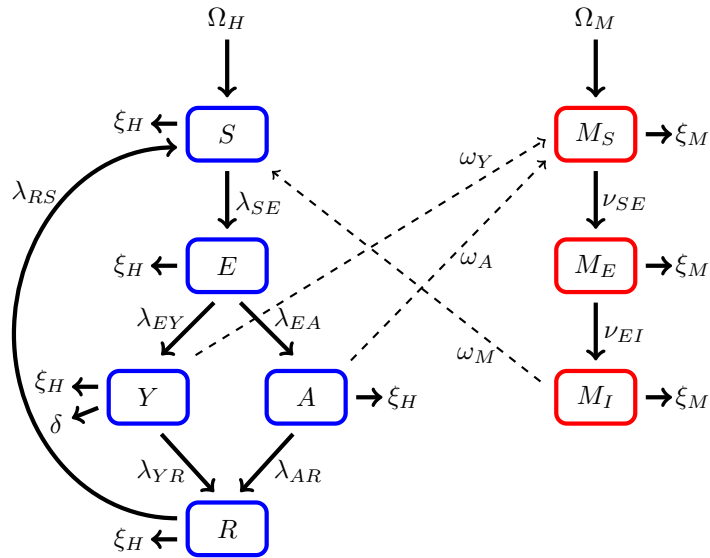
\begin{figure}[ht]
	\centering
	\begin{tikzpicture}[scale=0.7]
		%\draw [help lines] (0,0) grid (14,14);
		% left arrows
		\draw [->, ultra thick] (4.25, 11.3) to (4.25, 10.2);
		\draw [->, ultra thick] (4.25, 8.8) to (4.25, 7.7);
		\draw [->, ultra thick] (4, 6.3) to (3, 5.3);
		\draw [->, ultra thick] (4.5, 6.3) to (5.5, 5.3);
		\draw [->, ultra thick] (3, 3.8) to (4, 2.7); 
		\draw [->, ultra thick] (5.5, 3.8) to (4.5, 2.7); 
		% curvy arrow
		\draw [->, ultra thick] (3.3, 2.2) [in= 270, out= 180]  to (0.2,5.7)  [in=180, out=90] to (3.3, 9.25);
		% left horizonal arrow
		\draw [->, ultra thick] (3.3, 1.8) to (2.8, 1.8);
		\draw [->, ultra thick] (1.8, 4.7) to (1.3,4.7);
		\draw [->, ultra thick] (1.8, 4.3) to (1.3,4.1); % slate arrow
		\draw [->, ultra thick]  (6.7, 4.5) to (7.2, 4.5);
		\draw [->, ultra thick] (3.3, 7) to (2.8, 7);
		\draw [->, ultra thick] (3.3, 9.7) to (2.8, 9.7);
		% boxes on the left
		\draw [blue, ultra thick, rounded corners] (3.5, 1.5) rectangle (5,2.5);
		\draw [blue, ultra thick, rounded corners] (2, 4) rectangle (3.5, 5);
		\draw [blue, ultra thick, rounded corners] (5, 4) rectangle (6.5, 5);
		\draw [blue, ultra thick, rounded corners] (3.5, 6.5) rectangle (5, 7.5);
		\draw [blue, ultra thick, rounded corners] (3.5, 9) rectangle (5, 10);
		% right arrows 
		\draw [->, ultra thick] (11.25, 11.3) to (11.25, 10.2);
		\draw [->, ultra thick] (11.25, 8.8) to (11.25, 7.7);
		\draw [->, ultra thick] (11.25, 6.3) to (11.25, 5.2);
		% right horizontal arrows
		\draw [->, ultra thick] (12.2, 7) to (12.7, 7);
		\draw [->, ultra thick] (12.2, 9.5) to (12.7, 9.5);
		\draw [->, ultra thick] (12.2, 4.5) to (12.7, 4.5);
		% boxes on the right
		\draw [red, ultra thick, rounded corners] (10.5, 9) rectangle (12, 10);
		\draw [red, ultra thick, rounded corners] (10.5, 6.5) rectangle (12, 7.5);
		\draw [red, ultra thick, rounded corners] (10.5, 4) rectangle (12, 5);
		% middle arrows 
		\draw [dashed, thick,->] (10.2, 4.8) to (5.1, 8.8);
		\draw [dashed, thick, ->] (3.65, 5.3) to (10.3, 9.4);
		%\draw [dashed, thick, ->] (5.2, 2)  [out=30, in=245] to (10.5, 8.7);
		\draw [dashed, thick, ->] (6.6, 5.2) to (10.3, 9);
		% letters in the box
		%left
		\node at (4.25, 2) {$R$};
		\node at (2.75, 4.5) {$Y$};
		\node at (5.75, 4.5) {$A$};
		\node at (4.25, 7) {$E$};
		\node at (4.25, 9.5) {$S$};
		%right
		\node at (11.25, 7) {$M_E$};
		\node at (11.25, 9.5) {$M_S$};
		\node at (11.25, 4.5) {$M_I$};
		% letters for arrow
		\node at (2.4, 1.8) {$\xi_H$};
		\node at (0.9, 4.8) {$\xi_H$};
		\node at (2.4, 7) {$\xi_H$};
		\node at (2.4, 9.75) {$\xi_H$};
		\node at (7.6, 4.5) {$\xi_H$};
		\node at (13.1, 9.5) {$\xi_M$};
		\node at (13.1, 7) {$\xi_M$};
		\node at (13.1, 4.5) {$\xi_M$};
		\node at (1, 4) {$\delta$};
		%letters for v a rrows
		\node at (4.25, 11.7) {$\Omega_H$};
		\node at (11.25, 11.7) {$\Omega_M$};
		\node at (0.6, 8.4) {$\lambda_{RS}$};
		\node at (11.8, 8.2) {$\nu_{SE}$};
		\node at (11.8, 5.7) {$\nu_{EI}$};
		\node at (4.85, 8.2) {$\lambda_{SE}$};
		\node at (3, 6) {$\lambda_{EY}$};
		\node at (5.6, 6) {$\lambda_{EA}$};
		\node at (3, 3.1) {$\lambda_{YR}$};
		\node at (5.6, 3.1) {$\lambda_{AR}$};
		\draw [white] (0.5,1) rectangle (1,1.1); % this rectangle does not show, since I put the the color in white, the purpose is to create more space between the picture and the caption. You can change the size of the rectangle to adjust the space. 
		\node at (9,9) {$\omega_{Y}$};
		\node at (9,7.1) {$\omega_{A}$};
		\node at (9,5.3) {$\omega_{M}$};
	\end{tikzpicture}
	\caption{\small Schematic diagram of the malaria model including an asymptomatic compartment}
	\label{SEYARdiagram}
\end{figure}

%----------------------------
In Figure \ref{SEYARdiagram}, the solid lines represent progression from one compartment to the next, while the dotted stand for the human-mosquito interaction. Humans enter the susceptible compartment either through birth of migration and then progress through each additional compartment subject to the rates described above. These assumptions give rise to the following $SEY \hspace{-3pt}AR$ model IVP (\ref{SEYAR_DS}) describing the dynamics of malaria disease transmission in the human and mosquito populations
%----------------------------
\begin{equation}
\begin{cases}
		\dot{S} =\Omega_H + \lambda_{RS} R - \left(\sigma \beta_M \frac{M_I}{N_H} +  \xi_H \right)S,\nonumber \\ 
		\dot{E} =\sigma \beta_M \frac{M_I}{N_H}S - (\gamma +\xi_H)E,\nonumber \\            
		\dot{Y} =\gamma(1-u(\varepsilon)) E -  \left(\xi_H +\delta + \lambda_{YR}\right)Y,\nonumber \\               
		\dot{A} =\gamma u(\varepsilon) E - \left(\lambda_{AR}+\xi_H\right)A,\nonumber \\              
		\dot{R} =\lambda_{AR}A +  \lambda_{YR}Y - \left(\lambda_{RS}  +\xi_H\right) R,\nonumber  \\          
		\dot{M_S} =\Omega_M  - \left ( \xi_M +  \sigma \beta_Y \frac{Y}{N_H} +\sigma \beta_A \frac{A}{N_H}  \right )M_S,\nonumber \\        
		\dot{M_E} =\sigma \left ( \beta_Y \frac{Y}{N_H} +\beta_A \frac{A}	{N_H}\right )M_S - \left(\xi_M + \tau \right)M_E,\nonumber \\      
		\dot{M_I} =\tau M_E- \xi_M M_I,\nonumber \\	
\\
\left(S_0,E_0,Y_0,A_0,R_0,M_{S_0},M_{E_0},M_{I_0}\right)^T\in\mathbb{R}^8_+ \text{ such that } E_0 \leq \vartheta N_0,
\end{cases} 
	\label{SEYAR_DS}
\end{equation}
%----------------------------
where $u$ and $\vartheta$ are defined by equations (\ref{Irate}) and (\ref{IntCon}), respectively. 
%----------------------------
For convenience, the model parameters are summarized in Table \ref{table:nonlin} below. All of the parameters are strictly positive, with exception for the disease induced death rate $\delta$, which is allowed to be non-negative.
%----------------------------
{\footnotesize
\begin{table}[ht]
\caption{Model Parameters}
\centering
\begin{tabular}{l l l l r}
\textbf{Parameter} & \textbf{Description} & \textbf{Dimension}\\ [0.5ex] % inserts table %heading
\hline
\medskip
$\Omega_H$ & Recruitment rate of humans & humans $\times$ $\text{time}^{-1}$ \\
$\Omega_M$ & Recruitment rate of mosquitoes & mosquitoes $\times$ $\text{time}^{-1}$ \\
$\xi_H$ & Natural mortality rate of human & $\text{time}^{-1}$  \\
$\xi_M$ & Natural mortality rate of mosquito & $\text{time}^{-1}$ \\
$\beta_A$ & Probability of disease transmission from  & $\text{time}^{-1}$ \\
& asymptomatic human to a susceptible mosquito &  \\
$\beta_Y$ & Probability of disease transmission from  & $\text{time}^{-1}$  \\
& symptomatic human to a susceptible mosquito &   \\
$\beta_M$ & Probability of disease transmission from  & $\text{time}^{-1}$ \\
& infected mosquito to susceptible human &  \\
$\gamma$  & The intermediate host mean latent period & $\text{time}^{-1}$  \\
$\tau$ & The definitive host mean latent period & $\text{time}^{-1}$  \\
$\delta$ & Disease-induced death rate for humans & $\text{time}^{-1}$  \\ 
$\sigma$ & Biting rate of mosquito & $\text{time}^{-1}$  \\
$\lambda_{A\hspace{-1pt}R}$ & Asymptomatic human recovery rate   & $\text{time}^{-1}$   \\
$\lambda_{Y\hspace{-1pt}R}$ & Symptomatic human recovery rate   & $\text{time}^{-1}$   \\
$\lambda_{RS}$ & Temporary immunity loss rate in humans & $\text{time}^{-1}$  \\
$u(\varepsilon)$ & Exposure dependent NAI protected proportion & $\text{time}^{-1}$  \\
$u_{_{\low}\hspace{5pt}}$,$u_{_{\high}}$ & Lower and upper NAI protected thresholds & $\text{time}^{-1}$  \\
[1ex]
\hline
\end{tabular}
\label{table:nonlin}
\end{table}}
%----------------------------
%----------------------------
\begin{Remark}
The parameter value labeled $\phi$ in Figure 6 on \citep[p.~2575]{FLIPE} corresponds to the term $(1-u_{_{\low}\hspace{5pt}})$ in the SEYAR model formulation, where the baseline value is assumed to be $0.5$. The naturally acquired immune proportion will increase or decrease, depending on the rate that the human population is being exposed. Hypothetically, as the rate of exposure increases or decreases, this proportion should grow or shrink up to a threshold value. This epidemiological behavior is quantified by the solution to the IVP (\ref{NAI_IVP}). Assuming  $u_{_{\low}\hspace{5pt}}=0.5$, as above, and that it is possible for at most ninety percent of the population to acquire sufficient protection through means of natural exposure, i.e. $u_{_{\high}\hspace{5pt}}=0.9$. Then, the initial exposed proportion $\varepsilon_0$ can be assumed to take any value in the interval $\left[0,\ln\left(\frac{u_{_{\high}\hspace{5pt}}}{u_{_{\high}\hspace{5pt}}-u_{_{\low}\hspace{5pt}}}\right)\right]=[0,\ln\left(\frac{9}{4}\right)] \approx [0,0.81]$. In other words, one can assume at most $81\%$ of the human population to be initially exposed. On the other hand, if one modifies the above assumptions so that the initial naturally acquired immune proportion is $u_{_{\low}\hspace{5pt}}=0.5$, then 
$\varepsilon_0 \in [0,\ln\left(\frac{9}{8}\right)] \approx [0,0.12]$, so that at most $12\%$ of the human population can be assumed to be initially exposed.
\end{Remark}
%----------------------------

\section{Model Analysis}
\label{sec:ModelAnalysis}
%PRELIMINARY ANALYSIS OF SYSTEM
\subsection{Well-Posedness and Feasible Region} \label{sec:WellPosedness}
%----------------------------
Although assuming that $\Phi \in C^1$ provides sufficient regularity to ensure that system (\ref{DS_rewrite}) is well-posed, this work is primarily concerned with the stability of the system near equilibrium points. This requires additional regularity assumptions on the vector field $\Phi$ in order to invoke a variation of the Center Manifold Theorem, cf. \cite{CASTSONG}. Consequently, from now on it is necessary to assume that $\Phi \in C^2\subset C^1$, i.e. it is at least twice-continuously differentiable. Moreover, to be reasonable in an epidemiological sense, the functions under consideration should posses a bounded first derivative, i.e. they should be members of the class $C^1_b(\mathbb{R}_+)$. From this point on, the model is studied in the more regular (smaller) function space $C^2(\mathbb{R}^8_+) \cap C^1_b(\mathbb{R}^8_+)$. In light of the mathematical and epidemiological well-posedness of the IVP (\ref{NAI_IVP}) and structure of the underlying vector field, the additional $C^2$ regularity will be inherited by the solution $\mathbf{x}$. 

To this end, let $\mathbf{x}=\left(S,E,Y,A,R,M_S,M_E,M_I\right)^T$, so that $x_i$ is the $i^\text{th}$ component of the 8-dimensional vector $\mathbf{x} \in C^2(\mathbb{R}^8_+) \cap C^1_b(\mathbb{R}^8_+)$ and rewrite (\ref{SEYAR_DS}) in the following compact form:
%----------------------------
\begin{equation}
	\begin{cases}
		\dot{\mathbf{x}}(t)=\Phi \left(\mathbf{x}(t)\right), \\
		\mathbf{x}(0)=\mathbf{x}_0.
	\end{cases}
		\label{DS_rewrite}
\end{equation}
%################
%----------------------------
%################
\begin{theorem}{(Existence Theory of the $SEY \hspace{-3pt}AR$ Model).}
There exists a sufficiently regular unique solution $\mathbf{x}$ to the SEYAR model IVP (\ref{DS_rewrite}) that can be continued to a maximal time interval. Additionally, $\mathbf{x}$ depends continuously on the initial data $\mathbf{x}_0$ and model parameters involved. 
\label{thm:SEYARlwp}
\end{theorem}
%################
%----------------------------
%################
%----------------------------
The dynamics of the total population are given by the following decoupled system
%----------------------------
\begin{equation}
		\begin{cases}
			\dot{N}_H = \Omega_H-\xi_H N_H-\delta Y,\\
			\dot{N}_M =\Omega_M-\xi_M N_M.\\
		\end{cases} 
			\label{Total}
\end{equation}
%----------------------------
Due to the non-homogeneous term $\delta Y$, the asymptotic behavior of the human population is more delicate. For the human population, we have the following theorem which provides a tighter lower bound on the attracting region for the model. In the absence of infection, the long-time behavior of $N_H$ is trivial, i.e. the total human population converges to the equilibrium population density, as in the mosquito population. In the case of an infectious disease, one would expect the population to converge to a smaller quantity, as there is a disease induced death rate $\delta$ adding to the natural death rate $\xi_H$. In this case, the long-time population size will be smaller, as it has to account for the additional disease-induced deaths suffered by symptomatic individuals. Listed below is the theorem concerning the feasible region of model (\ref{SEYAR_DS}).
%################
%----------------------------
%################
\begin{theorem}{(Feasible Region of the $SEY \hspace{-3pt}AR$ Model).}
Let $\left(N_H, N_M\right)$ be the solution of system (\ref{Total}) emanating from Theorem \ref{thm:SEYARlwp}, with corresponding initial data $\left( N_H(0), N_M(0)\right) \geq 0$. Define the following compact sub-space 
%----------------------------
\begin{equation*}
\Gamma:=\left\lbrace \mathbf{x} \in C^2(\mathbb{R}^8_+) \cap C^1_b(\mathbb{R}^8_+) : N_H \in \left[ \frac{\Omega_H - \delta \| Y \|_\infty }{\xi_H} ,\frac{\Omega_H}{\xi_H}  \right ], N_M = \frac{\Omega_M}{\xi_M}\right\rbrace. 
\end{equation*}
%----------------------------
Then, $\Gamma$ is a forward invariant attractor for system (\ref{Total}).
\label{thm:Reg}
\end{theorem}
%################
%----------------------------
%################
Mathematically speaking, Theorem (\ref{thm:Reg}) reduces the complexity of the analysis involved regarding the long-term dynamics of the system by 
allowing the replacement of a potentially unbounded (epidemiologically unreasonable) space with the smaller (epidemiologically reasonable) compact sub-space $\Gamma$.
If we let $L_t=e^{\small{\xi_H t}}$ denote an exponential multiplier, then the solution for the total dynamics of the human population is given by 
%----------------------------
\begin{equation*}
		N_H(t)=L_{-t}N_H(0)+ \frac{\Omega_H }{\xi_H}\left(1-L_{-t}\right) -\delta L_{-t} \ast Y(t). 
\end{equation*}
%----------------------------
The instantaneous rate of occurrence of death, i.e. the force of mortality $\xi_H$ is assumed to be constant. As a result, the probabilities of living and dying up to $t$ days are given by $e^{-\xi_Ht}$ and $\left(1-e^{-\xi_Ht}\right)$, respectively. Assuming that $N_H(0)=0$, the above solution says that the total population $N_H(t)$ at time $t$ is given by the weighted product of the carrying capacity  $\frac{\Omega_H }{\xi_H}$ and the distribution of humans that are left after those that have died due to natural causes $\left(1-e^{\small{-\xi_H t}}\right)$, minus a weighted average of humans that have died due to symptomatic infections. The later quantity is captured by the non-homogeneous forcing term $-\delta L_{-t} \ast Y(t)$, given by a convolution with the inverse multiplier. Since convolution is a smoothing operation, this emphasizes the fact that we are subtracting a "smoothing average" over past time of the humans which have died from symptomatic infections. A straight-forward calculation yields the following differential inequality
%----------------------------
\begin{equation*}
			\dot{N}_H \leq 0 ,\quad \text{if} \quad N_H \geq \frac{\Omega_H}{\xi_H}.
\end{equation*}
%----------------------------
From an epidemiological view point, the above inequalities imply that if the total population $N_H$ breaches its carrying capacity, then the weighted average of fatal symptomatic infections must increase to stabilize the population back to a healthy level. 
%----------------------------
\begin{Remark}
The main feature of the above analysis is that it provides a tighter lower bound for the feasible region corresponding to the dynamical system. In an epidemiological setting, one can define the term $\delta \| Y \|_\infty$ to be the maximum disease-impact on the human population. In previous variants of malaria models appearing in literature, e.g. SIR, SIER, SIERS, etc., the quantity $0$ is listed as the lower bound, however this is unreasonable since the populations under consideration usually do not go extinct, unless $\Omega_H=\delta \| Y \|_\infty$. If the maximum impact the disease is capable of having on the population is less than the recruitment rate, then their will always be accumulation over long-time. For other members of the SIR model class, the lower bound would be the same except with $Y$ replaced by $I$. In the case of an infectious disease, there will be additional disease induced deaths, so that the total human population will not converge to the equilibrium population density. In reality, the total human population will converge to a quantity trapped in the interval $\left( \frac{\Omega_H - \delta \| Y \|_\infty }{\xi_H} ,\frac{\Omega_H}{\xi_H}  \right )$.
\end{Remark}
%----------------------------

%REPRODUCTIVE NUMBER AND DFE STABILITY ANALYSIS
\subsection{Reproductive Threshold and Disease-Free Equilibrium}\label{sec:Rnumber}
%----------------------------
This section is focused on deriving a threshold value that characterizes the stability of the underlying dynamical system (\ref{SEYAR_DS}). This value, called the reproductive threshold, provides a way to estimate the reduction in transmission intensity required to eliminate malaria through vector-based control \cite{smith2007revisiting}. The basic reproductive number $\mathcal{R}_0$ corresponding to a given model is a threshold value which represents the average amount of new infections produced by a typical infectious individual in a completely susceptible population, at a disease-free equilibrium. This quantity is equal to the reproductive threshold for a class of simplified population models. However, in the case of more complicated dynamical systems $\mathcal{R}_0$ is a threshold which determines whether the disease will eventually, over long time, die out or persist and become an epidemic. Thus, in theory this threshold value does not necessarily represent the average number of secondary infections in a given population. This fact does not undermine the extreme importance of this threshold, as we will see that the size of this particular value has grave repercussions for the region under consideration.

Disease-Free Equilibrium (DFE) points are solutions of a dynamical system corresponding to the case where no disease is present in the population. Define the diseased classes to be $E,Y,A,M_E$ and $M_I$. Notice that $R$ is not considered to be a diseased class, as the asymptomatic class $A$ has been effectively removed, cf. \cite{ROOP}. As a result, individuals in the $R$ compartment are considered to be temporarily immune but not infectious. After determining the DFE of system (\ref{SEYAR_DS}), this threshold value is used to address its local asymptotic stability. Upon equating the right-hand side of (\ref{SEYAR_DS}) to zero and solving, we arrive at the following unique DFE
%----------------------------
\begin{equation*}
		\mathbf{x}_{\small{dfe}}=\left(\frac{\Omega_H}	{\xi_H},0,0,0,0,\frac{\Omega_M}{\xi_M},0,0\right)^T.
	\label{DFE}
\end{equation*}
%----------------------------
%################
%----------------------------
%################
\begin{lemma}{(Local Asymptotic Stability of the DFE for the $SEY \hspace{-3pt}AR$ Model).}
Define the following quantity 
\begin{equation}
		\mathcal{R}_0 :=\sqrt{\frac{\sigma^2 \tau \gamma \Omega_M \xi_H \beta_M}{\xi_M^2(\gamma + \xi_H)(\tau + \xi_M)\Omega_H} \left(\frac{\beta_AU_{_{\low}\hspace{5pt}} }{\lambda_{AR} + \xi_H} - \frac{\beta_Y(U_{_{\low}}-1)}{\lambda_{YR} + \xi_H + \delta} \right)},
	\label{RnumSEYAR} 
\end{equation}
where $U_{_{\low}\hspace{2pt}}:=e^{\varepsilon_0}(u_{_{\low}}-u_{_{\high}\hspace{5pt}}) + u_{_{\high}\hspace{5pt}}$. Then, the DFE $\mathbf{x}_{\small{dfe}}$ for the SEYAR model (\ref{SEYAR_DS}) is locally asymptotically stable provided that $\mathcal{R}_0<1$ and unstable if $\mathcal{R}_0>1$.
\label{thm:LAS}
\end{lemma}
%################
%----------------------------
%################
%----------------------------
Lemma (\ref{thm:LAS}) is proven by utilizing the next generation method covered in \cite{VDW}. The threshold value (\ref{RnumSEYAR}) has major epidemiological implications on the underlying dynamical system (\ref{SEYAR_DS}). To gain a deeper insight into the qualitative information encoded in this important quantity, we decompose it in the form of an epidemiological meaningful product in order to analyze each factors contribution
%----------------------------
\begin{align*}
		\mathcal{R}_0 &=\sigma\sqrt{\frac{\Omega_M}{\Omega_H}}\sqrt{\frac{\tau}{\tau + \xi_M}}\sqrt{\frac{\gamma}{\gamma + \xi_H}}\sqrt{\frac{\xi_H}{\xi_M}}\sqrt{\frac{\beta_M}{\xi_M}}\sqrt{\frac{\beta_AU_{_{\low}\hspace{5pt}} }{\lambda_{AR} + \xi_H} - \frac{\beta_Y(U_{_{\low}}-1)}{\lambda_{YR} + \xi_H + \delta}}, \\
		&:=\sigma\prod_{i=1}^6 \sqrt{r_i}.
\end{align*}
%----------------------------
Due to the above lemma and in an epidemiological setting it is desirable to have the reproductive threshold below unity. Listed below is the size contribution and biological description for each of the factors. The first factor is $\sigma$, which stands for the man-biting rate. This factor is usually much less than unity due to the fact that female anopheline mosquitoes generally transmit fewer than 100 \textit{sporozoites} per bite, \cite{Ponnud}. As malaria is a mosquito borne disease, the agent \textit{Plasmodium} will spread at a much slower rate provided less vectors are introducing it into human hosts. Owning to the monotonicity of the square root function, it is sufficient focus on the size of each $r_i$.
%----------------------------
%----------------------------
\begin{enumerate}[label=\roman*] 
	\item The term $r_1=\frac{\Omega_M}{\Omega_H}$ is the ratio of the mosquito and human recruitment rates. Clearly, this quotient is greater than unity due to fact that in any given population there will be more mosquitoes than human hosts. As the human and mosquito recruitment rates rank high on the sensitivity hierarchy of many epidemic models appearing in the literature, it is no surprise that this term is problematic with respect to the overall size of the threshold. 	
	\item The term $r_2=\frac{\tau}{\tau + \xi_M}$ is the mosquito latent period $\tau$ divided by itself plus the mosquito mortality rate $\xi_M$. This quantity is bounded above by one and is monotonically decreasing with respect to $\xi_M$. It follows that larger the mosquito death rate is, the smaller $r_2$ will be.
	\item The first fully human-dependent term $r_3=\frac{\gamma}{\gamma + \xi_H}$, is comprised of the human latent $\gamma$ period divided by itself plus the human mortality rate $\xi_H$. As in the case of $r_2$, this quantity is always less than one and monotonically decreases with respect to the human mortality rate. This is consistent with the fact that the fewer hosts there are for the parasite to invade, the less infectives will arise. However, since increasing $\xi_H$ is not practical, this terms offers no control over $\mathcal{R}_0$.
	\item The term $r_4=\frac{\xi_H}{\xi_M}$ is the ratio of the human and mosquito death rates. This particular quantity is always less than one as the mosquito death rate is much higher than the human death rate. 
	\item The term $r_5=\frac{\beta_M}{\xi_M}$ is the ratio of the mosquito to human transmission probability $\beta_M$ and the mortality rate of the mosquito population $\xi_M$. This quantity will be less than one provided $\beta_M<\xi_M$, i.e. the transmission probability is less than the death rate. In the case of a population with a relatively high vector transmission probability, then the vector death rate must be large enough to make $r_5$ less than one. This implies a restriction on the size of $\beta_M$. It will be shown in Section \ref{sec:EE} that if $\beta_M$ breaches a certain threshold, then sub-threshold endemic equilibria can emerge. 
	\item The second fully human-dependent term is given by the following equation
	\begin{equation*}
	r_6=\frac{\beta_AU_{_{\low}\hspace{5pt}}}{\lambda_{AR} + \xi_H} - \frac{\beta_Y(U_{_{\low}}-1)}{\lambda_{YR} + \xi_H + \delta}.
\end{equation*}
The quantity, $r_6$ is a difference of ratios consisting of asymptomatic and symptomatic vital dynamics, along with transmission, recovery and disease-induced death rates; weighted with a distribution consisting of the lower and upper NAI-rate threshold of the human population scaled by an exponentiation of the initial exposure rate. For simplicity of exposition, we assume that the initial exposed proportion is zero, i.e. $\varepsilon_0=0$, so that $U_{_{\low}\hspace{3pt}}=u_{_{\low}\hspace{5pt}}$. If the initial exposed proportion is not equal to zero, then initially, $U_{_{\low}\hspace{3pt}}<u_{_{\low}\hspace{5pt}}$ and the reproductive threshold will be larger under parameter configurations to be specified shortly. Under such a configuration, upon initial exposure there will be more symptomatic individuals, but as exposure increases, less humans will die as naturally acquired immunity will begin to develop in the overall population. However, the following discussion is unaffected by this minor detail. As a result, the human-dependent factor is taken to be
\begin{equation*}	
r_6=\frac{u_{_{\low}\hspace{5pt}} \beta_A}{\lambda_{AR} + \xi_H} - \frac{(u_{_{\low}}-1) \beta_Y}{\lambda_{YR} + \xi_H + \delta}.
\end{equation*}	
Denote the low and high thresholds be such that $u_{_{\low}\hspace{5pt}}>0$ and $u_{_{\high}\hspace{5pt}}<1$, respectively and define the following compact sub-interval $T_u := [u_{_{\low}\hspace{5pt}},u_{_{\high}\hspace{5pt}}] \subset (0,1)$ consisting of various NAI protected proportions corresponding to a given population. When subjected to a certain parameter restriction, the sizes of the quantities $u_{_{\low}}=u(0)$ and $r_6$ are inversely related, i.e. the larger $u_{_{\low}\hspace{5pt}}$ is, the smaller $r_6$ will be; resulting in a relatively smaller $\mathcal{R}_0$. Hence, an additional way to control the size of $\mathcal{R}_0$ arises, provided the parameters are such that this inequality restriction, to be mentioned below, holds. However, as we will see, if the parameters are such that this inequality reversed, then they are directly related.

Let $C_0:=\sigma\prod_{i=1}^5 \sqrt{r_i}$, $C_1:=\frac{\beta_A}{\lambda_{AR} + \xi_H}$ and $C_2:=\frac{\beta_Y}{\lambda_{YR} + \xi_H + \delta}$ and define $T \subset \mathbb{R}^+ \cup \{+\infty \}$ to be an ordered subset of the non-negative extended real numbers. In addition, let $u$ solve equation (\ref{Irate}) and denote $\{u(t) \in T_u| t \in T \}$ to be the set of NAI protected proportions experienced by a population over a prescribed, possibly infinite, time interval indexed with $t$ and consider the following function 
%----------------------------
\begin{equation*}
		\mathcal{R}_0(u(t)) :=C_0 \sqrt{(C_1-C_2) u(t)+ C_2}.
\end{equation*}
%----------------------------
As a result, the type of monotonicity obeyed by $\mathcal{R}_0(u(t))$ is dependent on the sign of the non-zero combination of parameters $C_1-C_2$. These observations motivate the following definition which classifies the configuration space of the model based on how the asymptotic dynamics of the corresponding population responds with respect to the NAI accumulated in response to the rate of exposure.
	\label{RnumberDecomposition}
\end{enumerate}
%----------------------------
\begin{definition}{(Configuration Space).}
The $SEY \hspace{-3pt}AR$ model (\ref{SEYAR_DS}) is said to possess a \textit{$Y$-dominant configuration} if $C_1-C_2>0$ and an \textit{$A$-dominant configuration} provided that $C_1-C_2<0$. Upon the trivial case that $C_1-C_2=0$, the system is said to have a \textit{Null-configuration}. Additionally, we refer to a system possessing such a configuration as either \textit{$A$-}, \textit{$Y$-} or \textit{Null-configured}. A given human population is called \textit{$A$-},\textit{$Y$-}, or \textit{Null-dominate} provided that its corresponding configuration is.
	\label{ConfigDef}
\end{definition}
%----------------------------
It is worth mentioning that although $r_6$ is always positive, the combination of parameters $C_1-C_2$ need not be. The positiveness of $r_6$ is a result of how the transmission probabilities are weighted by the NAI protected proportion. If the system is \textit{Null-configured}, then $\mathcal{R}_0(u(t)) :=C_0 \sqrt{C_2}$ and it is locally asymptotically stable provided that $C_0<1$ and $\beta_Y<\lambda_{YR}+\xi_H+\delta$, i.e. a high vector death rate and the symptomatics are recovering or dying out at a faster rate than they are transmitting. Consider the case of an \textit{$A$-dominant configuration}, i.e.
%----------------------------
\begin{equation}
\frac{\beta_A}{\beta_Y} < \frac{\lambda_{AR} + \xi_H}{\lambda_{YR} + \xi_H + \delta}.
	\label{ine}       
\end{equation}
%-----------------------------
As mentioned in Section \ref{sec:Methods}, in the formulation of the $SEY \hspace{-3pt}AR$ model (\ref{SEYAR_DS}) we do not assume an ordering on the asymptomatic and symptomatic transmission probabilities $\beta_A$ and $\beta_Y$, respectively. However, under the assumption that asymptomatic carriers transmit at a lower rate than that of symptomatic, the left hand side of the above inequality is less than unity. Moreover, if we additionally assume that asymptomatic individuals recover faster than symptomatic, provided the disease induced death rate $\delta$ and recovery rates are such that $\lambda_{AR}>\lambda_{YR} + \delta$, then the right hand side of the above inequality is greater than unity and inequality (\ref{ine}) is satisfied. In an epidemiological setting, such a configuration corresponds to holoendemic regions across sub-Saharan Africa \cite{SubSah} in which the majority of people are continuously infected by \textit{P. falciparum}, but only a small proportion display clinical symptoms. The high level of naturally acquired immunity present in the population allows them to live their daily lives feeling healthy despite a relatively high blood-parasite density, \cite{AcqIM}. 

Analytically speaking, in the case of an \textit{$A$-dominant configuration}, provided that the mosquito mortality rate $\xi_M$ can be made large enough so that the term $C_0$ compromised of fractional multipliers is sufficiently small, then $\mathcal{R}_0(u_t)$ achieves its maximum $\mathcal{R}_0(u_{_{\low}\hspace{5pt}})$ at the low NAI threshold $u_{_{\low}\hspace{5pt}}$ and decreases to its infimum $\mathcal{R}_0(u_{_{\high}\hspace{5pt}})$ as the high NAI threshold $u_{_{\high}\hspace{5pt}}$ is approached. Moreover, since the ordered set $T$ is a subset of a separable metric space, we can extract an ordered countable subset and form the partition $\{u_{_{\low}\hspace{5pt}}=u(t_1)<\cdots <u(t_n)<u(t_{n+1})<\cdots<u_{_{\high}\hspace{5pt}} \}$ of the compact sub-interval $T_u$. In this case, the corresponding values of $\mathcal{R}_0(u(t))$ obey the following descending order $\{\mathcal{R}_0(u_{_{\low}\hspace{5pt}}) >\cdots>\mathcal{R}_0(u(t_n))>\mathcal{R}_0(u(t_{n+1}))>\cdots>\mathcal{R}_0(u_{_{\high}\hspace{5pt}}) \}$. This is consistent with the fact that as a given population acquires natural immunity through exposure, the disease will start to spread at a slower rate. Conversely, if the system has a \textit{$Y$-dominant configuration}, i.e. $C_1-C_2>0$, then the monotonicity is reversed.

In mathematical terminology, provided $\sigma<1$, one can always find a large enough $\xi_M$ to make $C_0<1$. In this case, the factor that will cause $\mathcal{R}_0$ to breach unity is $r_6$. From an epidemiological standpoint, regardless of the size of vector transmission probability $\beta_M$ or man-biting rate $\sigma$, if enough mosquitoes are dying to significantly slow the disease transmission dynamics, then $C_0<1$ and, as a result, the size of the reproductive threshold $\mathcal{R}_0$ will be determined by $r_6$ which depends on the human immune systems response to the \textit{Plasmodium} parasite. This attest to the fact that in such vector transmitted diseases, the reproductive threshold will lower in response to vector elimination up to a point and the factor allowing the disease to persist under such low vector activity lies in the intricate relationship between the parasite and host. This delicate relationship is captured by the term $r_6$.
%----------------------------
\begin{Remark}
A verification of the reproductive threshold $\mathcal{R}_0$ (\ref{RnumSEYAR}) arising from Lemma (\ref{thm:LAS}) above is provided in the electronic supplementary material.
\end{Remark}
%----------------------------

%THE IMAPCT OF THE ASYMPTOMATIC CLASS ON THE REPRODUCTIVE THRESHOLD
\subsection{The Impact of the Asymptomatic Class on the Reproductive Threshold}\label{sec:RnumberComparison}
%----------------------------
It would be informative to investigate how the reproductive threshold arising from the $SEY \hspace{-3pt}AR$ model behaves in the case that the long time dynamic effects of asymptomatic carriers are not taken into account. The natural parameter space of the $SEY \hspace{-3pt}AR$ model corresponding to the reproductive threshold is
%----------------------------
\begin{equation*}
\Theta:=\left\lbrace (\Omega_H,\Omega_M,\xi_H,\xi_M,\beta_A,\beta_Y,\beta_M,\gamma,\tau,\delta,\sigma,\lambda_{A\hspace{-1pt}R},\lambda_{Y\hspace{-1pt}R},U_{_{\low}\hspace{5pt}}) \in \mathbb{R}^{14}_{\tiny{>0}}\right\rbrace,
\end{equation*}
%----------------------------
where $\mathbb{R}^n_{\tiny{>0}}:=\{x \in \mathbb{R}^n \hspace{3pt}:\hspace{3pt} x_i>0 \hspace{3pt}\text{for}\hspace{3pt} i=1,\cdots,n\}$. The set $\Theta$ consists of all possible positive, epidemiologically reasonable, parameter values for the $SEY \hspace{-3pt}AR$ model in which the reproductive threshold depends upon. Although the disease induced death rate $\delta$ is allowed to be non-negative, the analysis presented in this section is unaffected by this minor detail. Under this formalization, neglecting the disease transmission and recovery rates of asymptomatic human hosts on the reproductive threshold formally corresponds to restricting the model to the following \textit{$A$-nullified} parameter configuration space $\tilde{\Theta}$, defined as 
%----------------------------
\begin{equation*}
\tilde{\Theta}:=\left\lbrace (\Omega_H,\Omega_M,\xi_H,\xi_M,\beta_Y,\beta_M,\gamma,\tau,\delta,\sigma,\lambda_{Y\hspace{-1pt}R},U_{_{\low}\hspace{5pt}}) \in \mathbb{R}^{12}_{\tiny{>0}}: \beta_A=\lambda_{A\hspace{-1pt}R}=0\right\rbrace. 
\end{equation*}
%----------------------------
Consider a typical element $\Theta_0 \in \Theta$ listed below
%----------------------------
\begin{equation*}
\Theta_0=\left(\Omega_{H_0},\Omega_{M_0},\xi_{H_0},\xi_{M_0},\beta_{A_0},\beta_{Y_0},\beta_{M_0},\gamma_{0},\tau_{0},\delta_0,\sigma_0,\lambda_{{A\hspace{-1pt}R}_0},\lambda_{{Y\hspace{-1pt}R}_0},U_{{_{\low}\hspace{5pt}\hspace{-1pt}}_{_0}}\right).
\end{equation*}
%----------------------------
In a similar fashion, the dual element $\tilde{\Theta}_0 \in \tilde{\Theta}$ corresponding to the symptomatic class $Y$ is given by
%----------------------------
\begin{equation*}
\tilde{\Theta}_0=\left(\Omega_{H_0},\Omega_{M_0},\xi_{H_0},\xi_{M_0},\beta_{Y_0},\beta_{M_0},\gamma_{0},\tau_{0},\delta_0,\sigma_0,\lambda_{{Y\hspace{-1pt}R}_0},U_{{_{\low}\hspace{5pt}\hspace{-1pt}}_{_0}}\right). 
\end{equation*}
%----------------------------
It is worth mentioning that the dual element $\tilde{\Theta}_0$ is comprised of the same fixed parameter configuration as described by $\Theta_0$, but with the asymptomatic progression rates specified above set equal to zero. To emphasize the asymptotic dynamic influence of the asymptomatic class $A$ on the reproductive threshold $\mathcal{R}_0$ (\ref{RnumSEYAR}) arising from the $SEY \hspace{-3pt}AR$ model  (\ref{SEYAR_DS}), subject to a given fixed parameter configuration, the following notation is employed $\mathcal{R}_A:=\mathcal{R}_0\Big |_{\Theta_0}$. By denoting $\mathcal{R}_Y:=\mathcal{R}_0\Big |_{\tilde{\Theta}_0}$, the size relationship between the two quantities $\mathcal{R}_A$ and $\mathcal{R}_Y$ is captured below in the following theorem. 
%################
%----------------------------
%################
\begin{theorem}{(Impact of the Asymptomatic Class on the Reproductive Threshold).}
Let $\mathcal{R}_0$ be the threshold quantity (\ref{RnumSEYAR}) arising from Lemma (\ref{thm:LAS}) and consider the following fixed parameter configuration vectors $\Theta_0 \in \Theta$ and $\tilde{\Theta}_0 \in \tilde{\Theta}$, corresponding to the $SEY \hspace{-3pt}AR$ model (\ref{SEYAR_DS}), defined as above.
Denote $\mathcal{R}_A:=\mathcal{R}_0\Big |_{\Theta_0}$ and $\mathcal{R}_Y:=\mathcal{R}_0\Big |_{\tilde{\Theta}_0}$, then it follows that $\mathcal{R}_Y<\mathcal{R}_A$.
\label{thm:R_Comp}
\end{theorem}
%################
%----------------------------
%################
%---------------------------------------------------------------------------------
One should observe that the above inequality is strict. Therefore, neglecting to account for asymptomatic carriers results in an underestimation of the reproductive threshold. To demonstrate the theoretical estimate provided in Theorem (\ref{thm:R_Comp}), the numerical values of $\mathcal{R}_A$ and $\mathcal{R}_Y$, along with the Entomological Inoculation Rate ($EIR$) and parameter configuration space classifications, introduced via definition (\ref{ConfigDef}) in Section \ref{sec:Rnumber} are presented in Table \ref{table:ThrQ} of Section \ref{sec:NumRes} below. These numerical values correspond to the following three high transmission sites: Kaduna in Nigeria, Namawala in Tanzania, and Butelgut in Papua New Guinea. The parameter values associated with each site are listed in Section \ref{sec:Pvalues}.

Previously, it was shown that $\mathcal{R}_0$ can be written in the following form
%----------------------------
\begin{equation}
		\mathcal{R}_0=\sigma\prod_{i=1}^6 \sqrt{r_i},
\label{RDecomp}
\end{equation}
%----------------------------
where each term $\sqrt{r_i}$ is defined as in Section \ref{sec:Rnumber}. Care needs to be taken when arbitrarily substituting zero parameter values into the reproductive threshold $\mathcal{R}_0$ (\ref{RDecomp}). Clearly, if the man-biting rate $\sigma=0$ or mosquito to human transmission probability $\beta_M=0$, then it follows that $\mathcal{R}_0=0$. These quantities effectively nullifying the reproductive threshold stands to epidemiological reason and corresponds to the following scenarios, respectively: $(\romannumeral 1)$ no mosquitoes are biting humans, and $(\romannumeral 2)$ they are not transmitting the disease. Furthermore, if both the asymptomatic $\beta_A$ and symptomatic $\beta_Y$ transmission probabilities are equal to zero, then the reproductive threshold will be identically zero, as infected humans are not transmitting the disease to susceptible mosquitoes. 

However, it is implied that while one (or possibly all) of such parameter values may be equal to zero, the other parameter values in which the threshold depends on will be such that it is well-defined. This problem is primarily due to the inclusion of vital dynamics for the human and mosquito populations into the dynamical system (\ref{SEYAR_DS}). For example, if one lets the human recruitment rate $\Omega_H=0$ or the mosquito mortality rate $\xi_M=0$, then a singularity occurs and $\mathcal{R}_0$ ceases to be well-defined. Although, the mosquito mortality rate being equal to zero in unreasonable in an epidemiological sense, as all mosquitoes experience death, it is informative to study how this parameters size effects $\mathcal{R}_0$, as in general it is desirable to increase such a parameter when introducing control measures into the system. Moreover, it is worth mentioning from a mathematical standpoint, as it attests to the subtle fact that after including vital dynamics into a given compartmentalized infectious disease model, one can not expect to obtain qualitative information about the reproductive threshold in the absence of vital dynamics by simply setting the corresponding terms equal to zero. To properly study the asymptotic behavior of such models without vital dynamics included, one would have to go back to the original model derivation and not include them from the beginning, then proceed to calculate the threshold arising from the modified system.

The formal approach corresponding to the configuration space presented above ensures that this does not happen and adds another level of mathematical precision to the analysis presented in this work. Some other noteworthy consequences of the above formal approach related to the configuration space of the $SEY \hspace{-3pt}AR$ model (\ref{SEYAR_DS}) are listed below. 
%----------------------------
\begin{enumerate}[label=\roman*] 	
	 \item Consider the following fixed parameter configuration vector $\Theta_1\in \mathbb{R}^{13}_{\tiny{>0}}$, defined by
%----------------------------
\begin{equation*}
\Theta_1:=\left(\Omega_{H_0},\Omega_{M_0},\xi_{H_0},\beta_{A_0},\beta_{Y_0},\beta_{M_0},\gamma_{0},\tau_{0},\delta_0,\sigma_0,\lambda_{{A\hspace{-1pt}R}_0},\lambda_{{Y\hspace{-1pt}R}_0},U_{{_{\low}\hspace{5pt}\hspace{-1pt}}_{_0}}\right).
\end{equation*}	 
%----------------------------
Then, by equation (\ref{RDecomp}) and the fact that the square root function is uniformly continuous on $[0,+\infty)$, (so that the limit can be taken inside), it follows that 
%---------------------------------------------------------------------------------
\begin{align*}
		\lim_{\xi_M \rightarrow +\infty} \mathcal{R}_0\Big |_{\Theta_1} &=\sigma\sqrt{r^0_1r^0_3r^0_6}\cdot \sqrt{\lim_{\xi_M \rightarrow +\infty}\frac{\tau_{0}\xi_{H_0}\beta_{M_0}}{\xi^2_M\left(\tau_{0} +\xi_M \right)}}, \\
		                &=\sigma\sqrt{r^0_1r^0_3r^0_6} \cdot 0,\\
		                 &=0,\\			
\end{align*}
%---------------------------------------------------------------------------------
where the superscripts appearing above each appropriate $i^\text{th}$ term denote that fact that these quantities are fixed and thus invariant under the limit operation. 

Due to the equivalency of the $\epsilon-\delta$ and sequential definitions of limits, there exist a natural number $\xi_{M_{n_0}}\in \mathbb{N}$ such that for all $\xi_{M_n}\geq \xi_{M_{n_0}}$, it follows that $\mathcal{R}_0 \in [0,1)$. This qualitative observation can be interpreted as follows: provided a scenario such that all of the associated model parameters are fixed epidemiologically reasonable quantities, if control measures sufficient to increase the vector death rate are introduced into the model, then the corresponding reproductive threshold arising from the model will be lowered and eventually fall below unity. This is consistent with the discussion in Section \ref{sec:Rnumber}.
	\item In a similar fashion, consider the following fixed vectors $\left(\Theta_2, \Theta_3\right) \in \mathbb{R}^{13}_{\tiny{>0}} \times \mathbb{R}^{13}_{\tiny{>0}}$, defined as follows
%----------------------------
\begin{equation*}
\Theta_2:=\left(\Omega_{H_0},\xi_{H_0},\xi_{M_0},\beta_{A_0},\beta_{Y_0},\beta_{M_0},\gamma_{0},\tau_{0},\delta_0,\sigma_0,\lambda_{{A\hspace{-1pt}R}_0},\lambda_{{Y\hspace{-1pt}R}_0},U_{{_{\low}\hspace{5pt}\hspace{-1pt}}_{_0}}\right)
\end{equation*}	 
%----------------------------	
and 
%----------------------------
\begin{equation*}
\Theta_3:=\left(\Omega_{M_0},\xi_{H_0},\xi_{M_0},\beta_{A_0},\beta_{Y_0},\beta_{M_0},\gamma_{0},\tau_{0},\delta_0,\sigma_0,\lambda_{{A\hspace{-1pt}R}_0},\lambda_{{Y\hspace{-1pt}R}_0},U_{{_{\low}\hspace{5pt}\hspace{-1pt}}_{_0}}\right).
\end{equation*}	 
%----------------------------
Then it follows that 
%---------------------------------------------------------------------------------
\begin{align*}
		\lim_{\Omega_M \rightarrow +\infty} \mathcal{R}_0\Big |_{\Theta_2} &=\sigma\prod_{i=2}^6 \sqrt{r^0_i}\cdot\sqrt{\lim_{\Omega_M \rightarrow +\infty}\frac{\Omega_M}{\Omega_{H_0}}}, \\
		                &=+\infty,\\			
\end{align*}
%---------------------------------------------------------------------------------	
and 	
%---------------------------------------------------------------------------------
\begin{align*}
		\lim_{\Omega_H \rightarrow 0^+} \mathcal{R}_0\Big |_{\Theta_3} &=\sigma\prod_{i=2}^6 \sqrt{r^0_i}\cdot\sqrt{\lim_{\Omega_H \rightarrow 0^+}\frac{\Omega_{M_0}}{\Omega_H}}, \\
		                 &=+\infty.\\			
\end{align*}
%---------------------------------------------------------------------------------	
As pointed out in Section \ref{sec:Rnumber}, the factor $r_1$ is always greater than unity. The above analysis demonstrates how $\mathcal{R}_0$ behaves with respect to the human and mosquito recruitment rates. As we will see in Section \ref{sec:SenAna}, if a given population, e.g. the three sites from which the parameter values are taken for this work, has a relatively small human recruitment rate and relatively large mosquito recruitment rate, then the factor $r_1$ will dramatically contribute to the size of $\mathcal{R}_0$. For example, in the case of the Kaduna site $r_1\approx 14762941.18$. The fractional multipliers $r_i$ for $i=2,3,4$ will reduce the resulting size of this quantity, as they are all strictly less than unity. Further reduction in the size of the remaining quantity in the decomposition of  $\mathcal{R}_0$ depends on the terms $\sigma$, $r_5$ and $r_6$. These remaining factors will lower the resulting threshold provided that the mosquitoes are biting a relatively small amount of humans per unit time and transmitting less than they are dying off. Additionally, the asymptomatic and symptomatic human hosts need to be transmitting at a sufficiently low rate. For this reason, we must introduce control measures which both reduce the various disease transmission probabilities involved and increase the vector death rate. 
%----------------------------
	\label{R_CompPara}
\end{enumerate}
%----------------------------
It is worth mentioning that setting the parameters $\tau$ and $\gamma$ equal to zero obviously results in $\mathcal{R}_0=0$. However, these scenarios are not considered as the mosquito and human latent periods $\tau$ and $\gamma$ are considered to be intrinsic properties of the vector and host, respectively. Moreover, these specific terms only appear in the factors $\sqrt{r_i}$ for $i=2,3$ which are strictly bounded above by unity. Additionally, we do not consider the bizarre cases that the human mortality rate $\xi_H=0$ or mosquito recruitment rate $\Omega_M=0$. Although both cases result in effectively nullifying the threshold quantity $\mathcal{R}_0$, humans are not immortal and the mosquito recruitment rate is usually relatively large. Letting this particular parameter value be equal to zero would imply the epidemiologically unreasonable scenario that there are no mosquitoes present in the region being considered. Neither do we consider the qualitative behavior of $\mathcal{R}_0$ if $\xi_H$ is sufficiently large, as introducing a human transmission blocking control measure which also increases the human death rate of a given population is not an ethical control method. It is important to note that emphasis is made on the terms which are related to utilized control measures, i.e. the measures which have an effect on the mosquito death rate and the various transmission probabilities involved. Control measures will be formally introduced and properly covered in Section \ref{sec:ControlMeas} below. 

%BIFURCATION ANALYSIS
\subsection{Endemic Equilibria and Bifurcation Analysis}\label{sec:EE}
%----------------------------
An endemic occurs when disease persists in the population, mathematically this corresponds to all of the state variables being positive. For this reason, endemic equilibrium (EE) points are equilibrium points where the state variables are positive. Most epidemic models exhibit a dichotomy in terms of bifurcations that occur at the threshold $\mathcal{R}_0=1$, namely: super-critical (forward) and sub-critical (backward). These have drastically different epidemiological implications. A forward bifurcation happens when $\mathcal{R}_0$ crosses unity from below and, as a result, a small positive asymptotically stable super-threshold equilibria appears and the disease-free equilibrium losses its stability. Backward bifurcation happens when $\mathcal{R}_0<1$ and a small positive unstable sub-threshold equilibrium appears, while the disease-free equilibrium and a larger positive equilibrium are locally asymptotically stable. 

From an epidemiological viewpoint, a forward bifurcation is more desirable as it results in the reproductive number being below unity to be sufficient to ensure that an epidemic does not occur. In the presence of a backward bifurcation, the reproductive number being below unity is no longer sufficient, as sub-threshold endemic equilibrium can arise in response to perturbations of specific parameters. In most cases, the parameters chosen are contact rates.

Due to the presence of the term $e^{-\varepsilon}=\sum_{n=0}^\infty \frac{(-1)^n}{n!}\left(\frac{E}{N_H}\right)^n$, one can not obtain a closed-form expression for the endemic equilibrium of system (\ref{SEYAR_DS}). However, we turn to a variant of the Center Manifold Theorem, introduced in \cite{CASTSONG}, to show the existence of non-trivial equilibrium solutions of the $SEY \hspace{-3pt}AR$ model (\ref{SEYAR_DS}) near the DFE. This section is focused on the nonlinear stability analysis corresponding to the $SEY \hspace{-3pt}AR$ model. More precisely, the following lemma concerning its bifurcation behavior is proven.
%----------------------------
%################
%----------------------------
%################
\begin{lemma}{(Bifurcation Analysis for the $SEY \hspace{-3pt}AR$ Model).}
Let the positive quantities $\eta_1$ and $\eta_2$ be defined as follows:
%----------------------------
{\small
\begin{align*}
		\eta_1:=&\, Z_6 Q_2 + \frac{\tau^2 Z_1 Q^2_1 Q^2_4}{K_1 \xi_M}\left(1+ \frac{K_2}{K_4} + \frac{K_3}{K_5}+ Q_0\right)\\
		& +\frac{\tau Z_2 			K_2}{K_4 K_7}\left(\frac{\lambda_{RS}  Q_0}{K_6} + 1 + Q_0\right) +\frac{\tau Z_3 K_3}{K_5 K_7}\left(\frac{\lambda_{RS}  Q_0}{K_6} + 1 + 					Q_0\right)\\ 
		&+\frac{\tau Z_7 Q_4 Q_1 K_2}{\xi_M K_4}+\frac{\tau Z_5 Q_4 Q_1 K_3}{\xi_M K_5} + \frac{2\tau Z_2 K^2_2}{K^2_4 				K_7}+ \frac{2\tau Z_3 K^2_3}{K^2_5 K_7},\\
        \eta_2:=& \, \frac{\tau Z_2 K_1 K_2}{K_4 K_6 K_7}+\frac{\tau Z_3 K_1 K_3}{K_5 K_6 K_7}+\frac{\tau Z_4 K_2 K_3}{K_4 K_5 K_7} + Z_6 		Q_3,\\
\end{align*}
}
%----------------------------
where the terms labeled $K_i$, $Q_i$ and $Z_i$, for $1=1,...,7$ are defined in Section (\ref{sec:A1}) below. If the parameter $\Lambda$ is defined as 
%----------------------------
\begin{equation}
		\Lambda:=\frac{\eta_2}{\eta_1},
	\label{Lamb}
\end{equation}
%----------------------------
\label{thm:Bif}
then the bifurcation for the $SEY \hspace{-3pt}AR$ model (\ref{SEYAR_DS}) is sub-critical provided that $\Lambda>1$ and super-critical provided $\Lambda<1$.
\end{lemma}
%################
%----------------------------
%################
%----------------------------
As previously mentioned, lemma (\ref{thm:Bif}) is proven by making use of an application of the Center Manifold Theorem \cite{CASTSONG}, adapted to the case of nonlinear dynamical systems. As in the case of most malaria models appearing throughout scientific literature, the type of bifurcation experienced by the system is completely determined by the sign of the $a$-term (\ref{a}), appearing in Theorem (\ref{thm:CM}). In the case of the $SEY \hspace{-3pt}AR$ model (\ref{SEYAR_DS}) $a \propto (\eta_2-\eta_1)$, resulting in a size constraint on $\Lambda$. In an epidemiological setting, it is desirable for the bifurcation, if it exists, to be super-critical, i.e. forward. Lemma (\ref{thm:Bif}) tells us that if one wants to avoid the case of a sub-critical bifurcation from occurring, we must demand the quantity $\Lambda<1$ in addition with $\mathcal{R}_0<1$. 
%----------------------------
\begin{Remark}
A verification of the entries of the Jacobian and Hessian evaluated at the DFE is provided in the electronic supplementary material.
	\label{Expla}
\end{Remark}
%----------------------------

\section{Parameter Estimation and Generalized Control Measures}\label{sec:ParamVacc}

%PARAMETER ESTIMATION
\subsection{A Unified Formal Approach to Parameter Estimation} \label{sec:ParamEst}
In order to use the $SEY \hspace{-3pt}AR$ model (\ref{SEYAR_DS}) to obtain biologically meaningful numerical estimates, it is necessary to obtain values for the parameters appearing in Table \ref{table:nonlin}. Provided these parameter values exist, they are usually difficult to obtain. The formulas appearing in this section provide relationships between the parameters in Table \ref{table:nonlin} and other well-known biological quantities measured in malaria epidemiology. These functional relationships enable a characterization of many of the static quantities used in malaria modeling and provide numerous ways to estimate them.  
%---------------------------------------------------------------------------------
\begin{enumerate}[label=\Roman*]
\item \textit{Probability of surviving the length of the \textit{Plasmodium} incubation period} ($\mathcal{P}$): Due to equation (\ref{survFUN}) appearing in Section \ref{sec:A1}, if time is measured in days, then the probability of a mosquito surviving one day is $\nu(1)=e^{-\xi_M}$, so that the mosquito mortality rate is $\xi_M=-\ln \nu(1)$. Define $\tau$ to be the length of the \textit{Plasmodium} incubation period, i.e. the duration of the intrinsic incubation period. Therefore, the probability of surviving the length of the \textit{Plasmodium} incubation period is $\mathcal{P}:=\nu(\tau)=e^{-\xi_M \tau}$. This quantity varies with respect to the parasite species and has temperature dependency.
\item \textit{Human Feeding Rate} ($\sigma$): The human feeding rate quantifies the average number of mosquito bites suffered by a human, per mosquito, per day. Define $m_f$ to be the mosquito feeding rate, so that the reciprocal is then the time interval between blood meals, and let $m_q$ denote the proportion of mosquito blood meals taken on humans. Then the expected number of bites on humans per mosquito, per day is given by the equation $\sigma=m_q m_f$.
\item \textit{Human Biting Rate} ($HBR$):  The human biting rate quantifies the number of mosquito bites suffered by humans, per human, per day. Let $m_\epsilon$ be the rate of mosquito emergence per human, per day, so that the equilibrium mosquito density per human is $m_0=\frac{m_\epsilon}{\xi_M}$. The $HBR$ is given by the product of the human feeding rate $\sigma$ and the equilibrium mosquito density, i.e. $HBR=\sigma m_0$.
 \item \textit{Stability Index} ($\mathcal{S}$): A generic female \textit{Anopheles} mosquito lives an average of $\frac{1}{\xi_M}$ days and bites a human once every $\sigma$ days. The expected number of mosquito bites on humans over a mosquito's lifetime, called the stability index, is defined to be $\mathcal{S}:=\frac{\sigma}{\xi_M}$. The mortality rate $\xi_M$ is constant and the mosquitoes lifespan is given by an exponential distribution. As a result, $\mathcal{S}$ can be understood as the approximate number of bites after a mosquito becomes infected. 
	\item \textit{Human Blood Index} ($HBI$): Following the terminology utilized in \cite{smith2004statics}, a mosquito of age $a$ is expected to have given $m_f a$ bites, of which $\sigma a=m_f m_q a$ are on humans. For a cohort of recently emerged mosquitoes, the proportion of surviving mosquitoes of age $a$ that have ever bitten a human is $\eta_\sigma(a)=1-e^{-\sigma a}$. Thus, the proportion of mosquitoes in a population that has survived to age $a$ and has bitten a human is $\eta_\sigma(a) \nu(a)$, so that after a normalization, it follows that the $HBI$ is \label{HBI}\\
	\begin{equation*}
HBI := \frac{\int_{0}^\infty \eta_\sigma(a) \nu(a) da}{\int_{0}^\infty \nu(a) da}.
\end{equation*}
Making use of the functional (\ref{G}) defined in Lemma (\ref{thm:Gkap}) below, we have  $HBI = \mathcal{G}_\nu \left( \eta_\sigma (a) \right)$. Therefore, by equation (\ref{Gkappa}) with $\omega=0$ and $\kappa=\sigma$, it follows that
%----------------------------
\begin{equation*}
		HBI=\frac{\sigma}{\sigma + \xi_M}.
	\label{GHBI}
\end{equation*}
%----------------------------
As pointed out in \cite{smith2004statics}, the above quantity can by interpreted as a ratio of two waiting times. The numerator is the waiting time to the first human bite or death $\frac{1}{\sigma + \xi_M}$ and the denominator is the waiting time to the first human bite among surviving mosquitoes $\frac{1}{\sigma}$.
\item \textit{Proportion of Infected Mosquitoes} ($M^\star$): Denote $I^\star=Y^\star \cup A^\star$ to be the portion of infected humans. The stars superscripts are meant to emphasize the fact that this is a static analysis and as a result, $I^\star$ is assumed to be constant. Furthermore, let $\beta_I:=\beta_A+\beta_Y$ stand for the probability of disease transmission from an infected human to a susceptible mosquito. As a result, mosquitoes acquired infection at rate $\sigma \beta_I I^\star$ and the quantity of surviving mosquitoes of age $a$ that ever become infected is $\eta_{\sigma \beta_I I^\star}(a)=1-e^{-\sigma \beta_I I^\star a}$. The proportion of the mosquito cohort that is both alive and infected at age $a$ is $\eta_{\sigma \beta_I I^\star}(a) \nu(a)$. In the same manner as (\ref{HBI}),  $M^\star=\mathcal{G}_\nu \left( \eta_{\sigma \beta_I I^\star}(a) \right)$. Therefore, by equation (\ref{Gkappa}) with $\omega=0$ and $\kappa=\sigma \beta_I I^\star$ it follows that
%----------------------------
\begin{equation*}
		M^\star =\frac{\sigma \beta_I I^\star}{\sigma \beta_I I^\star + \xi_M}.
	\label{Mstar}
\end{equation*}
%----------------------------
The quantity $M^\star$ can be viewed as the ratio of the waiting time to death or infection $\frac{1}{\sigma \beta_I I^\star + \xi_M}$ and the waiting time to infection among surviving mosquitoes $\frac{1}{\sigma \beta_I I^\star}$.
 \item \textit{Sporozoite Rate} ($\mathcal{Z}$):  The infective state of the malaria parasite that is passed on to human hosts from the salivary glands of the Anopheles mosquito is called \textit{Sporozoite} stage.  A related concept is the sporozoite rate, which is the probability that an individual mosquito ever becomes infectious, or equivalently, the proportion of infectious mosquitoes. If $\tau$ is the length of the \textit{Plasmodium} incubation period, then the proportion of mosquitoes of age $a$ that are infectious is 
%---------------------------------------------------------------------------------
\begin{equation*}
\eta^\tau_{\sigma \beta_I I^\star} (a)=
	\begin{cases}
		1-e^{-\sigma \beta_I I^\star(a-\tau)} \hfill & \text{if $a>\tau$}, \\
		0 \hfill  & \text{if $a \leq \tau$}. \\
	\end{cases}
\end{equation*}
%---------------------------------------------------------------------------------
Placing this together, the proportion of mosquitoes that is both infectious and alive at age $a$ is $\eta^\tau_{\sigma \beta_I I^\star} (a) \nu(a)$ and $\mathcal{Z}=\mathcal{G}_\nu \left( \eta^\tau_{\sigma \beta_I I^\star}(a) \right)$. 
Thus, by equation (\ref{Gkappa}) with $\omega=\tau$ and $\kappa=\sigma \beta_I I^\star$ it follows that
%----------------------------
\begin{equation*}
		\mathcal{Z} =\frac{\sigma \beta_I I^\star}{\sigma \beta_I I^\star + \xi_M}e^{-\tau \xi_M}.
	\label{SporRate}
\end{equation*}
%----------------------------
It is worth pointing out that $\mathcal{Z}$ is simply the product of the probabilities of ever becoming infected and the incubation period, i.e. $\mathcal{Z}=M^\star \mathcal{P}$.
\item \textit{Lifetime Transmission Potential} ($\beta$): The lifetime transmission potential is defined as the expected amount of new infections that would be generated by a newly emerged adult. If $\beta_M$ denotes the transmission probability from an infectious mosquito to an susceptible human, then the total expected output from a mosquito population at age $a$ is $\sigma \beta_M \eta^\tau_{\sigma \beta_I I^\star} (a) \nu(a)$. Lifetime transmission potential is obtained by integrating over the positive real $a$ axis 
%----------------------------
\begin{equation*}
		\beta =\sigma \beta_M \int_{0}^\infty \eta^\tau_{\sigma \beta_I I^\star} (a) \nu(a) da.
\end{equation*}
%----------------------------
By similar reasoning as above, we have that $\beta=\frac{\sigma \beta_M}{\xi_M}\mathcal{G}_\nu \left( \eta^\tau_{\sigma \beta_I I^\star}(a) \right)$. 
So that by equation (\ref{Gkappa}) with $\omega=\tau$ and $\kappa=\sigma \beta_I I^\star$ it follows that
%----------------------------
\begin{equation*}
		\beta =\frac{\sigma^2 \beta_M \beta_I I^\star}{\xi_M(\sigma \beta_I I^\star + \xi_M)}e^{-\tau \xi_M}.
	\label{LTP}
\end{equation*}
%----------------------------
Therefore, $\beta$ can be viewed as the product of the four quantities: (1) the probability that a mosquito becomes infected $M^\star$, (2) the probability that an infected mosquito lives to become infectious $\mathcal{P}$, (3) the transmission efficiency $\beta_M$ and the stability index $\mathcal{S}$.
	\item \textit{Entomological Inoculation Rate} ($EIR$): The $EIR$ serves as a measure of human exposure to infectious mosquitoes. By definition, the $EIR$ stands for the quantity of infectious bites received per human, per day. Mathematically, this is expressed as the product of the $HBR$ $\sigma$ and the Sporozoite Rate $\mathcal{Z}$. More specifically 
%----------------------------
\begin{align*}
		EIR &=\sigma m_0 \mathcal{Z}, \\
            &=\frac{m_0 \sigma^2 \beta_I I^\star}{\sigma \beta_I I^\star + \xi_M}e^{-\tau \xi_M}.
\end{align*}
%---------------------------- 
	 \item \textit{Vectorial Capacity} ($V_C$): Vectorial Capacity quantifies the total number of potentially infectious bites that would eventually arise from all the mosquitoes biting a single perfectly infectious (i.e., all mosquito bites result in infection) human on a single day. More concisely, $V_C$ is the expected number of humans infected, per human, per day, assuming ideal transmission efficiency , i.e. $\beta_I=\beta_M=1$. The vectorial capacity $V_C$ of a mosquito population can be obtained by taking the product of the following four quantities: (1) the vectorial competence $b$ of the female \textit{Anopheles} mosquito, which is its ability to transmit malaria. (2) the emergence rate of mosquitoes $m_\epsilon$, (3) the squared stability index $\mathcal{S}^2$ (this quantity is squared owning to the fact that two bites are necessary for malaria transmission: the one that infects the definite host and the one that infects the indefinite), and (4) the probability of a mosquito surviving the \textit{Plasmodium} incubation period 
$\mathcal{P}$. Taking these principles into account we arrive at the following formula for $V_C$ 
\begin{align*}
V_C &:= b m_\epsilon \mathcal{S}^2 \mathcal{P},\\
	 &= b m_0 \xi_M \left(\frac{\sigma}{\xi_M}\right)^2  e^{-\xi_M \tau}, \\
     &= \frac{b m_0 \sigma^2 \nu^\tau(1)}{-\ln \nu(1)}.
\end{align*}	
\item \textit{Individual Vectorial Capacity} ($V_I$): Individual Vectorial Capacity quantifies the total number of potentially infectious bites emanating from a single  mosquito after feeding on an infectious human host. For this reason, it is defined as the product of following three quantities: (1)  the probability of a mosquito surviving the \textit{Plasmodium} incubation period 
$\mathcal{P}$, (2) the stability index $\mathcal{S}$, and (3) the probability of disease transmission from an infected human to a susceptible mosquito $\beta_M$. Therefore, we arrive at the following formula for $V_I$:
%----------------------------
\begin{align*}
V_I &:= \beta_M \mathcal{P} \mathcal{S},\\
	 &= \frac{\sigma \beta_M e^{-\xi_M \tau}}{\xi_M}.
\end{align*}	
%----------------------------
\end{enumerate}
%################
%----------------------------
%################
\begin{theorem}{(Functional Equation for the $SEY \hspace{-3pt}AR$ Model).}
Define the functional $\mathcal{G}_\nu(\cdot):L^\infty(\mathbb{R}_+) \rightarrow \mathbb{R}_+$ by the following integral equation
%----------------------------
\begin{equation}
\mathcal{G}_\nu \left( \varphi(a) \right) := \xi_M \int_{0}^\infty \varphi(a) \nu(a) da,
\label{G}
\end{equation}
%----------------------------
where $\nu(a):=e^{-a\xi_M}$. For exponential distributions of the form 
%---------------------------------------------------------------------------------
\begin{equation*}
\eta^\omega_\kappa (a)=
	\begin{cases}
		1-e^{-\kappa(a-\omega)} \hfill & \text{if $a>\omega$}, \\
		0 \hfill  & \text{if $a \leq \omega$}, \\
	\end{cases}
\end{equation*}
%---------------------------------------------------------------------------------
it follows that
%----------------------------
\begin{equation}
		\mathcal{G}_\nu \left( \eta^\omega_\kappa (a) \right)=\frac{\kappa}{\kappa + \xi_M}e^{-\omega \xi_M},
	\label{Gkappa}
\end{equation}
%----------------------------
\label{thm:Gkap}
in the sense of improper Riemann integrals.
\end{theorem}
%################
%----------------------------
%################

%##############################################
%---------------------------------------------------------------------------------
%##############################################
%---------------------------------------------------------------------------------
%##############################################
%\subsubsection[Proof of the Lemma]{Proof of Lemma~\ref{thm:Gkap} (See page~%%\pageref{thm:Gkap})}
%----------------------------

%##############################################
%---------------------------------------------------------------------------------
%##############################################
%---------------------------------------------------------------------------------
%##############################################
%----------------------------
\begin{Remark}
The fact that $\int_{0}^\infty \nu(a) da=\xi^{-1}_M$ in combination with the discussion in item \ref{HBI} of Section \ref{sec:ParamEst} involving weighted averages motivates the definition of the above functional (\ref{G}).
\label{GfuncRem}
\end{Remark}
%----------------------------

%CONTROL MEASURES
\subsection{Incorporating Generalized Control Measures} \label{sec:ControlMeas}
%----------------------------
The entomological inoculation rate, discussed in Section \ref{sec:ParamEst}, is a meaningful epidemiologic predictor that serves as a good measure of malaria intensity in a given region, \cite{killeen2000simplified}. In 2007, David L. Smith estimated the reproductive number for 121 African populations. On figure 2, found in \citep[p.~0534]{smith2007revisiting} two numerical plots are displayed in which the reproductive number estimates are compared with the entomological inoculation rate of the populations under consideration. One plot corresponds to heterogeneous biting and transmission-blocking immunity taken into account in the parameter estimates and the other without. In both cases, the quantities where shown to be directly proportional, i.e. regions with a relatively large (small) $EIR$ also have a relatively large (small) reproductive number. In other words, regions with relatively large $EIR$ values of each region also have relatively large $\mathcal{R}_0$ values. In areas with large $\mathcal{R}_0$, it is unlikely that one single control measure will be sufficient to stop the disease expansion \cite{smith2007revisiting}. 

Concerning protection obtained through vaccination, Figure 6 in \citep[p.~0536]{smith2007revisiting} presents a numerical plot comparing the reproductive number $\mathcal{R}_0$ and the proportion of the population that must be neutralized through a vaccine, under ideal assumptions. In low transmission areas, malaria eradication seems to be a practical and achievable goal. However, in high transmission settings, the classic theory suggests that greater than $99\%$ of hosts would need to be protected in order to sufficiently reduce transmission to reach a state of stability. The large range of $\mathcal{R}_0$ values across Africa attest to the immense challenge faced by malaria control programs. 

In practice, vaccine-conveyed immunity is not one hundred percent. This fact is accounted for by the vaccine efficacy $V_f$, which denotes the percentage of protection each vaccinated individual has. If $V_p$ denotes the proportion of the population that has been vaccinated, i.e. the vaccine coverage. The product $V_f V_p$ stands for the fraction of the population under consideration that is protected, so that the remaining proportion $(1-V_f V_p)$ is not directly protected, with respect to vaccine-conveyed immunity. As a result, vaccination controls are incorporated into the model by defining the weight $\bar{v}:=(1-V_f V_p)$. Therefore, the control-modified progression rates are given by the following equations
%---------------------------------------------------------------------------------
\begin{align*} 
	\begin{cases}
      	\tilde{\lambda}_{EA} &:=\bar{v}\gamma u(\varepsilon),\\ 
	  	\tilde{\lambda}_{EY} &:=\bar{v}\gamma (1-u(\varepsilon)).\\    
	\end{cases}                   
\end{align*} 
%---------------------------------------------------------------------------------
Insecticide-treated nets (ITNs) are the most prominent malaria preventive measure for large-scale deployment in highly endemic areas, such as sub-Saharan Africa, \cite{lengeler2004insecticide}. ITNs are nets coated with synthetic pyrethroid insecticides. Due to their high arthropod toxicity and low mammalian toxicity, such insecticides are well suited for this purpose. Many studies have shown them to both kill and repel mosquitoes. In a recent study, a regression analysis of the protective efficacy on the transmission intensity, as measured by the $EIR$, was performed at the following four different endemic regions of Africa: Burkina Faso, The Gambia, Ghana, and Kenya. It was noted that the protective efficacy was lower in areas with a higher $EIR$, which was consistent with the original hypothesis that relative impact is lower in areas with higher entomological inoculation rates, \cite{lengeler2004insecticide}. Moreover, in the case of homogeneous biting $99.95\%$ ITN coverage was predicted to be necessary, \cite{smith2007revisiting}.

Regarding the ITN coverage, let the symbol $\rho_f$ denote the protective efficacy, i.e. the percentage reduction in malaria episodes due to bed net usage. Upon letting $\rho_p$ be the proportion of ITN usage, i.e. the percentage decrease in transmission due to the employment of ITNs, then the reduction in mosquito to human transmission is captured by the multiplier $(1-\rho_f\rho_p)$. Three separate scenarios are considered in which the population coverage of ITNs is assumed to be $33\%$ (low), $66\%$ (medium) and $99\%$ (high). Additionally, let 
$\xi_{_{\itn}\hspace{5pt}}$ denote the maximum ITN-induced death rate for the mosquito population. Following \cite{agusto2013impact}, it is assumed that ITN usage reduces the effective human to mosquito effective contact rates $\beta_A$ and $\beta_Y$ and increases the mosquito mortality rate $\xi_M$. Thus, the effects of ITN usage on the disease transmission dynamics of the $SEY \hspace{-3pt}AR$ model (\ref{SEYAR_DS}) are accounted for by the following modification.
%---------------------------------------------------------------------------------
\begin{align*} 
	\begin{cases}  
	  	\tilde{\beta}_{A} &:=(1-\rho_f\rho_p)\beta_A,\\  
      	\tilde{\beta}_{Y} &:=(1-\rho_f\rho_p)\beta_Y,\\      
      	\tilde{\xi}_M &:=\xi_M + \rho_f\rho_p \xi_{_{\itn}\hspace{5pt}}.\\   
	\end{cases}                   
\end{align*} 
%---------------------------------------------------------------------------------
To obtain the modified system and the reproductive threshold, one needs to invoke the direct substitution where the tilde in each term being dropped corresponds to the system in the natural variables. One should note that this differs form the change of variable technique. Therefore, the resulting control-modified variant of the $SEY \hspace{-3pt}AR$ model (\ref{SEYAR_DS}) expanded in the original model parameters is 
%----------------------------
\begin{equation}
\begin{cases}
		\dot{S} =\Omega_H + \lambda_{RS} R - \left(\sigma \beta_M \frac{M_I}{N_H} +  \xi_H \right)S,\nonumber \\ 
		\dot{E} =\sigma \beta_M \frac{M_I}{N_H}S - (\gamma +\xi_H)E,\nonumber \\            
		\dot{Y} =(1-V_f V_p)\gamma(1-u(\varepsilon)) E -  \left(\xi_H +\delta + \lambda_{YR}\right)Y,\nonumber \\               
		\dot{A} =(1-V_f V_p)\gamma u(\varepsilon) E - \left(\lambda_{AR}+\xi_H\right)A, \label{SEYAR_CM} \\              
		\dot{R} =\lambda_{AR}A +  \lambda_{YR}Y - \left(\lambda_{RS} +\xi_H\right) R,\nonumber  \\          
		\dot{M_S} =\Omega_M  - \left (\xi_M + \rho_f\rho_p \xi_{_{\itn}\hspace{5pt}}  +  \sigma (1-\rho_f\rho_p)\beta_Y \frac{Y}{N_H} +\sigma (1-\rho_f\rho_p)\beta_A \frac{A}{N_H}  \right )M_S,\nonumber \\        
		\dot{M_E} =\sigma (1-\rho_f\rho_p)\left ( \beta_Y \frac{Y}{N_H} +\beta_A \frac{A}	{N_H}\right )M_S - \left(\xi_M + \rho_f\rho_p \xi_{_{\itn}\hspace{5pt}} + \tau \right)M_E,\nonumber \\      
		\dot{M_I} =\tau M_E- (\xi_M + \rho_f\rho_p\xi_{_{\itn}\hspace{5pt}}) M_I.\nonumber   	
\end{cases}
\end{equation}
%----------------------------
Assuming that the initial exposed proportion is zero, i.e. $\varepsilon_0=0$, the corresponding vaccination control-modified reproductive threshold $\mathcal{R}^{\bar{v}}_0$ is given by the following formula
%----------------------------
%----------------------------
\begin{align}
\mathcal{R}^{\bar{v}}_0  = & \, \sqrt{\frac{\sigma^2 \tau \gamma \Omega_M \xi_H \beta_M(1-\rho_f\rho_p)\bar{v}}{(\xi_M + \rho_f\rho_p \xi_{_{\itn}\hspace{5pt}})^2(\gamma + \xi_H)(\tau + \xi_M + \rho_f\rho_p \xi_{_{\itn}\hspace{5pt}})\Omega_H} }  \nonumber \\
 & \, \times\sqrt{\left(\frac{\beta_Au_{_{\low}\hspace{5pt}}}{\lambda_{AR} + \xi_H} - \frac{\beta_Y(u_{_{\low}\hspace{5pt}}-1)}{\lambda_{YR} + \xi_H + \delta} \right)}.
\label{ConR0}
\end{align}

%----------------------------
\begin{Remark}
It is worth emphasizing that the ITN vector $(\rho_f,\rho_p) \in [0,1) \times [0,1)$ is a fixed quantity and the only controls are the vaccination terms $V_f,V_p$. This avoids the issue of identifiability of the model parameters with respect to the control parametrization.
\end{Remark}
%----------------------------
%----------------------------
The incorporation of the vaccination controls into the system implicitly introduces a restriction on the control parameter domain of $\mathcal{R}^{\bar{v}}_0$. Let $\epsilon>1$, then if one takes the control parameter domain to be the entire rectangle $[0,\epsilon) \times [0,\epsilon) \supset [0,1) \times [0,1)$, then $\mathcal{R}^{\bar{v}}_0$ may be nullified or complex valued. Let $\mathbb{C}:=\{\bar{z} \hspace{3pt}:\hspace{3pt} \bar{z}=a+bi \hspace{3pt}\text{where}\hspace{3pt} (a,b) \in \mathbb{R} \times \mathbb{R}\hspace{3pt}\text{and}\hspace{3pt} i=\sqrt{-1} \}$ denote the field of complex numbers. Making use of the definitions of $C_i$ for $i=1,2$, presented in Section (\ref{sec:Rnumber}), it follows that the vaccination control-modified reproductive threshold will not be equal to zero or complex valued provided that $\mathcal{R}^{\bar{v}}_0$ is restricted to the reciprocal sub-domain $\mathcal{H}$, as described in the following theorem.
%----------------------------
%----------------------------
%################
%----------------------------
%################
\begin{theorem}{(Reciprocal Sub-domain for the Parameter Space of the $SEY \hspace{-3pt}AR$ Model).}
Fix the vector $(\rho_f,\rho_p) \in [0,1) \times [0,1)$ and each $C_i$ for $i=1,2$, as defined in Section (\ref{sec:Rnumber}), and define the following reciprocal sub-domain of the rectangle $[0,1] \times [0,1]$
\begin{equation*}
\mathcal{H}:=\left\lbrace (V_f,V_p) \in [0,1] \times [0,1]: V_f \in \left(0, \frac{1}{V_p}\right )\right\rbrace. 
\label{HypParam}
\end{equation*}
Then $\mathcal{R}^{\bar{v}}_0 \notin \mathbb{C}-\left\lbrace 0 \right\rbrace$ if and only if $(V_f,V_p) \in \mathcal{H}$.
\label{thm:HypPara}
\end{theorem}
%################
%----------------------------
%################
%----------------------------
%----------------------------
Although, it is epidemiologically unreasonable to consider a vaccine efficacy $V_f$ or vaccine coverage $V_p$ control measure which equals to or exceeds $100\%$, this formal analysis is presented to provide a solid mathematical foundation for the numerical implementation of the model. More generally, if we consider $m$ transmission blocking human control measures, then the modification is given by
%---------------------------------------------------------------------------------
\begin{align*} 
	\begin{cases}
      	\lambda_{EA}^c &:=\prod_{j=1}^m\left(1- V_f^j V_p^j\right)\gamma u(\varepsilon),\\ 
	  	\lambda_{EY}^c &:=\prod_{j=1}^m\left(1- V_f^j V_p^j\right)\gamma (1-u(\varepsilon)).\\    
	\end{cases}                   
\end{align*} 
%---------------------------------------------------------------------------------
For the $n$ vector controls we have
%---------------------------------------------------------------------------------
\begin{align*} 
	\begin{cases}  
	  	\beta_{A}^c &:=\prod_{i=1}^n\left(1-\rho_f^i\rho_p^i\right)\beta_A,\\  
      	\beta_{Y}^c &:=\prod_{i=1}^n\left(1-\rho_f^i\rho_p^i\right)\beta_Y,\\      
      	\xi_M^c &:=\xi_M + \prod_{i=1}^k\rho_f^i\rho_p^i \| \xi_i\|_\infty,\\   
	\end{cases}                   
\end{align*} 
%---------------------------------------------------------------------------------
where $(\rho_f^i,\rho_p^i) \in [0,1) \times [0,1)$ for $i=1,\cdots,n$ and each $\| \xi_i\|_\infty$ denotes the maximum, possibly unachievable, vector death rate increase corresponding to each $i^\text{th}$ vector control method. It should be noted that it is assumed that $k \leq n$ vector control measures increase the mosquito death rate. Formally, one could induce a sub ordering on the vector transmission reduction factors and extract the $k$ terms such that the death rate increases. However, this formality is implied and would be unnecessary. The system including generalized controls is obtained via the same substitution as in the motivational example discussed above. For convenience the generalized control-modified reproductive threshold is listed below in explicit form 
%----------------------------

\begin{align}
\mathcal{R}^{c}_0 = & \, \sqrt{\frac{\sigma^2 \tau \gamma \Omega_M \xi_H \beta_M\prod_{i=1}^n\left(1-\rho_f^i\rho_p^i\right)}
{\Omega_H(\gamma + \xi_H)\left(\xi_M + \prod_{i=1}^k\rho_f^i\rho_p^i \| \xi_i\|_\infty\right)^2}}\nonumber  \\
& \, \times \sqrt{\frac{\prod_{j=1}^m\left(1- V_f^j V_p^j\right)\left(\frac{\beta_Au_{_{\low}\hspace{5pt}}}{\lambda_{AR} + \xi_H} - \frac{\beta_Y(u_{_{\low}\hspace{5pt}}-1)}{\lambda_{YR} + \xi_H + \delta} \right)}{\left(\tau + \xi_M + \prod_{i=1}^k\rho_f^i\rho_p^i \| \xi_i\|_\infty\right)}},
\label{ConR0GEN}
\end{align}

%----------------------------
%----------------------------
where $V_f^j \in \left(0, \frac{1}{V_p^j}\right )$ for $j=1,\cdots,m$, so that $\mathcal{R}^{c}_0 \notin \mathbb{C}-\left\lbrace 0 \right\rbrace$, as covered in Theorem \ref{thm:HypPara} above.

%SEYAR VIVAX VARIANT
\subsection{SEYAR Model Including Relapse Rates}\label{sec:SeyarViv}
%----------------------------
While in the \textit{hepatocytes}, unlike \textit{P. falciparum}, \textit{P. vivax} and \textit{P. Ovale} enter into an additional form called the \textit{hypnozoite}. In the \textit{hypnozoite} stage, the parasite is in a dormant state, resulting in no clinical symptoms displayed in the recovered patients after months or even years, until suddenly resurfacing. In order to properly account for such biological behavior, relapse rates are including into the $SEY \hspace{-3pt}AR$ model. The asymptomatic and symptomatic human relapse rates are denoted by $\lambda_{SA}$ and $\lambda_{SY}$, respectively. This slight adjustment yields the following \textit{P. vivax} version of the $SEY \hspace{-3pt}AR$ model (\ref{SEYAR_DS}). Listed below is the flow diagram, dynamical system and reproductive threshold corresponding to this model. 
%----------------------------
%SEYAR DIAGRAM
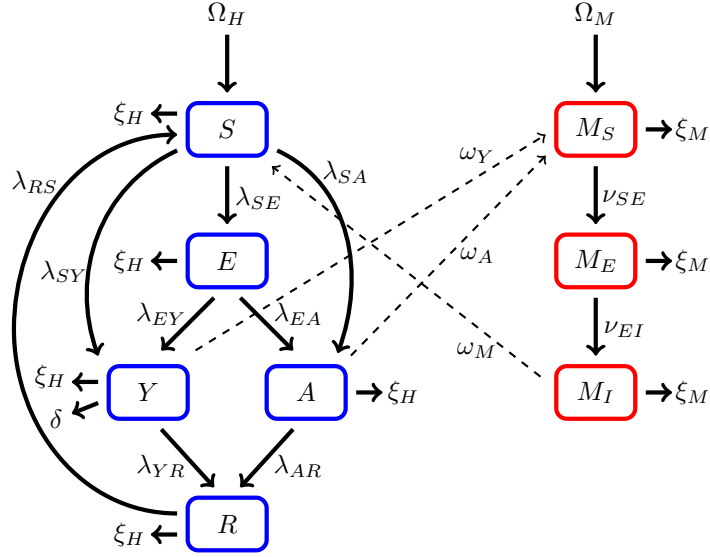
\begin{figure}[ht]
	\centering
	\begin{tikzpicture}[scale=0.7]
		%\draw [help lines] (0,0) grid (14,14);
		% left arrows
		\draw [->, ultra thick] (4.25, 11.3) to (4.25, 10.2);
		\draw [->, ultra thick] (4.25, 8.8) to (4.25, 7.7);
		\draw [->, ultra thick] (4, 6.3) to (3, 5.3);
		\draw [->, ultra thick] (4.5, 6.3) to (5.5, 5.3);
		\draw [->, ultra thick] (3, 3.8) to (4, 2.7); 
		\draw [->, ultra thick] (5.5, 3.8) to (4.5, 2.7); 
		% curvy arrow
		\draw [->, ultra thick] (3.3, 2.2) [in= 270, out= 180]  to (0.2,5.7)  [in=180, out=90] to (3.3, 9.4);
		%\draw [ultra thick, ->] (3.3,9.1)   [in=90, out=205 ] to (1.3,6.5) [in=130, out=270] to (1.8, 5.2) ;
		\draw [ultra thick, ->] (3.3,9.1)   [in=110, out=205 ]  to (1.8, 5.2) ;
        \draw [->, ultra thick] (5.2, 9.1) [in=70, out=335] to (6.35,5.25);		
		% left horizonal arrow
		\draw [->, ultra thick] (3.3, 1.8) to (2.8, 1.8);
		\draw [->, ultra thick] (1.8, 4.7) to (1.3,4.7);
		\draw [->, ultra thick] (1.8, 4.3) to (1.3,4.1); % slate arrow
		\draw [->, ultra thick]  (6.7, 4.5) to (7.2, 4.5);
		\draw [->, ultra thick] (3.3, 7) to (2.8, 7);
		\draw [->, ultra thick] (3.3, 9.8) to (2.8, 9.8);
		% boxes on the left
		\draw [blue, ultra thick, rounded corners] (3.5, 1.5) rectangle (5,2.5);
		\draw [blue, ultra thick, rounded corners] (2, 4) rectangle (3.5, 5);
		\draw [blue, ultra thick, rounded corners] (5, 4) rectangle (6.5, 5);
		\draw [blue, ultra thick, rounded corners] (3.5, 6.5) rectangle (5, 7.5);
		\draw [blue, ultra thick, rounded corners] (3.5, 9) rectangle (5, 10);
		% right arrows 
		\draw [->, ultra thick] (11.25, 11.3) to (11.25, 10.2);
		\draw [->, ultra thick] (11.25, 8.8) to (11.25, 7.7);
		\draw [->, ultra thick] (11.25, 6.3) to (11.25, 5.2);
		% right horizontal arrows
		\draw [->, ultra thick] (12.2, 7) to (12.7, 7);
		\draw [->, ultra thick] (12.2, 9.5) to (12.7, 9.5);
		\draw [->, ultra thick] (12.2, 4.5) to (12.7, 4.5);
		% boxes on the right
		\draw [red, ultra thick, rounded corners] (10.5, 9) rectangle (12, 10);
		\draw [red, ultra thick, rounded corners] (10.5, 6.5) rectangle (12, 7.5);
		\draw [red, ultra thick, rounded corners] (10.5, 4) rectangle (12, 5);
		% middle arrows 
		\draw [dashed, thick,->] (10.2, 4.8) to (5.1, 8.8);
		\draw [dashed, thick, ->] (3.65, 5.3) to (10.3, 9.4);
		%\draw [dashed, thick, ->] (5.2, 2)  [out=30, in=245] to (10.5, 8.7);
		\draw [dashed, thick, ->] (6.6, 5.2) to (10.3, 9);
		% letters in the box
		%left
		\node at (4.25, 2) {$R$};
		\node at (2.75, 4.5) {$Y$};
		\node at (5.75, 4.5) {$A$};
		\node at (4.25, 7) {$E$};
		\node at (4.25, 9.5) {$S$};
		
		%right
		\node at (11.25, 7) {$M_E$};
		\node at (11.25, 9.5) {$M_S$};
		\node at (11.25, 4.5) {$M_I$};
		% letters for arrow
		\node at (2.4, 1.8) {$\xi_H$};
		\node at (0.9, 4.8) {$\xi_H$};
		\node at (2.4, 7) {$\xi_H$};
		\node at (2.4, 9.8) {$\xi_H$};
		\node at (7.6, 4.5) {$\xi_H$};
		\node at (13.1, 9.5) {$\xi_M$};
		\node at (13.1, 7) {$\xi_M$};
		\node at (13.1, 4.5) {$\xi_M$};
		\node at (1, 4) {$\delta$};
		%letters for v a rrows
		\node at (4.25, 11.7) {$\Omega_H$};
		\node at (11.25, 11.7) {$\Omega_M$};
		\node at (0.6, 8.5) {$\lambda_{RS}$};
		\node at (11.8, 8.2) {$\nu_{SE}$};
		\node at (11.8, 5.7) {$\nu_{EI}$};
		\node at (4.85, 8.2) {$\lambda_{SE}$};
		\node at (3, 6) {$\lambda_{EY}$};
		\node at (5.6, 6) {$\lambda_{EA}$};
		\node at (1.15, 6.8) {$\lambda_{SY}$};
		\node at (6.5,8.7) {$\lambda_{SA}$};
		\node at (3, 3.1) {$\lambda_{YR}$};
		\node at (5.6, 3.1) {$\lambda_{AR}$};
		\draw [white] (0.5,1) rectangle (1,1.1); % this rectangle does not show, since I put the the color in white, the purpose is to create more space between the picture and the caption. You can change the size of the rectangle to adjust the space. 
		\node at (9,9) {$\omega_{Y}$};
		\node at (9,7.1) {$\omega_{A}$};
		\node at (9,5.3) {$\omega_{M}$};
	\end{tikzpicture}
	\caption{\small Schematic diagram of the $SEY \hspace{-3pt}AR$ model including relapse rates}
	\label{SEYARdiagramV}
\end{figure}

%----------------------------
In Figure \ref{SEYARdiagramV}, the solid lines represent progression from one compartment to the next, while the dotted stand for the human-mosquito interaction. Humans enter the susceptible compartment either through birth of migration and then progress through each additional compartment subject to the rates described above. These assumptions give rise to the following relapse rate modified $SEY \hspace{-3pt}AR$ model IVP (\ref{SEYAR_DSV}) describing the dynamics of malaria disease transmission in the human and mosquito populations
%----------------------------
\begin{equation}
\begin{cases}
		\dot{S} =\Omega_H + \lambda_{RS} R - \left(\sigma \beta_M \frac{M_I}{N_H} +  \xi_H \right)S-(\lambda_{SA}+\lambda_{SY})S,\nonumber \\ 
		\dot{E} =\sigma \beta_M \frac{M_I}{N_H}S - (\gamma +\xi_H)E,\nonumber \\            
		\dot{Y} =\gamma(1-u(\varepsilon)) E -  \left(\xi_H +\delta + \lambda_{YR}\right)Y + \lambda_{SY}S,\nonumber \\               
		\dot{A} =\gamma u(\varepsilon) E - \left(\lambda_{AR}+\xi_H\right)A +\lambda_{SA}S, \nonumber \\              
		\dot{R} =\lambda_{AR}A +  \lambda_{YR}Y - \left(\lambda_{RS} +\xi_H\right) R,\nonumber  \\          
		\dot{M_S} =\Omega_M  - \left ( \xi_M +  \sigma \beta_Y \frac{Y}{N_H} +\sigma \beta_A \frac{A}{N_H}  \right )M_S,\nonumber \\        
		\dot{M_E} =\sigma \left ( \beta_Y \frac{Y}{N_H} +\beta_A \frac{A}	{N_H}\right )M_S - \left(\xi_M + \tau \right)M_E,\nonumber \\      
		\dot{M_I} =\tau M_E- \xi_M M_I,\nonumber \\	
\\
\left(S_0,E_0,Y_0,A_0,R_0,M_{S_0},M_{E_0},M_{I_0}\right)^T\in\mathbb{R}^8_+ \text{ such that } E_0 \leq \vartheta N_0,
\end{cases}
	\label{SEYAR_DSV}
\end{equation}
%----------------------------
where $u$ and $\vartheta$ are defined by equations (\ref{Irate}) and (\ref{IntCon}), respectively. The model parameters and initial data restrictions are the same as in the original model. Moreover, the results of the mathematical analysis concerning the issues of well-posedness and nonlinear stability are similar to that of the original model. For example, listed below is a lemma regarding the LAS of the DFE for the above model variation. 
%----------------------------
%################
%----------------------------
%################
\begin{lemma}{(Local Asymptotic Stability of the DFE for the $SEY \hspace{-3pt}AR$ Model Including Relapse Rates).}
Define the following quantity 
\begin{equation}
		\mathcal{R}_0 :=\sqrt{\frac{\sigma^2 \tau \gamma \Omega_M \xi_H \beta_M}{\xi_M^2(\gamma + \xi_H)(\tau + \xi_M)\Omega_H} \left(\frac{\beta_AU_{_{\low}\hspace{5pt}} }{\lambda_{AR} + \xi_H} - \frac{\beta_Y(U_{_{\low}}-1)}{\lambda_{YR} + \xi_H + \delta} \right)},
	\label{RnumSEYAR_Vivax} 
\end{equation}
where $U_{_{\low}\hspace{2pt}}:=e^{\varepsilon_0}(u_{_{\low}}-u_{_{\high}\hspace{5pt}}) + u_{_{\high}\hspace{5pt}}$. Then, the DFE $\mathbf{x}_{\small{dfe}}$ for the SEYAR model including relapse rates (\ref{SEYAR_DSV}) is locally asymptotically stable provided that $\mathcal{R}_0<1$ and unstable if $\mathcal{R}_0>1$.
\label{thm:LAS_Vivax}
\end{lemma}
%################
%----------------------------
%################
%----------------------------
Therefore, the modified system has the same reproductive threshold as the original. This is due to the fact that the sub-matrices of the Jacobian $F$ and $V$, resulting from the block matrix partitioning technique utilized in the next generation method, are unaffected by this minor modification. Therefore, the results regarding the LAS of the DFE corresponding to the new system are inherited from the original. 
%##############################################
%---------------------------------------------------------------------------------
%##############################################
%---------------------------------------------------------------------------------
%##############################################
Additionally, the control-modified system is given by
%----------------------------
\begin{equation}
\begin{cases}
		\dot{S} =\Omega_H + \lambda_{RS} R - \left(\sigma \beta_M \frac{M_I}{N_H} +  \xi_H \right)S-(\lambda_{SA}+\lambda_{SY})S,\nonumber \\  
		\dot{E} =\sigma \beta_M \frac{M_I}{N_H}S - (\gamma +\xi_H)E,\nonumber \\            
		\dot{Y} =\prod_{j=1}^m\left(1- V_f^j V_p^j\right)\gamma(1-u(\varepsilon)) E -  \left(\xi_H +\delta + \lambda_{YR}\right)Y+ \lambda_{SY}S,\nonumber \\               
		\dot{A} =\prod_{j=1}^m\left(1- V_f^j V_p^j\right)\gamma u(\varepsilon) E - \left(\lambda_{AR}+\xi_H\right)A+\lambda_{SA}S, \label{SEYAR_CV} \\              
		\dot{R} =\lambda_{AR}A +  \lambda_{YR}Y - \left(\lambda_{RS} +\xi_H\right) R,\nonumber  \\          
		\dot{M_S} =\Omega_M  - \left (\xi_M + \prod_{i=1}^k\rho_f^i\rho_p^i \| \xi_i\|_\infty  +  \sigma \prod_{i=1}^n\big(1-\rho_f^i\rho_p^i\right)\beta_Y \frac{Y}{N_H}  \nonumber\\
			\quad\quad\,\,\,\, 	+\sigma \prod_{i=1}^n\left(1-\rho_f^i\rho_p^i\big)\beta_A \frac{A}{N_H}  \right )M_S,\nonumber \\        
	 \dot{M_E} =\sigma \prod_{i=1}^n\left(1-\rho_f^i\rho_p^i\right)\left ( \beta_Y \frac{Y}{N_H} +\beta_A \frac{A}	{N_H}\right )M_S \nonumber \\
	\quad\quad\,\,\,\,   - \left(\xi_M + \prod_{i=1}^k\rho_f^i\rho_p^i \| \xi_i\|_\infty + \tau\right)M_E,\nonumber \\      
		\dot{M_I} =\tau M_E- \left(\xi_M + \prod_{i=1}^k\rho_f^i\rho_p^i \| \xi_i\|_\infty\right)M_I.\nonumber   	
\end{cases}
\end{equation}
%----------------------------
\begin{Remark}
A verification of the reproductive threshold $\mathcal{R}_0$ (\ref{RnumSEYAR_Vivax}) arising from Lemma (\ref{thm:LAS_Vivax}) above is provided in the electronic supplementary material.
\end{Remark}
%----------------------------
\begin{Remark}
The above modification attests to the delicate relationship between the compartmentalized modeling of infectious diseases and the next generation approach. As we have seen in the above example, the reproductive threshold for a given compartmentalized model is not unique. Some of the most useful information one can obtain from such models resides in various threshold quantities. The reproductive threshold contains important information concerning the long-time behavior (stability) of the underlying model, not the local dynamics. Thus, it stands to reason that any two given compartmentalized epidemiological models may possess very different short-time (local) behavior, but can exhibit similar long-time (global) behavior. This troublesome fact suggests that good modeling lies in the delicate compromise between introducing terms which result in interesting local behavior that describe the disease dynamics accurately to some degree, while simultaneously resulting in the long-time behavior of the model under consideration containing new and useful information. For example, the modification introduced in this section which resulted in a variant of the $SEY \hspace{-3pt}AR$ model including relapse rates had no affect on the long-time dynamics, as the reproductive threshold was identical to the original version. However, clearly the new terms incorporated into the system will result in different local behavior. This attest to the fact that good modeling does not just depend on adding more terms and, as a result, over complicating the model. Some of the most important and useful information to be extrapolated from such models depends on certain threshold quantities that can be invariant under a given modification. As we have seen in the above example, the most important information, i.e. the reproductive threshold was shown to be invariant under the modification concerning relapse rates. To have an affect on the reproductive threshold, one would have to go back to the original epidemiological assumptions on which the model was built, e.g. re define what the infected states are. For example, if the recovered class $R$ was considered to be an infected state in the $SEY \hspace{-3pt}AR$ model formulation, then the reproductive threshold would be different. Extreme care should be taken when deriving or altering such a model, so that the new modification brings new and useful information to the foreground while simultaneously not over complicating the situation for no additional gain. 
\end{Remark}

\section{Sensitivity Analysis and Numerical Results}\label{sec:SAnumerical}
%SENSITIVITY ANALYSIS
\subsection{Sensitivity Analysis}\label{sec:SenAna}
Errors are usually present in data collection and presumed parameter values. In this section, a sensitivity analysis is applied to classify the parameters which have the highest impact on the reproductive threshold $\mathcal{R}_0$. This provides a way to determine which parameter values should be targeted by intervention strategies. A parameter with a relatively large sensitivity index should be estimated with precision, while a parameter with a relatively small sensitivity index does not require as much effort.  

Let $\Theta$ be defined as in Section \ref{sec:RnumberComparison}, then it is of trivial consequence that $\mathcal{R}_0 \in C^1(\Theta)$. Due to this fact and that we have an explicit expression for $\mathcal{R}_0$ of the $SEY \hspace{-3pt}AR$ model (\ref{SEYAR_DS}), we arrive at the following definition.
%----------------------------
\begin{definition}{(Sensitivity Index of the Reproductive Threshold \cite{chitnis2008determining}).}
Consider the reproductive threshold $\mathcal{R}_0 \in C^1(\Theta)$ given by equation (\ref{RnumSEYAR}) in Section \ref{sec:Rnumber} and let $\{e_i \hspace{3pt}:\hspace{3pt} 1\leq i \leq 14\}$ be the 
canonical basis in $\mathbb{R}^{14}$. For $\tilde{\rho} \in \Theta$ define $\rho_i :=\langle e_i,\tilde{\rho}\rangle$, where $\langle\cdot,\cdot\rangle$ denotes the inner product in $14$-dimensional Euclidean space, then the normalized forward sensitivity index of $\mathcal{R}_0$ with respect to the parameter $\rho_i$ is defined by the following differential equation 
%----------------------------
\begin{equation*}
\Upsilon^{\mathcal{R}_0}_{\rho_i}:=\frac{\partial \mathcal{R}_0}{\partial \rho_i} \times \frac{\rho_i}{\mathcal{R}_0}.
\label{SenInd}
\end{equation*}
%----------------------------
\end{definition}
%---------------------------
The normalized forward sensitivity index provides a way to quantify the relative change in the given expression when the parameter changes. The sensitivity index is well-defined provided that $\mathcal{R}_0$ is at least in $C^1$ with respect to each parameter $\rho_i$. The analytic formulas for the sensitivity indices are complex and do not offer much qualitative insight, as a result we evaluate the indices at the parameter values corresponding to each location. Listed below are tables containing the sensitivity indices of $\mathcal{R}_0$ for the $SEY \hspace{-3pt}AR$ model (\ref{SEYAR_DS}) evaluated at the parameter values given in Section \ref{sec:Pvalues}. The parameters are ordered form the most sensitive to the least.
%----------------------------
%----------------------------

\medskip

\begin{table}[ht]
\parbox{.3\linewidth}{
\centering
{\small
\begin{tabular}{l l l}
\textbf{Parameter} & \textbf{Sensitivity Index} \\ [0.5ex] % inserts table %heading
\hline
\medskip
$\xi_M$ & $-1.27500$\\
$\sigma$ &  $+1.00000$\\
$\Omega_M$ & $+0.50000$\\
$\beta_M$ & $+0.50000$\\
$\Omega_H$  & $-0.49999$\\
$\beta_Y$ & $+0.49369$\\
$U_{_{\low}}$ & $-0.48739$\\
$\lambda_{YR}$  & $-0.37495$\\
$\xi_H$ &  $+0.30742$\\
$\tau$ & $+0.27499$\\
$\gamma$  & $+0.07364$\\
$\beta_A$ &  $+0.00630$\\
$\lambda_{AR}$ & $-0.00610$\\
$\delta$ &  $-0.00001$\\
 \\ [1ex]
\hline
\end{tabular}}
\caption{Kaduna}
}
\hfill
\parbox{.3\linewidth}{
\centering
{\small
\begin{tabular}{l l l}
\textbf{Parameter} & \textbf{Sensitivity Index} \\ [0.5ex] % inserts table %heading
\hline
\medskip
$\xi_M$ & $-1.250000$\\
$\sigma$ &  $+1.000000$\\
$\Omega_M$ & $+0.500000$\\
$\beta_M$ & $+0.500000$\\
$\Omega_H$  & $-0.500000$\\
$\beta_Y$ & $+0.493906$\\
$U_{_{\low}}$ & $-0.487812$\\
$\lambda_{YR}$  & $-0.389920$\\
$\xi_H$ &  $+0.332370$\\
$\tau$ & $+0.250000$\\
$\gamma$  & $+0.063490$\\
$\beta_A$ &  $+0.006093$\\
$\lambda_{AR}$ & $-0.005935$\\
$\delta$ &  $-0.000007$\\
 \\ [1ex]
\hline
\end{tabular}}
\caption{Namawala}
}
\hfill
\parbox{.3\linewidth}{
\centering
{\small
\begin{tabular}{l l l}
\textbf{Parameter} & \textbf{Sensitivity Index} \\ [0.5ex] % inserts table %heading
\hline
\medskip
$\xi_M$ & $-1.291666$\\
$\sigma$ &  $+1.000000$\\
$\Omega_M$ & $+0.500000$\\
$\beta_M$ & $+0.500000$\\
$\Omega_H$  & $-0.499999$\\
$\beta_Y$ & $+0.493975$\\
$U_{_{\low}}$ & $-0.487951$\\
$\lambda_{YR}$  & $-0.395175$\\
$\xi_H$ &  $+0.341059$\\
$\tau$ & $+0.291666$\\
$\gamma$  & $+0.060000$\\
$\beta_A$ &  $+0.006024$\\
$\lambda_{AR}$ & $-0.005877$\\
$\delta$ &  $-0.000006$\\
 \\ [1ex]
\hline
\end{tabular}}
\caption{Butelgut}
}
\end{table}

\medskip

The sensitivity analysis conducted on the reproductive threshold of the model above shows that the most sensitive parameters are the mosquito mortality rate $\xi_M$ and the man biting rate $\sigma$. Conversely, the least sensitive is the disease-induced death rate for humans $\delta$. The sensitivity indices listed in the above tables can be viewed as growth measurements of the reproductive threshold with respect to the parameter under consideration. Without loss of generality, focus is turned to the Kaduna location. Concerning Kaduna, an increase in $\xi_M$ by $10\%$ will result in a decrease in $\mathcal{R}_0$ by $12.75\%$. Similarly, an increase in $\sigma$ by $10\%$ will cause a $10\%$ increase in $\mathcal{R}_0$. Also, it is worth noting the asymptomatic recovery and effective contact rates $\lambda_{AR}$ and $\beta_A$ are relatively less sensitive than the symptomatic human related terms $\lambda_{YR}$ and $\beta_Y$. Furthermore, an increase in $\Omega_M$ by $10\%$ results in an increase of $\mathcal{R}_0$ by approximately $5\%$ and a $10\%$ decrease in $\Omega_H$ will cause a $5\%$ increase in $\mathcal{R}_0$. These numerical observations are consistent with the qualitative analysis presented in Section \ref{sec:RnumberComparison}. 

It should be noted that definition (\ref{SenInd}) is pointwise, thus the ordering of the hierarchy is dependent on the relative sizes of the parameter values. The hierarchy of sensitivity that the $SEY \hspace{-3pt}AR$ model parameters obey is common among many compartmentalized homogeneous population malaria models appearing in literature, e.g. \cite{chitnis2008determining,ROOP}. The two most sensitive parameters correspond to the vector population. These parameters have the property that one is directly proportional to the reproductive number while one is inversely proportional. Increasing the mosquito death rate will also reduce the man biting rate, as the average mosquito life span is shortened. This is beneficial from a practical standpoint, as the parameter which is desirable to increase has the additional effect of decreasing the parameter that is desirable to decrease. Theoretically, this aids in the designing of programs for disease control, as it isolates the parameters that should be targeted for reduction by intervention strategies. Insecticide-treated bed nets and indoor residual spraying are among the most common methods used for such purposes. In Section \ref{sec:ControlMeas} these control measures are incorporated into the model in a generalized setting.
%----------------------------
\begin{Remark}
A verification of the numerical entries in the tables displayed above is provided in the electronic supplementary material.
\end{Remark}
%----------------------------

%NUMERICAL RESULTS
\subsection{Numerical Results}\label{sec:NumRes}
%----------------------------
Displayed below is a table containing numerical values for the following: $(\romannumeral 1)$ the parameter configuration space classification $C_1-C_2$, introduced via definition (\ref{ConfigDef}) in Section \ref{sec:Rnumber}, $(\romannumeral 2)$ the $EIR$ corresponding to each location, as discussed in Section \ref{sec:ParamEst}, $(\romannumeral 3)$ the reproductive threshold accounting for asymptomatic carriers $\mathcal{R}_A$, given by equation (\ref{RnumSEYAR}), and $(\romannumeral 4)$ the reproductive threshold neglecting asymptomatic carriers $\mathcal{R}_Y$, as discussed in Section \ref{sec:RnumberComparison}. These numerical values correspond to the following three high transmission sites: Kaduna in Nigeria, Namawala in Tanzania, and Butelgut in Papua New Guinea. The parameter values associated with each site are listed in Section \ref{sec:Pvalues}.
%----------------------------
{\footnotesize
\begin{table}[H]
\centering
\begin{threeparttable}
\caption{$EIR$ and Threshold Quantities}
\begin{tabular}{l l l l l}
\textbf{Site} & \textbf{Classification} & \textbf{$EIR$} & \textbf{$\mathcal{R}_A$} & \textbf{$\mathcal{R}_Y$} \\ [0.5ex] % inserts table %heading
\hline
\medskip
Kaduna & \textit{$A$-dominant} /$-6.00$\tnote{$a$} & $120$\tnote{$b$}  & $393.05$\tnote{$e$} & $390.56$\tnote{$f$}\\
Namawala & \textit{$A$-dominant} /$-6.24$\tnote{$a$} & $329$\tnote{$c$} & \hspace{4pt}$68.38$\tnote{$e$} & \hspace{4pt}$67.96$\tnote{$f$}\\
Butelgut & \textit{$A$-dominant} /$-6.32$\tnote{$a$} & $517$\tnote{$d$} & \hspace{4pt}$81.12$\tnote{$e$} & \hspace{4pt}$80.63$\tnote{$f$} \\
 \\ [1ex]
\hline
\end{tabular}
\label{table:ThrQ}
\begin{tablenotes}
	  \item $a$ These quantities were calculated from definition (\ref{ConfigDef}) in Section \ref{sec:Rnumber}. \\
      \item $b$ This value was taken from \cite{service1965some}. \\
      \item $c$ This value was taken from \cite{smith1993absence}. \\
      \item $d$ This value was calculated in \cite{killeen2000simplified}, using data obtained from the following sources \cite{graves1990estimation,rates1988human}. \\
      \item $e$ These quantities were calculated from equation (\ref{RnumSEYAR}) assuming that the initial exposed proportion is zero, i.e. $\varepsilon_0=0$. \\
      \item $f$ These quantities were calculated by setting the asymptomatic progression rates equal to zero.
    \end{tablenotes}
  \end{threeparttable}
\end{table}}
%----------------------------
One should observe that the size of the reproductive thresholds for the three sites under consideration are consistent with the sizes of the corresponding entomological inoculation rate values.  Regions with a relatively large $EIR$ value also have relatively large $\mathcal{R}_0$ values. As mentioned in Section \ref{sec:ControlMeas}, in areas with large reproductive thresholds it is unlikely that one single control measure will be sufficient to stop the disease expansion. Additionally the threshold quantities $\mathcal{R}_A$ and $\mathcal{R}_Y$ obey the theoretical estimate provided in Theorem (\ref{thm:R_Comp}). Therefore neglecting to account for asymptomatic carriers results in an underestimation of the reproductive threshold corresponding to each location. The theoretical estimate provided in Theorem (\ref{thm:R_Comp}) holds for all possible positive epidemiologically reasonable quantities. Furthermore, the numerical entries displayed in the tables of Section \ref{sec:SenAna} and those listed in Table \ref{table:ThrQ} are pointwise evaluations.  As mentioned in Section \ref{sec:Rnumber}, there have been reports which indicate that asymptomatic carriers in a given population may transmit the disease at a higher rate than the symptomatic. In this case, the sensitivity hierarchy will possess a different ordering and the size differences in the threshold quantities displayed above will be larger. 
%----------------------------
\begin{Remark}
A verification of the numerical threshold quantities presented in Table \ref{table:ThrQ} is provided in the electronic supplementary material.
\end{Remark}
%----------------------------

%CONCLUSIONS
\section{Conclusions and Discussions}\label{sec:Conclusion}
%----------------------------
This study clearly shows that the existence of asymptomatic individuals results in a strict underestimation of $\mathcal{R}_0$, and provides the means to quantify this influence. It also provides the means to study NAI as the factor that drives asymptomaticity. As mentioned in \cite{AcqIM}, the exploration of NAI is key to the rational development and deployment of vaccines and other malaria control methods corresponding to any given population at risk. Therefore, it is a necessary foundation upon to build strategies of eradication by any means. 

The $SEY \hspace{-3pt}AR$ model (\ref{SEYAR_DS}) accounts for the impact that the exposure dependent naturally acquired immune proportion has on asymptomatic carriers and malaria disease transmission dynamics. Through making use of the IVP (\ref{NAI_IVP}), the infected compartment $I$ is effectively decomposed into two mutually disjoint sub-compartments accounting for symptomatic and asymptomatic individuals. This results in a model which does not fall into a sub-class of the type studied in \cite{FLIPE}. Current asymptomatic models appearing in literature are formed by inserting a state invariant constant control parameter or a sum of transcendental expressions involving various state invariant immunity acquisition rates. The $SEY \hspace{-3pt}AR$ model is derived by a separation through means of the NAI proportion of a population which depends on exposure through equation (\ref{Irate}). After deriving the model and addressing the issues of well-posedness and stability analysis, a nonlinear stability analysis is performed in which the bifurcation behavior of the model is characterized. A sensitivity analysis is carried out and generalized control measures are introduced in the model. Moreover, numerical values of various quantities discussed throughout this work are provided for the following three high transmission sites: Kaduna in Nigeria, Namawala in Tanzania, and Butelgut in Papua New Guinea. A brief summary of highlights drawn from the conclusions of this work is presented in the form of a list below.
%----------------------------
\begin{enumerate}%[label=\Roman*]%
	\item In Section \ref{sec:WellPosedness} it was shown that the $SEY \hspace{-3pt}AR$ model (\ref{SEYAR_DS}) is mathematically and epidemiologically well-posed provided the initial data satisfied suitable regularity assumptions. Additionally, in theorem (\ref{thm:Reg}) a mathematically precise and epidemiologically reasonable lower bound for the feasible region of the model was provided.
	\item In Lemma (\ref{thm:LAS}) of Section \ref{sec:Rnumber} the $SEY \hspace{-3pt}AR$ model (\ref{SEYAR_DS}) was shown to satisfy the threshold condition. Moreover, $\mathcal{R}_0$ was decomposed into a product to properly analyze the size contribution of each factor involved. Provided that the mosquito mortality rate $\xi_M$ can be made large enough so that the term $C_0$ compromised of fractional multipliers is sufficiently small, i.e. $C_0<1$, then the size of $\mathcal{R}_0$ is completely determined by the human-dependent factor $r_6$, defined in remark (\ref{RnumberDecomposition}), which consists of a weighted difference of vital dynamics with the NAI proportion, recovery and death rates. Motivated by the monotonic behavior of $\mathcal{R}_0$, a formal characterization of the parameter configuration space was introduced via definition (\ref{ConfigDef}).
	\item In Section \ref{sec:RnumberComparison} an estimate is provided which characterizes the impact that the asymptomatic class has on the reproductive threshold. More precisely, it was shown that neglecting asymptomatic carriers results in an underestimation of the threshold.	
	\item In Section \ref{sec:EE}, use is made of Theorem (\ref{thm:CM}) to show the existence of nontrivial sub-threshold equilibrium solutions near the DFE. More precisely, it was shown that the bifurcation experienced by the $SEY \hspace{-3pt}AR$ model (\ref{SEYAR_DS}) is forward or backward depending on the size of  an auxiliary threshold parameter $\Lambda$ defined by (\ref{Lamb}). As it is desirable for no endemic equilibrium states to arise while $\mathcal{R}_0<1$, we impose the additional requirement that $\Lambda<1$.
	\item In Section \ref{sec:SenAna}, a sensitivity analysis is conducted on the reproductive number of the model for parameter configurations corresponding to high transmission settings. Additionally, tables are provided containing the sensitivity indices of $\mathcal{R}_0$ with respect to each parameter. An ordering of the parameters from the most sensitive to least revealed that the most sensitive parameters were the mosquito mortality rate $\xi_M$ and man biting rate $\sigma$. The least sensitive was the disease-induced human death rate $\delta$. Additionally, the asymptomatic recover and effective contact rates $\lambda_{AR}$ and $\beta_A$ were shown to be relatively less sensitive than the symptomatic human related terms $\lambda_{YR}$ and $\beta_Y$.
	\item In Section \ref{sec:NumRes} numerical results corresponding to the three high transmission sites mentioned above are provided. The numerical values of the threshold quantities were shown to be consistent with the theory presented in Section \ref{sec:RnumberComparison}. As the sites of interest are high transmission areas with relatively high entomological inoculation rates, the reproductive thresholds were shown to be comparably large. 
\end{enumerate}
%----------------------------
As directions of future research, it will be interesting to apply the method of decomposing the infected compartment by means of the related rate IVP (\ref{NAI_IVP}) in Section \ref{sec:Methods} to other infectious diseases where asymptomatic individuals play a fundamental role on the dynamical behavior. Additionally, it will be informative to consider extensions of the $SEY \hspace{-3pt}AR$ model formed by incorporating the other kinds of immunity mentioned in the introduction. Furthermore, it will be beneficial to introduce additional control measures into the models, such as aerial fogging and a time-dependent treatment rate of symptomatic carriers. In conclusion, the $SEY \hspace{-3pt}AR$ model (\ref{SEYAR_DS}) has provided us with a precise mathematical understanding of the relationship between the exposure dependent nature of NAI and asymptomatic malaria disease transmission dynamics.

\section*{Acknowledgments}

Research reported in this publication was supported by a National Institute of Allergy and Infectious Diseases (NIAID) cooperative agreement of the National Institutes of Health (NIH) under award number U19AI089702-01 and the Malaria Host-Pathogen Interaction Center - MaHPIC, NIH’s NIAID contract HHSN272201200031C.

\section{Appendix}\label{sec:appendix}

%APPENDIX 1-PROOFS 
\label{sec:A1}
This section of the appendix contains formal proofs of the lemmas and theorems presented in Section \ref{sec:ModelAnalysis}. Listed below is an overview of the notation and mathematical framework utilized in the proofs. For an in-depth look into the function classes utilized in this work, and many other closely related topics in the field of harmonic analysis, the reader is referred to \cite{fabec,grafakos}. 
%----------------------------
\subsection{Notation and Mathematical Framework}
Firstly, let $\mathbb{R}_+:=\{x \in \mathbb{R} \hspace{3pt}:\hspace{3pt} x\geq 0\}$ be the space of non-negative real numbers and $dt$ denote the Lebesgue measure for any complex-valued measurable function $\varphi$. The $L^1(\mathbb{R}_+)$ norm of $\varphi$ is defined as   
%----------------------------
\begin{equation*}
	\| \varphi\|_1:= \int_0^\infty | \varphi | dt.
\end{equation*}
%---------------------------
Extensive use is made of the space of essentially bounded functions $L^\infty(\mathbb{R}_+)$ characterized by the following norm
%----------------------------
\begin{equation*}
	\| \varphi\|_\infty= \esssup\limits_{t \in \mathbb{R}_+} | \varphi |:=\inf \left\lbrace B>0 : dt( \left\lbrace t \geq 0: | \varphi |>B \right				\rbrace)=0 \right\rbrace.  
\end{equation*}
%---------------------------
Analogously to the case of the essential supremum, the essential infimum is defined as  
%----------------------------
\begin{equation*}
	\essinf\limits_{t \in \mathbb{R}_+} \| \varphi \|:=\sup \left\lbrace B>0 : dt( \left\lbrace t \geq 0: | \varphi |<B \right\rbrace)=0 \right\rbrace. 
\end{equation*}
%----------------------------
Some of the proofs require at least a bounded first derivative, however a general $\varphi \in C^1(\mathbb{R}_+)$ is not bounded. Due to this fact, we need to restrict our analysis to a smaller space possessing a higher degree of regularity. The most natural space to turn to is the sub-space denoted $C^1_b(\mathbb{R}_+) \subset  C^1(\mathbb{R}_+)$. This is the Banach space of bounded continuous functions whose first derivative is also bounded, endowed with the norm
%----------------------------
\begin{equation*}
	C^1_b(\mathbb{R}_+):=\{\varphi \in C^1(\mathbb{R}_+) \hspace{3pt}:\hspace{3pt} \| \varphi \|_{C^1_b} :=\|\varphi\|_\infty +\|\dot{\varphi}\|_\infty <+\infty\}.
\end{equation*}
%----------------------------
The convolution of two functions is defined as $\varphi_1 \ast \varphi_2(t)= \int_{0}^t \varphi_1(\tau)\varphi_2(t-\tau)d\tau$. Whenever $\varphi_1 \ast \varphi_2(t) \in L^1(\mathbb{R}_+)$ the integral is finite and, as a result, well-defined. 
%##############################################
%---------------------------------------------------------------------------------
%##############################################
%---------------------------------------------------------------------------------
%##############################################
\subsection{Survival Analysis}
Let $T$ be a non-negative continuous random variable,  standing for survival time, with probability density function $z(t)$ and cumulative distribution function $Z(t):=\text{P} \left \lbrace T<t \right \rbrace$. Define the survival function to be
%----------------------------
\begin{align*}
\nu(t) &:=\text{P} \left \lbrace T \geq t \right \rbrace, \\
     &= 1 - Z(t), \\
     &= \int_{t}^\infty z(x)dx.
\end{align*}
%----------------------------
In the above, $Z$ represents the probability that the event of interest has occurred by duration $t$. In the context of mosquito survivorship, the event of interest is death and the waiting time is survival time. As a result, $T$ stands for the waiting time until death occurs. It then stands to reason to define the survival function $\nu$ as the complement of $Z$. It follows that $\nu$ gives the probability of being alive just before duration $t$, or in other words, the probability that death has not occurred by duration $t$. The probability of dying on or before $t+\Delta t$, given survival to age $t$ is $\text{P} \left \lbrace T < t + \Delta t | T \geq t \right \rbrace$. An alternative characterization the distribution $T$ is given in terms of the hazard function $\xi_M(t)$ (or force of mortality), which quantifies the instantaneous rate of occurrence of death
%----------------------------
\begin{equation*}
\xi_M(t):=\lim_{\Delta t \rightarrow 0^+}  \frac{\text{P} \left \lbrace T < t + \Delta t | T \geq t \right \rbrace}{\Delta t}.
\end{equation*}
%----------------------------
The mosquito mortality rate $\xi_M$ was chosen for convenience and the derivation for the corresponding human rate is similar. The numerator is the conditional probability that death will occur on or before time $t+\Delta t$, provided that it has not previously occurred, and the denominator is an infinitesimal increment of time. A straightforward calculation yields 
%----------------------------
\begin{align*}
\xi_M(t) &:=\lim_{\Delta t \rightarrow 0^+}  \frac{\text{P} \left \lbrace T < t + \Delta t | T \geq t \right \rbrace}{\Delta t},\\
	 &=\lim_{\Delta t \rightarrow 0^+}  \frac{\text{P} \left \lbrace t \leq T < t + \Delta t \right \rbrace}{\text{P} \left \lbrace T \geq t \right \rbrace \Delta t},  \\
	 &=\lim_{\Delta t \rightarrow 0^+} \frac{Z(t+\Delta t)-Z(t)}{\nu(t) \Delta t},        \\
     &= \frac{z(t)}{\nu(t)}, \\
     &= -\frac{d}{dt} \ln \nu(t).
\end{align*}
%----------------------------
The above equation implies that the rate of occurrence of death at duration $t$ is equivalent to the density of deaths at $t$, divided by the probability of surviving to that duration without dying. Upon integrating from $0$ to $t$ and imposing the boundary condition $\nu(0)=1$, which makes sense, since death surely has not occurred by time $0$. Placing this together, we arrive at an explicit formula for the probability of being alive up to duration $t$ which has functional dependence on the hazard function for all duration up to $t$:
%----------------------------
\begin{equation*}
\nu(t):=e^{-\int_{0}^t \xi_M(s)ds}.
\end{equation*}
%----------------------------
The hazard function is constant, i.e. $\xi_M(t) \equiv \xi_M$, under the assumption of constant risk over time. Thus, the survival function corresponding to the mosquito population is 
%----------------------------
\begin{equation}
\nu(t)=e^{-\xi_M t}.
\label{survFUN}
\end{equation}
%----------------------------
This is a temporally dependent exponential distribution with mosquito mortality parameter $\xi_M$. 
%----------------------------
Additionally, the expected value of $T$ is given by the integral equation 
%----------------------------
\begin{equation*}
\mathbb{E}(T):=\int_{0}^\infty t z(t) dt.
\end{equation*}
%----------------------------
By definition, we must have $\int_{0}^\infty z(t) dt=1$. Weighting an average by a function which does not integrate to unity, i.e. $\int_{0}^\infty z(t) dt \neq 1$, is accomplished by forcing the weighting function to behave like a probability measure by the common normalization technique, similar to that used in \ref{HBI} of Section \ref{sec:ParamEst}. More precisely, 
%----------------------------
\begin{equation*}
\tilde{\mathbb{E}}(T):=\frac{\int_{0}^\infty t z(t) dt}{\int_{0}^\infty z(t) dt}.
\end{equation*}
%----------------------------
%##############################################
%---------------------------------------------------------------------------------
%##############################################
%---------------------------------------------------------------------------------
%##############################################
\subsection{Proofs}
%---------------------------------------------------------------------------------
\subsubsection[Proof of the Existence Theory of SEYAR Model]{Proof of Theorem~\ref{thm:SEYARlwp} (See page~\pageref{thm:SEYARlwp})}
%---------------------------------------------------------------------------------
\begin{proof}
The functions under consideration represent bounded, continuous, smoothly-varying, non-negative biological quantities. Consequently, it is reasonable to assume that $\mathbf{x} \in C^1_b(\mathbb{R}_+)$. This reasoning combined with the fact that the vector field $\Phi \in C^1$ is locally Lipschitz implies the existence of a unique solution $\mathbf{x}$. Moreover, the solution $\mathbf{x}$ depends continuously on the initial data and model parameters and can be continued to a maximal time interval, cf. \cite{HSD}. 
\end{proof}
%##############################################
%---------------------------------------------------------------------------------
%##############################################
%---------------------------------------------------------------------------------
%##############################################
\subsection{Proofs}
%---------------------------------------------------------------------------------
\subsubsection[Proof of the Initial Data Constraint of SEYAR Model]{Proof of Theorem~\ref{thm:SEYARlwp} (See page~\pageref{thm:IntData})}
%---------------------------------------------------------------------------------
\begin{proof}
In view of the well-posedness of the SEYAR model (\ref{SEYAR_DS}), it follows that the continuous composition of functions $u(t)=(u \circ \varepsilon)(t)$ solves the IVP (\ref{NAI_IVP}) for all $\varepsilon \in C^2(\mathbb{R}_+)$ in the solution space. By the non-negativity of $\varepsilon$, if the the initial data $\varepsilon_0$ is such that
\begin{equation*}
e^{\varepsilon_0}(u_{_{\low}}-u_{_{\high}\hspace{5pt}}) + u_{_{\high}\hspace{5pt}}\geq 0,
\end{equation*}
then the solution $u \in C^2(\mathbb{R}_+)$ is non-negative for all $t \in \mathbb{R}_+$. The above inequality is satisfied provided $\varepsilon_0 \leq \vartheta$ for $\vartheta$ defined by equation (\ref{IntCon}) in Section \ref{sec:Methods}. 
\end{proof}
%---------------------------------------------------------------------------------
%##############################################
%---------------------------------------------------------------------------------
%##############################################
%---------------------------------------------------------------------------------
%##############################################
\subsubsection[Proof of the Feasible Region of SEYAR Model]{Proof of Theorem~\ref{thm:Reg} (See page~\pageref{thm:Reg})}
%---------------------------------------------------------------------------------
\begin{proof}
The case for the mosquito population is trivial, viz. 
%---------------------------------------------------------------------------------
\begin{equation*}
		\lim_{t \rightarrow +\infty}N_M(t) = \frac{\Omega_M}{\xi_M}.
\end{equation*}
%---------------------------------------------------------------------------------
For the human population, let $L_t=e^{\int_{0}^t\xi_H ds}=e^{\small{\xi_H t}}$, so that
%---------------------------------------------------------------------------------
\begin{align*} 
		\dot{(L_tN_H)} &=  L_t \Omega_H-\delta L_t Y, \\ 
	        L_t N_H(t) &= N_H(0) + \int_{0}^t L_s\Omega_H ds- \delta \int_{0}^t L_s Y(s)ds,\\    
	                     &= N_H(0) + \Omega_H \left( \frac{e^{\small{\xi_H t}}}{\xi_H} - \frac{1}{\xi_H} \right)
	                 		 - \delta \int_{0}^t e^{\small{\xi_Hs}}  Y(s)ds.\\                     
\end{align*} 
%---------------------------------------------------------------------------------
Therefore, the explicit solution is given by 
%---------------------------------------------------------------------------------
\begin{align*}
		N_H(t) &= e^{\small{-\xi_H t}} N_H(0) + \frac{\Omega_H }{\xi_H}\left(1-e^{\small{-\xi_H t}}\right)
		- \delta \int_{0}^t e^{\small{-\xi_H(t-s)}} Y(s)ds, \\ 
\end{align*}
%---------------------------------------------------------------------------------
or in operator form
%---------------------------------------------------------------------------------
\begin{equation*}
		N_H(t)=L_{-t}N_H(0)+ \frac{\Omega_H }{\xi_H}\left(1-L_{-t}\right) -\delta L_{-t} \ast Y(t). 
\end{equation*}
%---------------------------------------------------------------------------------
After passing to the limit, it follows that
%---------------------------------------------------------------------------------
\begin{equation}
		\lim_{t \rightarrow +\infty}N_H(t) = \frac{\Omega_H}{\xi_H}-\delta \lim_{t \rightarrow +\infty}L_{-t} \ast Y(t).
	\label{Total_longterm}
\end{equation}  
%---------------------------------------------------------------------------------
In the absence of symptomatic infection, i.e. $Y=0$, it follows that $L_{-t} \ast Y(t)=0$, so that
$$\lim_{t \rightarrow +\infty}N_H(t) = \frac{\Omega_H}{\xi_H}.$$
As expected, $N_H$ asymptotes to the equilibrium population density of the human population. If symptomatic infection occurs in the population, by H$\ddot{o}$lders inequality it follows that 
%---------------------------------------------------------------------------------
\begin{align*}
		L_{-t} \ast Y(t) &\leq \| Y \|_\infty \| L_{-(t-s)}\|_1, \\
		&= \| Y \|_\infty\int_{0}^t e^{\small{-\xi_H(t-s)}}ds, \\
		&=\| Y \|_\infty \left[ \frac{1}{\xi_H} -\frac{e^{\small{-\xi_H t}}}{\xi_H}\right].
	\label{hhh}
\end{align*}
%---------------------------------------------------------------------------------
Where the right-hand side of the above inequality is finite due to the assumption that $Y \in C^2(\mathbb{R}_+) \cap C^1_b(\mathbb{R}_+) \subset L^\infty(\mathbb{R}_+)$. By letting $t\rightarrow +\infty$ we arrive at the following estimate for the forcing term
%---------------------------------------------------------------------------------
\begin{equation*}
		\lim_{t \rightarrow +\infty} L_{-t} \ast Y(t)  \leq  \frac{\| Y \|_\infty}{\xi_H}.
\end{equation*}
%---------------------------------------------------------------------------------
The above estimate in combination with (\ref{Total_longterm}) yields the following lower bound 
$$\lim_{t \rightarrow +\infty}N_H(t) \geq \frac{\Omega_H}{\xi_H}- \delta \frac{\| Y \|_\infty}{\xi_H}.$$
Therefore,
$$\lim_{t \rightarrow +\infty}N_H(t) \in \left[ \frac{\Omega_H - \delta \| Y \|_\infty }{\xi_H} ,\frac{\Omega_H}{\xi_H}  \right ].$$
Furthermore, for system (\ref{Total}) it follows that
%---------------------------------------------------------------------------------
\begin{equation*}
		\dot{N}_H \leq \Omega_H-\xi_H N_H = \xi_H \left(\frac{\Omega_H}{\xi_H} -	N_H\right).
\end{equation*}
%---------------------------------------------------------------------------------
Combining the above observation with a similar argument for $N_M$ yields 
%---------------------------------------------------------------------------------
\begin{equation}
		\begin{cases}
			\dot{N}_H \leq 0 ,\quad \text{if} \quad N_H \geq \frac{\Omega_H}{\xi_H},\\
			\dot{N}_M \leq 0 ,\quad \text{if} \quad N_M \geq \frac{\Omega_M}{\xi_M}.\\
		\end{cases} 
	\label{Total_ineqs}
\end{equation}
%---------------------------------------------------------------------------------
Inequalities (\ref{Total_ineqs}) imply that if the solution leaves the region $\Gamma$, then its derivative will instantaneously become negative, forcing it back to $\Gamma$. Moreover, if $x_i(0)=0$ for any $1 \leq i \leq 8$ in system (\ref{DS_rewrite}), then it directly follows that $\dot{x}_i \geq 0$. Therefore, all trajectories tend to $\Gamma$ and are forward invariant. Due to this fact it is sufficient to study the dynamics of the system on the smaller compact sub-space $\Gamma$.
\end{proof}
%##############################################
%---------------------------------------------------------------------------------
%##############################################
%---------------------------------------------------------------------------------
%##############################################
\subsubsection[Proof of the Local Asymptotic Stability of DFE for SEYAR Model]{Proof of Lemma~\ref{thm:LAS} (See page~\pageref{thm:LAS})}
%---------------------------------------------------------------------------------
\begin{proof}
Firstly, we order the compartments so that the first five correspond to infected individuals and denote $\mathbf{w}=\left(E,Y,A,M_E,M_I,R,S,M_S\right)^T$. The corresponding DFE is
%---------------------------------------------------------------------------------
\begin{equation*}
		\mathbf{w}_{\small{dfe}}=\left(0,0,0,0,0,0,\frac{\Omega_H}{\xi_H},\frac{\Omega_M}{\xi_M}\right)^T.
		\label{DFE}
\end{equation*}
%---------------------------------------------------------------------------------
Following the next generation method, system (\ref{SEYAR_DS}) is rewritten in the following form
%---------------------------------------------------------------------------------
\begin{equation*}
		\dot{\mathbf{w}}=\Phi \left(\mathbf{w}\right)=\mathcal{F}\left(\mathbf{w}\right)-\mathcal{V}\left(\mathbf{w}\right),
\end{equation*}
%---------------------------------------------------------------------------------
where $\mathcal{F}:=\left(\mathcal{F}_1,\dots,\mathcal{F}_8\right)^T$ and $\mathcal{V}:=\left(\mathcal{V}_1,\dots,\mathcal{V}_8\right)^T$, or more explicitly
%---------------------------------------------------------------------------------
\begin{displaymath}
		\begin{pmatrix}
			\dot{E}\\
			\dot{Y}\\
			\dot{A}\\
			\dot{M}_E\\
			\dot{M}_I\\
			\dot{R}\\
			\dot{S}\\
			\dot{M}_S\\
 		\end{pmatrix} 
 		=
   		\begin{pmatrix}
			\sigma \beta_M \frac{M_I}{N_H}S  \\
			0 \\
			0 \\
			\frac{\sigma\left ( \beta_Y Y + \beta_A A\right )}{N_H} M_S\\
			0 \\
			0 \\
			0 \\
			0 \\
 		\end{pmatrix} 
		 -
   		\begin{pmatrix}
			(\gamma +\xi_H)E\\
			-\gamma\left(1-u(\varepsilon)\right) E + \left(\xi_H +\delta + \lambda_{YR}\right)Y\\
			-\gamma u(\varepsilon) E + \left(\lambda_{AR}+\xi_H\right)A \\
 			\left(\xi_M + \tau \right)M_E\\
			-\tau M_E + \xi_M M_I\\
			-\lambda_{AR}A -  \lambda_{YR}Y +\left(\lambda_{RS} +\xi_H\right) R \\
			-\Omega_H - \lambda_{RS} R +\left(\sigma \beta_M \frac{M_I}{N_H} + \xi_H \right)S\\
			-\Omega_M  + \left ( \xi_M +  \frac{\sigma\left ( \beta_Y Y + \beta_A 	A\right )}{N_H} \right )M_S \\
 		\end{pmatrix}.
\end{displaymath}
%---------------------------------------------------------------------------------
In addition, the matrix $\mathcal{V}$ admits the decomposition $\mathcal{V}=\mathcal{V}^{-}-\mathcal{V}^{+}$, where the component-wise definition is inherited. Biologically speaking: $\mathcal{F}_i$ is the rate of appearance of new infections in compartment $i$, $\mathcal{V}^{+}_i$ stands for the rate of transfer of individuals into compartment $i$ by any other means and $\mathcal{V}^{-}_i$ is the rate of transfer of individuals out of compartment $i$. It is easy to see that $\mathcal{F},\mathcal{V}^{-},\mathcal{V}^{+}$ satisfy assumptions (\ref{10})-(\ref{50}) in Theorem (\ref{Rtheo}). As mentioned in the beginning of Section \ref{sec:WellPosedness}, to study the stability of the equilibrium points we assume that each of the above vector fields is at least twice-continuously differentiable. Define $U_{_{\low}\hspace{2pt}}:=e^{\varepsilon_0}(u_{_{\low}}-u_{_{\high}\hspace{5pt}}) + u_{_{\high}\hspace{5pt}}$ and let $F$ and $V$ be the following sub-matrices of the Jacobian of the above system, evaluated at the solution $\mathbf{w}_{\small{dfe}}$             
{\small
%---------------------------------------------------------------------------------
\begin{displaymath}
	F=
		\begin{pmatrix}
 			\frac{ \partial \mathcal{F}_i}{\partial x_j} \Big |_{\mathbf{w}_{\small{dfe}}} \\[5pt]
 		\end{pmatrix}_{1 \leq i,j \leq 5}
	 	=
  		\begin{pmatrix}
			0 & 0 & 0 & 0 & \sigma \beta_M \\
			0 & 0 & 0 & 0 & 0 \\
			0 & 0 & 0 & 0 & 0 \\
			0 & \sigma \beta_Y \frac{\Omega_M}{\Omega_H} \frac{\xi_H}{\xi_M}& \sigma \beta_A \frac{\Omega_M}{\Omega_H} \frac{\xi_H}	   		    		{\xi_M} & 0 & 0 \\
			0 & 0 & 0 & 0 & 0 \\
 		\end{pmatrix} 
\end{displaymath}
%---------------------------------------------------------------------------------
}
and
{\small
%---------------------------------------------------------------------------------
\begin{displaymath}
	V=
  		\begin{pmatrix}
			\frac{ \partial \mathcal{V}_i}{\partial x_j} \Big 	|_{\mathbf{w}_{\small{dfe}}} \\	
 		\end{pmatrix}_{1 \leq i,j \leq 5}
		 =
  		\begin{pmatrix}
			(\gamma +\xi_H) & 0 & 0 & 0 & 0\\
			\gamma \left(U_{_{\low}} -1 \right) & \left(\xi_H +\delta + \lambda_{YR}\right) & 0 & 0 & 0 \\
			-\gamma U_{_{\low}}  & 0 & \left(\lambda_{AR}+\xi_H\right) & 0 & 0 \\
			0 & 0 & 0 & (\xi_M+\tau) & 0 \\
			0 & 0 & 0 & -\tau & \xi_M \\
 		\end{pmatrix}. 
\end{displaymath}
%---------------------------------------------------------------------------------
}
A straight-forward calculation shows that
{\small
%---------------------------------------------------------------------------------
\begin{displaymath}
		V^{-1 }=
		\begin{pmatrix}
			(\gamma +\xi_H)^{-1} & 0 & 0 & 0 & 0\\
			-\frac{\gamma \left(U_{\low \hspace{5pt}}-1 \right)}{(\gamma +\xi_H)\left(\xi_H +\delta + \lambda_{YR}\right)} &\left(\xi_H +\delta + 				\lambda_{YR}\right)^{-1} & 0 & 0 & 0 \\
			\frac{\gamma U_{\low}\hspace{5pt}}{(\gamma +\xi_H) \left(\lambda_{AR}+\xi_H\right)}  & 0 & \left(\lambda_{AR}+\xi_H\right)^{-1} & 				0 & 0 \\
			0 & 0 & 0 & (\xi_M+\tau)^{-1} & 0 \\
			0 & 0 & 0 & \frac{\tau}{\xi_M(\xi_M+\tau)}& \xi_M^{-1} \\
 		\end{pmatrix}
\end{displaymath}
%---------------------------------------------------------------------------------
}
and $FV^{-1}$ is given by the following matrix
{\footnotesize
%---------------------------------------------------------------------------------
\begin{displaymath}
  		\begin{pmatrix}
			0 & 0 & 0 & \frac{\sigma \beta_M \tau}{\xi_M(\xi_M+\tau)}& \frac{\sigma 
			\beta_M}{\xi_M}\\
			0 & 0 & 0 & 0 & 0\\
			0 & 0 & 0 & 0 & 0\\
			\frac{\sigma \gamma \Omega_M \xi_H}{(\gamma + \xi_H)\Omega_H 	\xi_M}\left(\frac{\beta_A U_{_{\low}}}{\lambda_{AR}+\xi_H} -						\frac{\beta_Y\left(U_{_{\low}\hspace{5pt}} - 1\right)}{\xi_H +\delta + \lambda_{YR}}\right) & \frac{\sigma \beta_Y \Omega_M \xi_H}					{(\xi_H + \delta + \lambda_{YR})\Omega_H \xi_M} & \frac{\sigma \beta_A 	\Omega_M \xi_H}{(\xi_H + \lambda_{AR})\Omega_H \xi_M}				& 0 & 0 \\
			0 & 0 & 0 & 0 & 0\\
 		\end{pmatrix}.
\end{displaymath}
%---------------------------------------------------------------------------------
}
Let $\mathcal{I}$ denote the $5 \times 5$ identity matrix, so that the characteristic polynomial $P(\lambda)$ of the matrix $FV^{-1}$ is given by
%---------------------------------------------------------------------------------
\begin{align*}
	P(\lambda)&= \det{\left(FV^{-1}-\lambda  \mathcal{I} \right)},\\
	&=\lambda^3 \left( \lambda^2 - \left( \frac{\sigma^2 \tau \gamma \Omega_M 	\xi_H \beta_M}{\xi_M^2(\gamma + \xi_H)(\tau + \xi_M)				\Omega_H} \left(\frac{\beta_AU_{_{\low}\hspace{5pt}} }{\lambda_{AR} + \xi_H} - \frac{\beta_Y(U_{_{\low}}-1)}{\lambda_{YR} + \xi_H + 				\delta} \right) \right) \right).
\end{align*}
%---------------------------------------------------------------------------------
The solution set $\lbrace \lambda_i \rbrace_{1 \leq i \leq 5}$ is given by
%---------------------------------------------------------------------------------
\begin{equation*}
	 \left\lbrace 0,0,0, \pm\sqrt{\frac{\sigma^2 \tau \gamma \Omega_M \xi_H \beta_M}						{\xi_M^2(\gamma + 	 \xi_H)(\tau + \xi_M)\Omega_H} \left(\frac{\beta_AU_{_{\low}\hspace{5pt}} }{\lambda_{AR} + \xi_H} - \frac{\beta_Y(U_{_{\low}}-1)}{\lambda_{YR} + \xi_H + 				\delta} \right)}\right\rbrace.
\end{equation*}
%---------------------------------------------------------------------------------
Therefore, the reproductive threshold for the $SEY \hspace{-3pt}AR$ model (\ref{SEYAR_DS}) is given by
%---------------------------------------------------------------------------------
\begin{align*}
	\mathcal{R}_0 & :=\rho\left(FV^{-1}\right),   \\
	&=\max\limits_{1 \leq i \leq 5} {\lbrace \lambda_i \rbrace },  \\ 
	&=\sqrt{\frac{\sigma^2 \tau \gamma \Omega_M \xi_H \beta_M}{\xi_M^2(\gamma + \xi_H)(\tau + \xi_M)\Omega_H} \left(\frac{\beta_AU_{_{\low}\hspace{5pt}} }{\lambda_{AR} + \xi_H} - \frac{\beta_Y(U_{_{\low}}-1)}{\lambda_{YR} + \xi_H + \delta} \right)}.\\
\end{align*}
%---------------------------------------------------------------------------------
The proof of the lemma regarding the local asymptotic stability (LAS) of the DFE $\mathbf{w}_{\small{dfe}}$ corresponding to the $SEY \hspace{-3pt}AR$ Model (\ref{SEYAR_DS}) is now complete after invoking Theorem (\ref{Rtheo}) in Section \ref{sec:A2}.
\end{proof}
%##############################################
%---------------------------------------------------------------------------------
%##############################################
%---------------------------------------------------------------------------------
%##############################################
\subsubsection[Impact of Asymptomatic Class on Reproductive Threshold]{Proof of Theorem~\ref{thm:R_Comp} (See page~\pageref{thm:R_Comp})}
%---------------------------------------------------------------------------------
\begin{proof}
Let $\Theta_0$, $\tilde{\Theta}_0$, $\mathcal{R}_A$ and $\mathcal{R}_Y$ be defined as in the statement of the theorem. Additionally, denote the following quantities $r^0_6:=r_6\Big |_{\Theta_0}$, $C^0_1:=C_1\Big |_{\Theta_0}$, $C^0_2:=C_2\Big |_{\tilde{\Theta}_0}$ and $C^0_0:=C_0\Big |_{\Theta_0}\hspace{-3pt}=C_0\Big |_{\tilde{\Theta}_0}$, where the terms $r_6$ and $C_i$ for $i=0,1,2$ are defined as in Section \ref{sec:Rnumber}. By the monotonicity of the square foot function, it follows that 
%---------------------------------------------------------------------------------
{\small
\begin{align*}
		\mathcal{R}_A &:=\mathcal{R}_0\Big |_{\Theta_0}= C^0_0 \sqrt{r^0_6}> C^0_0\sqrt{r^0_6-U_{{_{\low}\hspace{5pt}\hspace{-1pt}}_{_0}}C^0_1}= C^0_0\sqrt{\left(1-U_{{_{\low}\hspace{5pt}\hspace{-1pt}}_{_0}}\right)C^0_2}=\mathcal{R}_0\Big |_{\tilde{\Theta}_0}:=\mathcal{R}_Y. \\		
\end{align*}
}
%---------------------------------------------------------------------------------
\end{proof}
%---------------------------------------------------------------------------------
%##############################################
%---------------------------------------------------------------------------------
%##############################################
%---------------------------------------------------------------------------------
%##############################################
\subsubsection[Proof of the Bifurcation for SEYAR Model]{Proof of Lemma~\ref{thm:Bif} (See page~\pageref{thm:Bif})}
%---------------------------------------------------------------------------------
\begin{proof}
If we consider the dynamical system $\dot{x}=g(x,\omega)$, where $\omega$ is a bifurcation parameter and the vector field $g$ is $C^2$ in both $x$ and $\omega$, then the disease-free equilibrium can be viewed as the manifold $(\mathbf{x}_{\small{dfe}};\omega)$ where the local stability of $\mathbf{x}_{\small{dfe}}$ changes at the point $(\mathbf{x}_{\small{dfe}};\omega^\star)$. Now we shall investigate the existence and stability of non-trivial equilibrium states in a neighborhood of the bifurcation point. We focus on the disease-free equilibrium $\mathbf{x}_{\small{dfe}}$ and study the occurrence of a transcritical bifurcation at $\mathcal{R}_0=1$. Since $\mathcal{R}_0$ consists of the square root of a complicated combination of parameters, it is not practical to use as a bifurcation parameter. However, observe that $\mathcal{R}_0=1$ if and only if 
%---------------------------------------------------------------------------------
\begin{align*}
	\beta_M &=\frac{\xi_M^2 \Omega_H (\gamma + \xi_H)(\tau + \xi_M)(\lambda_{AR} + \xi_H)(\lambda_{YR} + \xi_H + \delta)}{\sigma^2 \tau  			\xi_H \Omega_M  \left(\gamma U_{_{\low}\hspace{5pt}} \beta_A (\lambda_{YR} + \xi_H + \delta)- \gamma 	(U_{_{\low}}-1) \beta_Y(\lambda_{AR} + \xi_H)				\right)},\\ 
	&:= \beta_M^\star.  
\end{align*}
%---------------------------------------------------------------------------------
In lieu of Lemma (\ref{thm:LAS}) it follows that $\mathbf{x}_{\small{dfe}}$ is locally asymptotically stable when $\beta_M<\beta_M^\star$ and unstable if $\beta_M>\beta_M^\star$. Thus, the combination of parameters $\beta_M^\star$ is a bifurcation value. To simplify the notation, we rewrite system (\ref{SEYAR_DS}) as $\dot{x}=g(x,\beta_M)$ where $\mathbf{x}=\left(S,E,Y,A,R,M_S,M_E,M_I\right)^T$ so that $x_i$ is the $i^\text{th}$ component of $\mathbf{x}$ and $g=(g_1,g_2,g_3,g_4,g_5,g_6,g_7,g_8)^T$, or more explicitly
%---------------------------------------------------------------------------------
\begin{align}
		g_1&=\Omega_H + \lambda_{RS} x_5 - \left(\sigma \beta_M \frac{x_8}{\sum_i^5 x_i} + \xi_H \right)x_1,\nonumber \\ 
		g_2 &=\sigma \beta_M \frac{x_8}{\sum_i^5 x_i}x_1- (\gamma +\xi_H)x_2,\nonumber \\            
		g_3 &=\gamma\left(1-u(\varepsilon)\right) x_2 -  \left(\xi_H +\delta + 	\lambda_{YR}\right)x_3,\nonumber \\               
		g_4 &=\gamma u(\varepsilon) x_2 - \left(\lambda_{AR}+\xi_H\right)x_4, 	\label{rewriteG} \\              
		g_5 &=\lambda_{AR}x_4 +  \lambda_{YR}x_3 - \left(\lambda_{RS} +\xi_H\right) 	x_5,\nonumber  \\          
		g_6 &=\Omega_M  - \left ( \xi_M +  \sigma \beta_Y \frac{x_3}{\sum_i^5 x_i} +\sigma \beta_A \frac{x_4}{\sum_i^5 x_i}  \right )x_6,					\nonumber \\        
		g_7 &=\sigma \left ( \beta_Y \frac{x_3}{\sum_i^5 x_i} +\beta_A \frac{x_4}	{\sum_i^5 x_i}\right )x_6 - \left(\xi_M + \tau \right)x_7,					\nonumber \\      
		g_8 &=\tau x_7- \xi_M x_8.\nonumber   			 
\end{align}
%---------------------------------------------------------------------------------
Denote $J(\mathbf{x}_{\small{dfe}},\beta_M^\star)$  to be the Jacobian of $g$ evaluated at the DFE $\mathbf{x}_{\small{dfe}}$ and threshold $\beta_M^\star$. Thus $J(\mathbf{x}_{\small{dfe}},\beta_M^\star)$  is given by 
{\footnotesize
%---------------------------------------------------------------------------------
\begin{displaymath}
  	\begin{pmatrix}
		-\xi_H & 0 & 0 & 0 & \epsilon & 0 & 0 & -\sigma \beta_M^\star\\
		0 & -(\gamma + \xi_H) & 0 & 0 & 0 & 0 & 0 & \sigma \beta_M^\star\\
		0 & \gamma (1-U_{_{\low}\hspace{5pt}}) & -(\delta + \lambda_{YR} + \xi_H) & 0 & 0 & 0 & 0 & 0\\
		0 & \gamma U_{_{\low}} & 0 & -(\lambda_{AR} + \xi_H)& 0 & 0 & 0 & 0\\
		0 & 0 & \lambda_{YR} & \lambda_{AR} & -(\lambda_{RS} + \xi_H) & 0 & 0 & 0\\
		0 & 0 & -\frac{\sigma \beta_Y \xi_H \Omega_M}{\Omega_H \xi_M} & -	\frac{\sigma \beta_A \xi_H \Omega_M}{\Omega_H \xi_M} & 0 & -	        \xi_M & 0 & 0\\
		0 & 0 & \frac{\sigma \beta_Y \xi_H \Omega_M}{\Omega_H \xi_M} & \frac{\sigma \beta_A \xi_H \Omega_M}{\Omega_H \xi_M} & 0 & 0 & 	        -(\tau + \xi_M) & 0\\
		0 & 0 & 0 & 0 & 0 & 0 &\tau & -\xi_M\\
 	\end{pmatrix}.
\end{displaymath}
%---------------------------------------------------------------------------------
}
If we invoke the following positive change of variables
%---------------------------------------------------------------------------------
\begin{align*} 
	\begin{cases}
      	K_1 &=  \gamma + \xi_H,\\ 
	  	K_2 &= \gamma (1-U_{_{\low}\hspace{5pt}}),\\    
	  	K_3 &= \gamma U_{_{\low}\hspace{5pt}},\\  
      	K_4 &= \delta + \lambda_{YR} + \xi_H,\\      
      	K_5 &=  \lambda_{AR} + \xi_H,\\ 
	  	K_6 &= \lambda_{RS} + \xi_H,\\    
	  	K_7 &= \tau + \xi_M,\\    
	\end{cases}                   
\end{align*} 
%---------------------------------------------------------------------------------
then $\beta_H^\star$ can be written as
%---------------------------------------------------------------------------------
\begin{equation*}
	\beta_M^\star =\frac{\xi_M^2 \Omega_H K_1 K_4 K_5 K_7}{\sigma^2 \tau \xi_H 	\Omega_M  \left(\beta_A K_3 K_4 + \beta_Y K_2 K_5 			\right)}.  \\ 
\end{equation*}
%---------------------------------------------------------------------------------
As a result, $J(\mathbf{x}_{\small{dfe}},\beta_M^\star)$ can be written as 
{\footnotesize
%---------------------------------------------------------------------------------
\begin{displaymath}
  	\begin{pmatrix}
		-\xi_H & 0 & 0 & 0 & \epsilon & 0 & 0 & -\frac{\xi_M^2 \Omega_H K_1 K_4 K_5 K_7}{\sigma \tau \xi_H \Omega_M  \left(\beta_A K_3        		K_4 + \beta_Y K_2 	K_5 \right)}\\
		0 & -K_1 & 0 & 0 & 0 & 0 & 0 & \frac{\xi_M^2 \Omega_H K_1 K_4 K_5 K_7}	{\sigma \tau \xi_H \Omega_M  \left(\beta_A K_3 K_4 +    		\beta_Y K_2 K_5 \right)}\\
		0 & K_2 & -K_4 & 0 & 0 & 0 & 0 & 0\\
		0 & K_3 & 0 & -K_5 & 0 & 0 & 0 & 0\\
		0 & 0 & \lambda_{YR} & \lambda_{AR} & -K_6 & 0 & 0 & 0\\
		0 & 0 & -\frac{\sigma \beta_Y \xi_H \Omega_M}{\Omega_H \xi_M} & -\frac{\sigma \beta_A \xi_H \Omega_M}{\Omega_H \xi_M} & 0 & -        		\xi_M & 0 & 0\\
		0 & 0 & \frac{\sigma \beta_Y \xi_H \Omega_M}{\Omega_H \xi_M} & \frac{\sigma \beta_A \xi_H \Omega_M}{\Omega_H \xi_M} & 0 & 0 &    		-K_7 & 0\\
		0 & 0 & 0 & 0 & 0 & 0 &\tau & -\xi_M\\
 	\end{pmatrix}.
\end{displaymath}
%---------------------------------------------------------------------------------
}
Due to the equivalency of the two conditions $\mathcal{R}_0=1$ and $\beta_M=\beta_M^\star$, it follows that $J(\mathbf{x}_{\small{dfe}},\beta_M^\star)$ contains information of the linearized system evaluated at the disease-free equilibrium and threshold value. Utilizing the machinery covered in Section \ref{sec:A2} below, the Jacobian evaluated at the threshold value, i.e. $J(\mathbf{x}_{\small{dfe}},\beta_M^\star)$ has a zero simple eigenvalue with all others having negative real parts. Therefore, the hypothesis of Theorem (\ref{thm:CM}) is satisfied. We proceed by calculating the $a$ and $b$ terms (\ref{a}) and (\ref{b}) appearing in Theorem (\ref{thm:CM}). In observance of the conclusions in the theorem, it follows that the $SEY \hspace{-3pt}AR$ model (\ref{SEYAR_DS}) will undergo a super-critical bifurcation if $a>0$ and $b>0$ and a sub-critical bifurcation if $a<0$ and $b>0$. The main ingredients in calculating $a$ and $b$ are the generalized right and left eigenvectors of the matrix $J(\mathbf{x}_{\small{dfe}},\beta_M^\star)$ and their corresponding non-zero Hessian entries, evaluated at the DFE $\mathbf{x}_{\small{dfe}}$. In this fashion we let $w =\left(w_1,w_2,w_3,w_4,w_5,w_6,w_7,w_8\right)$ and $v^T=\left(v_1,v_2,v_3,v_4,v_5,v_6,v_7,v_8\right)^T$ be right and left generalized eigenvectors of $J(\mathbf{x}_{\small{dfe}},\beta_M^\star)$, respectively. Since $J$ is not symmetric, the left and right generalized eigenspaces are not equivalent. Solving the equations $Jw=0$ and $\left(v^TJ\right)^T=J^Tv=0$, where $D_i \in \mathbb{R}_{\tiny{>0}}$ for $i=1,2$, yields 
%---------------------------------------------------------------------------------
\begin{displaymath}
  	\begin{pmatrix}
			w_1\\
			w_2\\
			w_3\\
			w_4\\
			w_5\\
			w_6\\
			w_7\\
			w_8\\
 	\end{pmatrix} 
	 =
  	D_1\begin{pmatrix}
			\frac{\epsilon \lambda_{AR} K_3 K_4 +\epsilon \lambda_{YR} K_2 K_5 -	K_1 K_6 K_4 K_5}{K^2_6 K_4 K_5}\\
			1\\
			\frac{K_2}{K_4}\\
			\frac{K_3}{K_5}\\
			\frac{\lambda_{YR} K_2 K_5 + \lambda_{AR} K_3 K_4}{K_4 K_5 K_6}\\
			-\frac{\sigma \xi_H \Omega_M(\beta_Y K_2 K_5 + \beta_A K_3 K_4)}{\Omega_H \xi^2_M  K_4 K_5}\\
			\frac{\sigma \xi_H \Omega_M \left(\beta_A K_3 K_4 + \beta_Y K_2 K_5\right)}{\xi_M \Omega_H K_4 K_5 K_7}\\
			\frac{\sigma \tau \xi_H \Omega_M \left(\beta_A K_3 K_4 + \beta_Y K_2 K_5\right)}{\xi^2_M \Omega_H K_4 K_5 K_7}\\
 	\end{pmatrix} 
 	\end{displaymath}
 	\begin{displaymath}
  	\begin{pmatrix}
			v_1\\
			v_2\\
			v_3\\
			v_4\\
			v_5\\
			v_6\\
			v_7\\
			v_8\\
 	\end{pmatrix} 
 	=
  	D_2\begin{pmatrix}
			0\\
			\frac{\sigma \tau \xi_H \Omega_M \left(\beta_A K_3 K_4 + \beta_Y K_2 K_5\right)}{\xi_M \Omega_H  K_1 K_4 K_5 K_7}\\
			\frac{\sigma \tau \beta_Y \xi_H \Omega_M}{\Omega_H \xi_M K_4 K_7}\\
			\frac{\sigma \tau \beta_A \xi_H \Omega_M}{\Omega_H \xi_M K_5 K_7}\\
			0\\
			0\\
			\frac{\tau}{K_7}\\
			1\\
 		\end{pmatrix}.
\end{displaymath}
%---------------------------------------------------------------------------------
For system (\ref{rewriteG}), the non-zero partial derivatives of $g$ evaluated at $\mathbf{x}_{\small{dfe}}$ are 
%---------------------------------------------------------------------------------
\begin{align*}
		\frac{\partial^2 g_1}{\partial x_8 \partial x_2} &= \frac{\partial^2 g_1}{\partial x_8 \partial x_3}=\frac{\partial^2 g_1}{\partial x_8 \partial 			x_4}=\frac{\partial^2 g_1}{\partial x_8 \partial x_5}=\frac{\sigma \beta_M \xi_H}{\Omega_H},\\
		\frac{\partial^2 g_2}{\partial x_8 \partial x_2} &= \frac{\partial^2 g_2}{\partial x_8 \partial x_3}=\frac{\partial^2 g_2}{\partial x_8 \partial 			x_4}=\frac{\partial^2 g_2}{\partial x_8 \partial x_5}=-\frac{\sigma \beta_M \xi_H}{\Omega_H},\\
		\frac{\partial^2 g_3}{\partial x_2 \partial x_2} &= \frac{2 \gamma \xi_H e^{\varepsilon_0}}{\Omega_H}(u_{_{\low}}-u_{_{\high}\hspace{5pt}}), \quad
		\frac{\partial^2 g_4}{\partial x_2 \partial x_2} = \frac{2 \gamma \xi_H e^{\varepsilon_0}}{\Omega_H} (u_{_{\high}}-u_{_{\low}\hspace{5pt}}),\\
		\frac{\partial^2 g_6}{\partial x_3 \partial x_1} &= \frac{\partial^2 g_6}	{\partial x_3 \partial x_2}=\frac{\partial^2 g_6}{\partial x_3 					\partial x_5}=\frac{\sigma \beta_Y \xi^2_H \Omega_M}{\Omega^2_H \xi_M},\\
		\frac{\partial^2 g_6}{\partial x_4 \partial x_1} &= \frac{\partial^2 g_6}	{\partial x_4 \partial x_2}=\frac{\partial^2 g_6}{\partial x_4 					\partial x_5}=\frac{\sigma \beta_A \xi^2_H \Omega_M}{\Omega^2_H \xi_M},\\
		\frac{\partial^2 g_6}{\partial x_3 \partial x_3} &= 2\frac{\partial^2 	g_6}{\partial x_3 \partial x_1},\quad
		\frac{\partial^2 g_6}{\partial x_4 \partial x_4} = 2\frac{\partial^2 g_6}	{\partial x_4 \partial x_1},\\
		\frac{\partial^2 g_6}{\partial x_6 \partial x_3} &= -\frac{\sigma \beta_Y 	\xi_H}{\Omega_H},\quad
		\frac{\partial^2 g_6}{\partial x_6 \partial x_4} = -\frac{\sigma \beta_A 	\xi_H}{\Omega_H},\\
			\end{align*}
		\begin{align*}
		\frac{\partial^2 g_6}{\partial x_3 \partial x_4} &= \frac{\sigma \xi^2_H 	\Omega_M (\beta_Y +\beta_A)}{\xi_M \Omega^2_H},\\			
		\frac{\partial^2 g_7}{\partial x_3 \partial x_1} &= \frac{\partial^2 g_7}	{\partial x_3 \partial x_2}=\frac{\partial^2 g_7}{\partial x_3 					\partial x_5}=-\frac{\sigma \beta_Y \xi^2_H \Omega_M}{\Omega^2_H \xi_M},\\
		\frac{\partial^2 g_7}{\partial x_4 \partial x_1} &= \frac{\partial^2 g_7}		{\partial x_4 \partial x_2}=\frac{\partial^2 g_7}{\partial x_4 				\partial x_5}=-\frac{\sigma \beta_A \xi^2_H \Omega_M}{\Omega^2_H \xi_M},\\
		\frac{\partial^2 g_7}{\partial x_3 \partial x_3} &= 2\frac{\partial^2 	g_7}{\partial x_3 \partial x_1},\quad
		\frac{\partial^2 g_7}{\partial x_4 \partial x_4} = 2\frac{\partial^2 g_7}	{\partial x_4 \partial x_1},\\
		\frac{\partial^2 g_7}{\partial x_6 \partial x_3} &= \frac{\sigma \beta_Y 	\xi_H}{\Omega_H},\quad
		\frac{\partial^2 g_7}{\partial x_6 \partial x_4} = \frac{\sigma \beta_A 	\xi_H}{\Omega_H},\\
		\frac{\partial^2 g_7}{\partial x_3 \partial x_4} &= -\frac{\sigma \xi^2_H 	\Omega_M (\beta_Y +\beta_A)}{\xi_M \Omega^2_H},\\
		\frac{\partial^2 g_1}{\partial x_8 \partial \beta_H} &= -\sigma,\quad
		\frac{\partial^2 g_2}{\partial x_8 \partial \beta_H} = \sigma.
\end{align*}
%---------------------------------------------------------------------------------
Due to the fact that the parameter $\beta_M$ only appears in system (\ref{rewriteG}) twice, the calculation of the $b$ term is less involved. Indeed, for $b$ we have 
%---------------------------------------------------------------------------------
\begin{align}
		b &=\sum_{i,k=1}^{8} v_k w_i \frac{\partial^2 g_k}{\partial x_i \partial 	\beta_H}\left(\mathbf{x}_{\small{dfe}},\beta_M^\star \right), 					\label{bterm} \\
 		&=v_1 w_8 \frac{\partial^2 g_1}{\partial x_8 \partial 	\beta_M}\left(\mathbf{x}_{\small{dfe}},\beta_M^\star \right) + v_2 w_8 							\frac{\partial^2 g_2}{\partial x_8 \partial \beta_M}\left(\mathbf{x}_{\small{dfe}},\beta_M^\star \right), \nonumber \\
  		&= \sigma v_2 w_8, \nonumber \\ 
  		&= \frac{\sigma^2 \tau \xi_H \Omega_M \left(\beta_A K_3 K_4 + \beta_Y K_2  K_5\right)}{\xi_M \Omega_H K_1 K_4 K_5 K_7}v_8 w_8, 		\nonumber \\    
		&= \frac{\sigma^3 \tau^2 \xi^2_H \Omega^2_M \left(\beta_A K_3 K_4 + \beta_Y K_2 K_5\right)^2}{\xi^3_M \Omega^2_H K_1 K^2_4 				K^2_5 K^2_7} w_2 	v_8, \nonumber \\  
 		&= \frac{\sigma^3 \tau^2 \xi^2_H \Omega^2_M \left(\beta_A K_3 K_4 + \beta_Y K_2 K_5\right)^2}{\xi^3_M \Omega^2_H K_1 K^2_4 				K^2_5 K^2_7} D_1 	D_2, \nonumber \\ 
		&>0.  \nonumber \\  \nonumber         
\end{align}
%---------------------------------------------------------------------------------
Since $D_i \in \mathbb{R}_{\tiny{>0}}$ for $i=1,2$, it follows that $b$ is always positive, hence the bifurcation behavior of system (\ref{rewriteG}) is completely determined by the sign of $a$. To simplify the pending calculation we invoke the following change of variables
%---------------------------------------------------------------------------------
\begin{align*} 
	\begin{cases}
      Z_1 &=  \frac{\sigma \beta_M \xi_H}{\Omega_H},\\ 
	  Z_2 &= \frac{\sigma \beta_Y \xi^2_H \Omega_M}{\Omega^2_H \xi_M},\\    
	  Z_3 &= \frac{\sigma \beta_A \xi^2_H \Omega_M}{\Omega^2_H \xi_M},\\  
      Z_4 &= \frac{\sigma \xi^2_H \Omega_M (\beta_Y +\beta_A)}{\xi_M \Omega^2_H},\\      
      Z_5 &=  \frac{\sigma \beta_A \xi_H}{\Omega_H},\\ 
	  Z_6 &= \frac{2 \gamma \xi_H e^{\varepsilon_0}}{\Omega_H} (u_{_{\high}}-u_{_{\low}\hspace{5pt}}),\\    
	  Z_7 &= \frac{\sigma \beta_Y \xi_H}{\Omega_H},\\
	  Q_0 &= \frac{\lambda_{YR} K_2 K_5 + \lambda_{AR} K_3 K_4}{K_4 K_5 K_6},\\     
	  Q_1 &= \frac{\beta_Y K_2 K_5 + \beta_A K_3 K_4}{K_4 K_5 K_7},\\ 
	  Q_2 &= \frac{\sigma \tau \xi_H \Omega_M \beta_Y}{\Omega_H \xi_M K_4 K_7},\\ 
      Q_3 &= \frac{\sigma \tau \xi_H \Omega_M \beta_A}{\Omega_H \xi_M K_5 K_7},\\ 
      Q_4 &= \frac{\sigma \xi_H \Omega_M}{\Omega_H \xi_M},\\ 
	\end{cases}                 
\end{align*}
%---------------------------------------------------------------------------------
so that the generalized eigenvectors can be written as:
%---------------------------------------------------------------------------------
\begin{displaymath}
  	\begin{pmatrix}
			w_1\\
			w_2\\
			w_3\\
			w_4\\
			w_5\\
			w_6\\
			w_7\\
			w_8\\
 	\end{pmatrix} 
	 =
  	D_1\begin{pmatrix}
			\frac{\lambda_{RS} Q_0-K_1}{K_6}\\
			1\\
			\frac{K_2}{K_4}\\
			\frac{K_3}{K_5}\\
			Q_0\\
			-\frac{Q_4 Q_1 K_7}{\xi_M}\\
			Q_4 Q_1\\
			\frac{\tau Q_4 Q_1}{\xi_M}\\
 	\end{pmatrix} 
		\quad and \quad
  	\begin{pmatrix}
			v_1\\
			v_2\\
			v_3\\
			v_4\\
			v_5\\
			v_6\\
			v_7\\
			v_8\\
 	\end{pmatrix} 
 	=
  	D_2\begin{pmatrix}
			0\\
			\frac{\tau Q_4 Q_1}{K_1}\\
			Q_2\\
			Q_3\\
			0\\
			0\\
			\frac{\tau}{K_7}\\
			1\\
 		\end{pmatrix}.
\end{displaymath}
%---------------------------------------------------------------------------------
In view of the relatively strong regularity assumptions mentioned in the beginning of Section \ref{sec:WellPosedness} and natural symmetry of the system, it follows that $\frac{\partial^2 g_k}{\partial x_i \partial x_j} v_k w_i w_j=\frac{\partial^2 g_k}{\partial x_j \partial x_i} v_k w_j w_i$ for all $1 \leq i,j \leq 8$. Furthermore, $\frac{\partial^2 g_k}{\partial x_i \partial x_j} v_k w_i w_j=0$ for $k=1,5,6,8$, since $\frac{\partial^2 g_8}{\partial x_i \partial x_j}=0$ for all $1 \leq i,j \leq 8$ and $v_k=0$ for $k=1,5,6$. As a result, the terms that contribute to the sum correspond to $k=2,3,4,7$. Thus, $a$ can be written as 
%---------------------------------------------------------------------------------
\begin{align} 
	a = & \sum_{i,j,k=1}^{8} v_k w_i w_j \frac{\partial^2 g_k}{\partial x_i \partial x_j}\left(\mathbf{x}_{\small{dfe}},\beta_M^\star \right) 				\nonumber\\
	= & 2\Bigg[v_2 w_8 w_2 \frac{\partial^2 g_2}{\partial x_8\partial x_2}\left(\mathbf{x}_{\small{dfe}},\beta_M^\star \right) + v_2 w_8 w_3 \frac{\partial^2 g_2}{\partial x_8\partial x_3}\left(\mathbf{x}_{\small{dfe}},\beta_M^\star \right) \nonumber \\
	& + v_2 w_8 w_4 \frac{\partial^2 g_2}{\partial x_8\partial x_4}\left(\mathbf{x}_{\small{dfe}},\beta_M^\star \right)+ v_2 w_8 w_5 \frac{\partial^2 g_2}{\partial x_8\partial x_5}\left(\mathbf{x}_{\small{dfe}},\beta_M^\star \right)  \nonumber \\
 	&  + v_3 w^2_2 	\frac{\partial^2 g_3}{\partial x_2\partial x_2}\left(\mathbf{x}_{\small{dfe}},\beta_M^\star \right) +v_4 w^2_2 \frac{\partial^2 g_4}{\partial x_2\partial x_2}\left(\mathbf{x}_{\small{dfe}},\beta_M^\star \right) \nonumber \\
	&+ v_7 w_3 w_1 \frac{\partial^2 g_7}{\partial x_3\partial x_1}\left(\mathbf{x}_{\small{dfe}},\beta_M^\star \right) + v_7 w_3 w_2 					\frac{\partial^2 g_7}{\partial x_3\partial x_2}\left(\mathbf{x}_{\small{dfe}},\beta_M^\star \right) \nonumber \\
	& +  v_7 w_3 w_5 	\frac{\partial^2 g_7}	{\partial x_3\partial x_5}\left(\mathbf{x}_{\small{dfe}},\beta_M^\star \right) +  v_7 w_4 w_1 \frac{\partial^2 g_7}{\partial x_4\partial x_1}\left(\mathbf{x}_{\small{dfe}},\beta_M^\star \right) \nonumber \\
	& + v_7 w_4 w_2 \frac{\partial^2 g_7}{\partial x_4\partial x_2}\left(\mathbf{x}_{\small{dfe}},\beta_M^\star \right) + v_7 w_4 w_5 \frac{\partial^2 g_7}{\partial 		x_4\partial x_5}\left(\mathbf{x}_{\small{dfe}},\beta_M^\star \right) \nonumber \\
 	&+ v_7 w^2_3\frac{\partial^2 g_7}{\partial x_3\partial x_3}\left(\mathbf{x}_{\small{dfe}},\beta_M^\star \right) +  v_7 w^2_4 						\frac{\partial^2 g_7}{\partial x_4\partial x_4}\left(\mathbf{x}_{\small{dfe}},\beta_M^\star \right)  	\label{aterm}  \\
 	& + v_7 w_6 w_3 \frac{\partial^2 g_7}{\partial 		x_6\partial x_3}\left(\mathbf{x}_{\small{dfe}},\beta_M^\star \right) +v_7 w_6 w_4 \frac{\partial^2 g_7}{\partial x_6\partial x_4}\left(\mathbf{x}_{\small{dfe}},\beta_M^\star \right)\nonumber \\
 	&  + v_7 w_3 w_4 					\frac{\partial^2 g_7}{\partial x_3\partial x_4}\left(\mathbf{x}_{\small{dfe}},\beta_M^\star \right) \Bigg]\nonumber \\
	= &2\Bigg[-Z_1 v_2 w_8 w_2 - Z_1 v_2 w_8 w_3 - Z_1 v_2 w_8 w_4 - Z_1 v_2 w_8 w_5 - Z_6 v_3 w^2_2 \nonumber \\
	& \quad + Z_6 v_4 w^2_2
 	- Z_2 v_7 w_3 w_1 -Z_2 v_7 w_3 w_2 - Z_2 v_7 w_3 w_5-Z_3 v_7 w_4 w_1 \nonumber \\
	 &\quad - Z_3 v_7 w_4 w_2 - Z_3 v_7 w_4 w_5 - 2Z_2 v_7 w^2_3 - 2Z_3 v_7 w^2_4 +Z_7 v_7 w_6 w_3 \nonumber\\
	 &\quad + Z_5 v_7 w_6 w_4 + Z_4 v_7 w_3 w_4 \Bigg]\nonumber \\ 	
	= &2\Bigg[Z_7 v_7 w_6 w_3 + Z_5 v_7 w_6 w_4 + Z_4 v_7 w_3 w_4+Z_6 v_4 w^2_2-\Big(Z_1 v_2 w_8 w_2 \nonumber \\
	& \quad + Z_1 v_2 w_8 w_3 + Z_1 			v_2 w_8 w_4 + Z_1 v_2 w_8 w_5 + Z_6 v_3 w^2_2 +Z_2 v_7 w_3 w_1 \nonumber \\
	&\quad + Z_2 v_7 w_3 w_2 + Z_2 v_7 w_3 w_5 + Z_3 v_7 w_4 w_1 + Z_3 v_7 w_4 w_2\nonumber \\
	& \quad + Z_3 v_7 w_4 w_5 + 2Z_2 v_7 w^2_3 + 2Z_3 v_7 w^2_4\Big)\Bigg]. \nonumber \\ \nonumber
\end{align}    
%---------------------------------------------------------------------------------
Upon grouping positive terms and simplifying, the right-hand side of the above expression can be written as: 
%---------------------------------------------------------------------------------
{\small
\begin{align*} 
	& 2 D_1^2 D_2 \Bigg[-\frac{\tau Z_7 Q_4 Q_1 K_2}{\xi_M K_4}-\frac{\tau Z_5 Q_4 Q_1 K_3}{\xi_M K_5}+\frac{\tau Z_4 K_2 K_3}{K_4 K_5 	K_7} + Z_6 Q_3 \\
	& -\Bigg(Z_6 Q_2 + \frac{\tau^2 Z_1 Q^2_1 Q^2_4}{K_1 \xi_M}\left(1+ \frac{K_2}{K_4} + \frac{K_3}{K_5}+ Q_0\right)+\frac{\tau Z_2 K_2}{K_4 K_7}\left(\frac{\lambda_{RS} Q_0-K_1}{K_6} + 1 + Q_0\right) \\
	&+\frac{\tau Z_3 K_3}{K_5 K_7}\left(\frac{\lambda_{RS}Q_0-	K_1}{K_6} + 1 + Q_0\right) + \frac{2\tau Z_2 K^2_2}{K^2_4 K_7}+ \frac{2\tau Z_3 K^2_3}{K^2_5 K_7}\Bigg) \Bigg]\\
 	=\, &2 D_1^2 D_2 \Bigg[\frac{\tau Z_2 K_1 K_2}{K_4 K_6 K_7}+\frac{\tau Z_3 K_1 K_3}{K_5 K_6 K_7}+\frac{\tau Z_4 K_2 K_3}{K_4 K_5 			K_7} + Z_6 Q_3 -\Bigg(Z_6 Q_2 \nonumber \\
 	&+ \frac{\tau^2 Z_1 Q^2_1 Q^2_4}{K_1 \xi_M}\left(1+ \frac{K_2}{K_4} + \frac{K_3}{K_5}+ Q_0\right)+\frac{\tau Z_2 K_2}{K_4 K_7}\left(\frac{\lambda_{RS} Q_0}{K_6} + 1 + Q_0\right) \\
	&+\frac{\tau Z_3 K_3}{K_5 K_7}\left(\frac{\lambda_{RS} Q_0}{K_6} 	+ 1 + Q_0\right) +\frac{\tau Z_7 Q_4 Q_1 K_2}{\xi_M K_4}+\frac{\tau Z_5 Q_4 Q_1 K_3}{\xi_M K_5} + \frac{2\tau Z_2 K^2_2}{K^2_4 			K_7}+ \frac{2\tau Z_3 K^2_3}{K^2_5 K_7}\Bigg) \Bigg]\\    
 =\,  &2 D_1^2 D_2 (\eta_2-\eta_1)\\
  =\, &2 D_1^2 D_2 (\frac{\eta_2}{\eta_1}-1)\eta_1,
\end{align*}
}
where 
\begin{align*}
		\eta_1:=&\, Z_6 Q_2 + \frac{\tau^2 Z_1 Q^2_1 Q^2_4}{K_1 \xi_M}\left(1+ \frac{K_2}{K_4} + \frac{K_3}{K_5}+ Q_0\right)\\
		& +\frac{\tau Z_2 			K_2}{K_4 K_7}\left(\frac{\lambda_{RS}  Q_0}{K_6} + 1 + Q_0\right) +\frac{\tau Z_3 K_3}{K_5 K_7}\left(\frac{\lambda_{RS}  Q_0}{K_6} + 1 + 					Q_0\right)\\ 
		&+\frac{\tau Z_7 Q_4 Q_1 K_2}{\xi_M K_4}+\frac{\tau Z_5 Q_4 Q_1 K_3}{\xi_M K_5} + \frac{2\tau Z_2 K^2_2}{K^2_4 				K_7}+ \frac{2\tau Z_3 K^2_3}{K^2_5 K_7},\\
	&	\\
        \eta_2:=& \, \frac{\tau Z_2 K_1 K_2}{K_4 K_6 K_7}+\frac{\tau Z_3 K_1 K_3}{K_5 K_6 K_7}+\frac{\tau Z_4 K_2 K_3}{K_4 K_5 K_7} + Z_6 		Q_3.\\
\end{align*}
%---------------------------------------------------------------------------------
Since $\left(D_i,\eta_i\right) \in \mathbb{R}_{\tiny{>0}} \times \mathbb{R}_{\tiny{>0}}$ for $i=1,2$, it follows that the sign of $a$ is completely dependent on the size of the quantity $\frac{\eta_2}{\eta_1}$. Therefore, define 
%---------------------------------------------------------------------------------
\begin{equation*}
	\Lambda:=\frac{\eta_2}{\eta_1} 
\end{equation*}
%---------------------------------------------------------------------------------
to arrive at the following dichotomy
%---------------------------------------------------------------------------------
\begin{equation*}
	\begin{cases}
		a<0 \iff \Lambda<1, \\
		a>0 \iff \Lambda>1. \\
	\end{cases}
\end{equation*}
%---------------------------------------------------------------------------------
This completes the proof concerning the bifurcation analysis of model (\ref{SEYAR_DS}).
%---------------------------------------------------------------------------------
\end{proof}
%##############################################
%---------------------------------------------------------------------------------
%##############################################
%---------------------------------------------------------------------------------
%##############################################
\subsubsection[Proof of the Functional Equation for SEYAR Model]{Proof of Theorem~\ref{thm:Gkap} (See page~\pageref{thm:Gkap})}
%---------------------------------------------------------------------------------
\begin{proof}
%----------------------------
\begin{align*}
\mathcal{G}_\nu \left( \eta^\omega_\kappa (a) \right) &= \xi_M \int_{0}^\infty \eta^\omega_\kappa (a) \nu(a) da,\\
 &= \xi_M \int_{\omega}^\infty \eta^\omega_\kappa (a) \nu(a) da,\\
&= \xi_M \lim_{\zeta \rightarrow +\infty} \int_{\omega}^\zeta e^{-a\xi_M}-e^{-a(\kappa + \xi_M)+\kappa \omega} da,\\
&= \xi_M \lim_{\zeta \rightarrow +\infty} \left[\frac{e^{-a(\kappa + \xi_M)+\kappa \omega}}{\kappa + \xi_M} -\frac{e^{-a\xi_M}}{\xi_M} \Bigg |_{[\omega,\zeta]} \right], \\
&= \xi_M \lim_{\zeta \rightarrow +\infty} \left[\frac{e^{-\zeta(\kappa + \xi_M)+\kappa \omega}}{\kappa + \xi_M} -\frac{e^{-\zeta\xi_M}}{\xi_M} + e^{-\omega \xi_M}\left(\frac{1}{\xi_M}-\frac{1}{\kappa + \xi_M}\right)\right], \\
&= \xi_M e^{-\omega \xi_M}\left(\frac{1}{\xi_M}-\frac{1}{\kappa + \xi_M}\right), \\
&= \frac{\kappa}{\kappa + \xi_M}e^{-\omega \xi_M}.
\end{align*}
%----------------------------
\end{proof}
%---------------------------------------------------------------------------------
%##############################################
%---------------------------------------------------------------------------------
%##############################################
%---------------------------------------------------------------------------------
%##############################################
\subsubsection[Reciprocal Sub-domain for Parameter Space SEYAR Model]{Proof of Theorem~\ref{thm:HypPara} (See page~\pageref{thm:HypPara})}
%---------------------------------------------------------------------------------
\begin{proof}
Fix the vector $(\rho_f,\rho_p) \in [0,1) \times [0,1)$ and each $C_i$ for $i=1,2$, as defined in Section (\ref{sec:Rnumber}). It follows that the sign of the term under the square root of $\mathcal{R}^v_0$, as defined by equation (\ref{ConR0}) in Section (\ref{sec:ControlMeas}), is determined by the sign of the following expression 
\begin{equation*}
((C_1- C_2)U_{_{\low}\hspace{5pt}}+C_2)(1-\rho_f\rho_p)(1-V_fV_p).
\end{equation*}
Thus, the control-modified reproductive threshold will not be nullified or complex-valued provided the vaccination terms satisfy the following reciprocal inequality $V_fV_p<1$. Therefore, $\mathcal{R}^v_0 \notin \mathbb{C}-\left\lbrace 0 \right\rbrace $ if and only if $(V_f,V_p) \in \mathcal{H}$, where $\mathcal{H}$ is defined as in the statement of Lemma (\ref{HypParam}).
\end{proof}
%---------------------------------------------------------------------------------
%##############################################
%---------------------------------------------------------------------------------
%##############################################
%---------------------------------------------------------------------------------
%##############################################
\subsubsection[Proof of the Local Asymptotic Stability of DFE for SEYAR Model Including Relapse Rates]{Proof of Lemma~\ref{thm:LAS_Vivax} (See page~\pageref{thm:LAS_Vivax})}
%---------------------------------------------------------------------------------
\begin{proof}
A straight-forward calculation shows that sub-matrices $F$ and $V$ of the Jacobian evaluated at the DFE $\mathbf{w}_{\small{dfe}}$ corresponding to the dynamical system (\ref{SEYAR_DSV}) are given by            
{\small
%---------------------------------------------------------------------------------
\begin{displaymath}
	F=
		\begin{pmatrix}
 			\frac{ \partial \mathcal{F}_i}{\partial x_j} \Big |_{\mathbf{w}_{\small{dfe}}} \\[5pt]
 		\end{pmatrix}_{1 \leq i,j \leq 5}
	 	=
  		\begin{pmatrix}
			0 & 0 & 0 & 0 & \sigma \beta_M \\
			0 & 0 & 0 & 0 & 0 \\
			0 & 0 & 0 & 0 & 0 \\
			0 & \sigma \beta_Y \frac{\Omega_M}{\Omega_H} \frac{\xi_H}{\xi_M}& \sigma \beta_A \frac{\Omega_M}{\Omega_H} \frac{\xi_H}	   		    		{\xi_M} & 0 & 0 \\
			0 & 0 & 0 & 0 & 0 \\
 		\end{pmatrix} 
\end{displaymath}
%---------------------------------------------------------------------------------
}
and
{\small
%---------------------------------------------------------------------------------
\begin{displaymath}
	V=
  		\begin{pmatrix}
			\frac{ \partial \mathcal{V}_i}{\partial x_j} \Big 	|_{\mathbf{w}_{\small{dfe}}} \\	
 		\end{pmatrix}_{1 \leq i,j \leq 5}
		 =
  		\begin{pmatrix}
			(\gamma +\xi_H) & 0 & 0 & 0 & 0\\
			\gamma \left(U_{_{\low}} -1 \right) & \left(\xi_H +\delta + \lambda_{YR}\right) & 0 & 0 & 0 \\
			-\gamma U_{_{\low}}  & 0 & \left(\lambda_{AR}+\xi_H\right) & 0 & 0 \\
			0 & 0 & 0 & (\xi_M+\tau) & 0 \\
			0 & 0 & 0 & -\tau & \xi_M \\
 		\end{pmatrix}. 
\end{displaymath}
%---------------------------------------------------------------------------------
}
The reproductive threshold $\mathcal{R}_0 :=\rho\left(FV^{-1}\right)$ for any given compartmentalized infectious disease model is completely determined by the matrices $F$ and $V$. Therefore, it is of trivial consequence that the models (\ref{SEYAR_DS}) and (\ref{SEYAR_DSV}) possess identical reproductive thresholds, given by (\ref{RnumSEYAR}) arising from Lemma (\ref{thm:LAS}).
\end{proof}
%---------------------------------------------------------------------------------

%APPENDIX 2-SUMMARY OF STABILITY THEOREMS
\subsection{Summary of Stability Theorems}\label{sec:A2}
The goal of this section of the appendix is to provide the reader with a brief collection and overview of the theorems that are widely used in determining the stability of the equilibrium points for nonlinear dynamical systems. The first theorem will be needed in order to determine the local asymptotic stability of the DFE corresponding to the $SEY \hspace{-3pt}AR$ model, while the second is used to investigate the existence of non-trivial sub-threshold equilibrium states of the model. In the setting of dynamical systems one cannot usually pinpoint a solution exactly, but only approximately. As a result, an equilibrium point must be stable to be physically meaningful. A stable equilibrium point of a system is a solution $\mathbf{x}^\star$ with the property that if for every open ball $B(\mathbf{x}^\star,\epsilon)$ of radius $\epsilon$, centered at $\mathbf{x}^\star$, there is a $\delta<\epsilon$, such that if every solution $\mathbf{x}$ with initial data $\mathbf{x}(0) \in B(\mathbf{x}^\star,\delta)$, remains in $B(\mathbf{x}^\star,\epsilon)$ for $t>0$. In other words, if the initial data starts in $B(\mathbf{x}^\star,\delta)$, then the flow map $\phi(t,\mathbf{x})$ of the model remains in  $B(\mathbf{x}^\star,\epsilon)$ for eternity. An equilibrium point $\mathbf{x}^\star$ is said to be asymptotically stable, if in addition to the above, there is a $\delta>0$ such that  
%---------------------------------------------------------------------------------
\begin{equation*}
	\lim_{t \rightarrow +\infty} \mathbf{x}(t)=\mathbf{x}^\star.
\end{equation*}
%---------------------------------------------------------------------------------
 Provided an epidemiological model can be grouped into $n$ homogeneous compartments, the local asymptotic stability of the equilibrium states can be established by utilizing the next generation method, appearing in \cite{VDW}. By making use of the Center Manifold Theorem \cite{CMT}, Van den Driessche and Watmough provided a simple prescription for determining the local asymptotic stability of DFE points of a given system. This criterion is given in terms of the reproductive number $\mathcal{R}_0$ of the system which acts as a threshold value. This effectively relates $\mathcal{R}_0$ to the DFE of the system. As a result, this has proven to be very useful in disease control. To cast the above discussion into a mathematical framework, we let $\mathbf{x}=\left(x_1,\cdots,x_k\right)^T \in \mathbb{R}^k_+$ and define the space of disease free states for the compartmental model to be $X:=\left\lbrace x \in \mathbb{R}^k_+ \hspace{3pt}:\hspace{3pt} x_i=0 \hspace{3pt}\text{for}\hspace{3pt} i=1,\cdots,m, \ m<n\right\rbrace$. Then for $\Phi \in C^2(\mathbb{R}^k)$, we form the following dynamical system
%---------------------------------------------------------------------------------
\begin{equation}
	\dot{\mathbf{x}}(t)=\Phi \left(\mathbf{x}(t)\right)=\mathcal{F}\left(\mathbf{x}(t)\right)-\mathcal{V}\left(\mathbf{x}	(t)\right),
		\label{DS_R}
\end{equation}
%---------------------------------------------------------------------------------
where $\mathcal{V}=\mathcal{V}^{-}-\mathcal{V}^{+}.$
%---------------------------------------------------------------------------------
%##############################################
%---------------------------------------------------------------------------------
%##############################################
%---------------------------------------------------------------------------------
%##############################################
\begin{theorem}{(\cite{VDW})}
	\label{Rtheo}
Define  $\mathcal{R}_0=\rho\left(FV^{-1}\right)$ and consider the disease transmission model given by (\ref{DS_R}) such that $\Phi$ satisfies the follow criteria:
%------------------------------------------------------------------------------
\begin{enumerate}[label=\roman*]
	\item If $\mathbf{x} \geq 0$, then  so are $\mathcal{F}$, $\mathcal{V}^{+}$, and $\mathcal{V}^{-}$,\label{10}
	\item If $\mathbf{x} = 0$, then $\mathcal{V}^{-}=0$,
	\item If $i>m$, then $\mathcal{F}_i=0$,
	\item If $\mathbf{x} \in X$, then $\mathcal{F}_i=\mathcal{V}^{+}=0$, for all $0 \leq i \leq m$,
	\item If $\mathcal{F}(\mathbf{x})=0$, then all eigenvalues of $D\Phi(\mathbf{x}^\star)$ have negative real parts. \label{50}
\end{enumerate}
%------------------------------------------------------------------------------
If $\mathbf{x}^\star$ is a DFE for (\ref{DS_R}), then $\mathbf{x}^\star$ is locally asymptotically stable provided $\mathcal{R}_0<1$ and unstable if $\mathcal{R}_0>1$. 
\end{theorem}
%---------------------------------------------------------------------------------
The above theorem is proved by making use of the following lemma.
%---------------------------------------------------------------------------------
\begin{lemma}
If $\mathbf{x}^\star$ is a DFE of system (\ref{DS_R}) and $\Phi$ satisfies assumptions (\ref{10})-(\ref{50}), then the derivatives $D\mathcal{F}(\mathbf{x}^\star)$ and $D\mathcal{V}(\mathbf{x}^\star)$ can be partitioned as follows:
%---------------------------------------------------------------------------------
\begin{displaymath}
	D\mathcal{F}(\mathbf{x}^\star)=
  	\begin{pmatrix}
  		F &  0 \\       
    	0 & 0 \\
  	\end{pmatrix} 
		\quad
		\text{and}
		\quad
	D\mathcal{V}(\mathbf{x}^\star)=
  	\begin{pmatrix}
		V & 0  \\      
		J_3 & J_4 \\
 	\end{pmatrix}. 
\end{displaymath}
%---------------------------------------------------------------------------------
Where $F$ and $V$ are defined as:
%---------------------------------------------------------------------------------
\begin{displaymath}
		F=
  	\begin{pmatrix}
 		\frac{ \partial \mathcal{F}_i}{\partial x_j} \Big |_{\mathbf{x}^\star} \\		
 	\end{pmatrix}_{1 \leq i,j \leq m}
		\quad
		\text{and}
		\quad
		V=
  	\begin{pmatrix}
 		\frac{ \partial \mathcal{V}_i}{\partial x_j} \Big |_{\mathbf{x}^\star} \\		
    \end{pmatrix}_{1 \leq i,j \leq m}.
\end{displaymath}
%---------------------------------------------------------------------------------
\label{matrixPartition}
\end{lemma}
%---------------------------------------------------------------------------------
\begin{Remark}
In the above matrix partitioning, $F$ is non-negative, $V$ is an invertible $M$-matrix and $J_3$, $J_4$ are sub-matrices of the Jacobian associated with various transmission terms. This theorem provides a convenient epidemiological interpretation of the reproductive threshold $\mathcal{R}_0$ corresponding to a given dynamical system in the SIR family. Additionally, due to the partitioning of the Jacobian mentioned above, the stability of the system is determined by $\det{\left(FV^{-1}-\lambda  \mathcal{I} \right)}$, where $\mathcal{I}$ is the identity matrix. If the matrix $F$ containing transmission probabilities and contact rates is set to zero, then all eigenvalues of $-V$ have negative real part. As a result, the stability, or lack of, experienced by the system in question depends on the entries of $F$. Due to this reason, the bifurcation parameters are chosen from the entries of $F$. In the case that the transmission probabilities are relatively high, an endemic could occur and additional non-trivial equilibrium can arise. Regarding the $SEY \hspace{-3pt}AR$ model, besides the man-biting rate $\sigma$, the only possible parameters are the transmission probabilities $\beta_Y$, $\beta_A$ and $\beta_M$. From a epidemiological perspective, we have more control over the mosquito to human transmission probability $\beta_M$. Due to this reason, $\beta_M$ is chosen as the parameter for the bifurcation analysis. One could perform a similar analysis for $\beta_Y$ and $\beta_A$, however all of the transmission probabilities involved are related through the reproductive number $\mathcal{R}_0$ of the model. As a result, the analysis would be similar.   
\label{matrixPartitionRemark}
\end{Remark}
%---------------------------------------------------------------------------------
Let $s(A)$ and $\rho(A)$ stand for the spectral abscissa and radius, respectively. The proof of Theorem \ref{Rtheo} hinges on M-matrix theory. A matrix $B$ is said to have the $Z$-sign pattern provided all of its off diagonal entries are non-positive. If $B=s \mathcal{I}-P$, where  $P \geq 0$. If $s>\rho(P)$, then $B$ is a non-singular M-matrix, if $s=\rho(P)$, then $B$ is a singular M-matrix. This observation is then combined with a linear algebra lemma, which we restate below for convenience of the reader. To this end, they make use of the following argument: define $J_1:=F-V$, then $V$ is a non-singular M-matrix and $F$ is non-negative. It follows that $-J_1:=V-F$ has the $Z$-sign pattern. By the non-negativity of $FV^{-1}$, it is a direct consequence that $-J_1V^{-1}:=\mathcal{I}-FV^{-1}$ also has the $Z$-sign pattern. Now we apply Lemma \ref{matrixLemma} with $H=V$ and $-J_1:=V-F$, to conclude that $-J_1$ is a non-singular M-matrix if and only if $\mathcal{I}-FV^{-1}$. Also, since all of its eigenvalues have magnitude bounded above by $\rho(FV^{-1})$, we have that $\mathcal{I}-FV^{-1}$ is a non-singular M-matrix if and only if $\rho(FV^{-1})<1$. Therefore, $s(J_1)<0$ if and only if $\mathcal{R}_0<1$. Through making use of a similar argument in combination with Lemma 6 of Appendix A, found in \cite{VDW}, they arrive at the following trichotomy, establishing $\mathcal{R}_0$ as a threshold parameter
%---------------------------------------------------------------------------------
\begin{equation*}
	\begin{cases}
		s(J_1)<0 \iff \mathcal{R}_0<1,\\
		s(J_1)=0 \iff \mathcal{R}_0=1,\\
		s(J_1)>0 \iff \mathcal{R}_0>1.\\
	\end{cases}
\end{equation*}
%---------------------------------------------------------------------------------
%---------------------------------------------------------------------------------
\begin{lemma}
Let $H$ be a non-singular M-matrix and assume that $B$ and $BH^{-1}$ have the $Z$ sign pattern. Then $B$ is a non-singular M-matrix if and only if $BH^{-1}$ is a non-singular M-matrix. 
\label{matrixLemma}
\end{lemma}
%##############################################
%---------------------------------------------------------------------------------
%##############################################
%---------------------------------------------------------------------------------
%##############################################
In Section (\ref{sec:EE}) use is made of the following variant of the Center Manifold Theorem specifically adapted to the case of bifurcation analysis for nonlinear systems. The utility of this theorem resides in the classification of the bifurcation due to the sign of the parameters $a$ and $b$, defined below.
%##############################################
%---------------------------------------------------------------------------------
%##############################################
%---------------------------------------------------------------------------------
%##############################################
\begin{theorem}{(\cite{CASTSONG})}
Consider the following dynamical system with real parameter $\omega$ 
%---------------------------------------------------------------------------------
\begin{displaymath}
	\dot{x}=f(x,\omega),\quad f:\mathbb{R}^n \times \mathbb{R} \rightarrow \mathbb{R},\quad \text{and} \quad f \in C^2(\mathbb{R}^n \times 			\mathbb{R}).
\end{displaymath}
%---------------------------------------------------------------------------------
Without loss of generality, we assume that $x=0$ is an equilibrium point for the above system for all $\omega$, i.e. $f(0,\omega) \equiv 0$ for all $\omega$. Provided that the following assumptions are satisfied:
%\begin{enumerate}[label=\Roman*]
%---------------------------------------------------------------------------------
\begin{enumerate}[label=\Roman*]
	\item $A=D_x f(0,0)=\Big ( \frac{\partial f_i}{\partial x_j} (0,0)\Big)$ is the linearization matrix of the system around the equilibrium point $0$ 			with $\omega$ evaluated at $0$. Zero is a simple eigenvalue of $A$ and all remaining eigenvalues have negative real part. Non trivial null 				space of dimension one. 
	\item The matrix $A$ has a right eigenvector $w$ and left eigenvector $v$ corresponding to the zero eigenvalue. 
\end{enumerate}
%---------------------------------------------------------------------------------
Let $f_k$ denote the $k^\text{th}$ component of $f$ and 
%---------------------------------------------------------------------------------
\begin{align}
	a &= \sum_{i,j,k=1}^{n} v_k w_i w_j \frac{\partial^2 f_k}{\partial x_i \partial x_j}\left(0,0\right),\label{a}\\
	b &=\sum_{i,k=1}^{n} v_k w_i \frac{\partial^2 f_k}{\partial x_i \partial \omega}\left(0,0\right). \label{b}	
\end{align}
%---------------------------------------------------------------------------------
\label{thm:CM}
\end{theorem}
%---------------------------------------------------------------------------------
Then the local dynamics around the equilibrium point $0$ are completely determined by the signs of $a$ and $b$. More precisely, 
%---------------------------------------------------------------------------------
\begin{enumerate}[label=\roman*]
	\item $a>0$, $b>0$. When $\omega<0$ such that $|\omega | \ll 1$, $0$ is locally asymptotically stable and there exists a positive unstable 				equilibrium; when $0<\omega \ll 1$, $0$ is unstable and there exists a negative locally asymptotically stable equilibrium. 
	\item $a<0$, $b<0$. When $\omega<0$ such that $|\omega | \ll 1$, $0$ is unstable; when $0<\omega \ll 1$, $0$ is locally asymptotically 				stable and there exists a positive unstable equilibrium. 
	\item $a>0$, $b<0$. When $\omega<0$ such that $|\omega | \ll 1$, $0$ is unstable and there exists a locally asymptotically stable negative 				equilibrium; when $0<\omega \ll 1$, $0$ is stable and a positive  unstable equilibrium exists.  
	\item $a<0$, $b>0$. When $\omega$ changes sign, $0$ changes its stability from stable to unstable. As a result, a negative unstable 						equilibrium becomes positive and locally asymptotically stable. In particular, if $a>0$ and $b>0$, then a backward bifurcation occurs at $					\omega=0$.
\end{enumerate}
%##############################################
%---------------------------------------------------------------------------------
%##############################################
%---------------------------------------------------------------------------------
%##############################################
\begin{Remark}
It is worth mentioning that the $a$ and $b$ terms appearing in the above theorem depend on generalized eigenvectors, i.e. zero entries are allowed. In the proof of Theorem (\ref{thm:CM}), the $a$ and $b$ terms arise from a differential equation obtained from a parameterization of a one-dimensional center manifold $c(t)$, given by
%---------------------------------------------------------------------------------
\begin{equation*}
	\dot{c}=\frac{a}{2}c^2+b\omega c.
\end{equation*}
%---------------------------------------------------------------------------------
Observe how a transcritical bifurcation occurs in the above equation at $\omega=0$ and can be classified according the signs of the $a$ and $b$ terms, defined above. These terms depend on the Kronecker product of generalized eigenvectors and entries of the Hessian evaluated at the DFE. As pointed out in \citet{CASTSONG}, negative components of the generalized eigenvectors are permitted through a modification of Theorem (\ref{thm:CM}). The essence of the argument hinges on the fact that the theorem can still be applied, one only has to compare the negative entries of the eigenvectors with their corresponding entries of the non-negative equilibrium of interest. Therefore, one has to consider the original parameterization of the center manifold, prior to the coordinate change where the DFE is assumed to be zero. Notice how the negative entries in the generalized eigenvectors calculated in Section \ref{sec:EE} correspond to the positive entries of the DFE of interest. 
\label{CMRemark}
\end{Remark}
%---------------------------------------------------------------------------------

%APPENDIX 3-NUMERICAL DATA
\subsection{Parameter Values}
\label{sec:Pvalues}
%----------------------------
Presented below are tables of numerical rates corresponding to the following three high transmission sites: Kaduna in Nigeria, Namawala in Tanzania and Butelgut in Papua New Guinea. The data has been collected from multiple sources, all of which are noted in the footnotes. Although the source of the data is heterogeneous, care has been taken to introduce as much accuracy as possible. As mentioned in \cite{killeen2000simplified}, all of the sites mentioned above are areas of intense \textit{P. falciparum} transmission. Due to the additional prevalence of \textit{P. vivax} in Butelgut, the corresponding data represents combined estimates of both species, \cite{killeen2000simplified}. The data for the other sites exclusively corresponds to the \textit{P.falciparum} species. There are a variety of vector species that inhabit these sites. The data listed here corresponds to the dominant vector species of the area being considered. The dominant vector species of each site is listed in the footnotes. Furthermore, as mentioned in Section \ref{sec:Methods} and reported in \cite{laishram2012complexities}, the parasites carried by asymptomatic human hosts can be more infectious than those of symptomatic. For the parameter values displayed in Section \ref{sec:Pvalues}, we invoke the assumption that the asymptomatic carriers corresponding to each site transmit at a lower rate than that of the symptomatic and that asymptomatic individuals recover faster than symptomatic, i.e. $\beta_A<\beta_Y$ and $\lambda_{AR}>\lambda_{YR}$.

Table \ref{table:POPULATIONhuman} below contains demographic data for the countries that each site is contained in. These numerical values are used to calculate the recruitment rates of the human populations $\Omega_H$ corresponding to each region under consideration. These quantities in combination with the parameter values listed in the subsequent tables, appearing on the following pages, will be used to calculate the reproductive thresholds associated with each location and the various quantities covered in Sections \ref{sec:Rnumber} and \ref{sec:RnumberComparison}.
%----------------------------
\vspace{-0.3cm}
{\footnotesize
\begin{table}[ht]
\caption{Human Population Data}
\label{table:POPULATIONhuman}
\begin{tabular}{l l l l l}
\textbf{Country} & \textbf{Life expectancy} & \textbf{Birth rate} & \textbf{Migration rate} & \textbf{Source}  \\ [0.5ex] % inserts table %heading
\toprule
\medskip
Nigeria & $53.02$ & $37.64$ & $-0.22$ & \cite{central2015world}\\
Tanzania & $61.71$ & $36.39$ & $-0.54$ & \cite{central2015world}\\
Papua New Guinea & $67.03$ & $24.38$ & \hspace{6pt}$0.00$ & \cite{central2015world}  \\ [1ex]
\bottomrule
\end{tabular}
\end{table}}

\vspace{-0.4cm}
The human population data displayed in table \ref{table:POPULATIONhuman} is from the Central Intelligence Agency (CIA). Denote $\Omega_{_{\lex}\hspace{5pt}}$, $\Omega_{_{\bir}\hspace{5pt}}$ and $\Omega_{_{\mig}\hspace{5pt}}$ to be the life expectancy, birth rate and migration rate of the population under consideration. Life expectancy is measured in years. The birth rate entries appearing in column 3 of the above table are crude birth rates. The Crude birth rate measures the average quantity of live births during a year, per $1,000$ people and is given in units of total births per $1,000$ people per year. The migration rates are net migration rates, which measure the difference of immigrants and emigrants in a given population over the span of a year and are given in units of humans per year, per $1,000$ people.

{\footnotesize
\begin{table}[H]
\label{table:Nigeria}
\begin{threeparttable}
\caption{Kaduna}
\begin{tabularx}{.9\textwidth}{lXXr}
\textbf{Parameter} & \textbf{Dimension} & \textbf{Value} & \textbf{Source}  \\ [0.5ex] % inserts table %heading
\hline
\medskip
$\Omega_H$ & humans $\times$ $\text{time}^{-1}$ & $1.02 \times 10^{-4}$ & \cite{central2015world}\tnote{$a$}\\
$\Omega_M$ & mosquitoes $\times$ $\text{time}^{-1}$&$1505.82$& \cite{killeen2000simplified}\tnote{$b$}\\
$\xi_H$ &  $\text{time}^{-1}$ & $0.019$ & \cite{central2015world}\tnote{$c$}\\
$\xi_M$ &  $\text{time}^{-1}$ & $0.11$ & \cite{molineaux1979assessment,chitnis2008determining}\tnote{$d$}\\
$\beta_A$ &  $\text{time}^{-1}$ & $0.048$ & assumed\tnote{$e$}\\
$\beta_Y$ & $\text{time}^{-1}$ & $0.48$ & \cite{chitnis2008determining}\tnote{$f$} \\
$\beta_M$ & $\text{time}^{-1}$ & $0.032$ & \cite{killeen2000simplified}\tnote{$g$}\\
$\gamma$  & $\text{time}^{-1}$ & $0.11$ & \cite{chitnis2008determining}\tnote{$h$}\\
$\tau$ & $\text{time}^{-1}$ & $0.09$ &\cite{killeen2000simplified,craig1999climate}\tnote{$i$}\\
$\delta$ & $\text{time}^{-1}$ & $1.66 \times 10^{-6}$ &  \cite{WLE}\tnote{$j$} \\
$\sigma$ & $\text{time}^{-1}$ & $0.42$ & \cite{killeen2000simplified}\tnote{$k$}\\
$\lambda_{AR}$ & $\text{time}^{-1}$ & $0.6$ & assumed\tnote{$l$}\\
$\lambda_{YR}$ & $\text{time}^{-1}$ & $0.06$ & assumed\tnote{$m$}\\
$\lambda_{RS}$ & $\text{time}^{-1}$ & $5.48 \times 10^{-4}$ & \cite{chitnis2008determining}\tnote{$n$}\\
$u_{_{\low}}$ & n/a &  $0.5$ & assumed\tnote{$o$}\\
 \\ [1ex]
\hline
\end{tabularx}
\begin{tablenotes}
	\item $a$ The human recruitment rate is obtained from the entries in the first row of Table \ref{table:POPULATIONhuman} and the weighted sum formula $\Omega_H=(\Omega_{_{\bir}\hspace{5pt}} +\Omega_{_{\mig}\hspace{5pt}})/365.25/1,000=(37.64-0.22)/365.25/1,000 \approx 1.02 \times 10^{-4}$. \\
	\item $b$ The daily vector emergence rate is measured in units of new adult female mosquitoes per day. Thus, it is given by dividing the entry in row six column four of Table 3 in \cite[p.~541]{killeen2000simplified} by the normalizing quantity $365.25$, i.e. $\Omega_M=(0.55 \times 10^{6})/365.25 \approx 1505.82$. \\
	\item $c$ The average natural human mortality rate is calculated by dividing the entry listed in row one column two of Table \ref{table:POPULATIONhuman} into unity, i.e. $\xi_H=1/\Omega_{_{\lex}\hspace{5pt}} =1/53.02 \approx 0.019$. \\
	\item $d$ The dominant vector species of Kaduna at the time the field measurements were taken was \textit{An. gambiae}, \cite{service1965some}. Let $\Omega_{_{\lm}\hspace{5pt}}$ stand for the average \textit{An. gambiae} life expectancy, which is dependent upon the region under consideration. The data provided in \cite{molineaux1979assessment} and appearing in row nine column one of Table A.3 in \cite[p.~19]{chitnis2008determining} corresponds to \textit{An. gambiae} activity in Nigeria. Due to this, we select the entry contained in row one column nine, so that $\Omega_{_{\lm}\hspace{5pt}}=9$. Therefore, the average daily \textit{An. gambiae} mosquito mortality rate is $\xi_M=1/\Omega_{_{\lm}\hspace{5pt}}=1/9 \approx 0.11$. \\
	\item $e$ The assumption is made that the probability of transmission from asymptomatic humans to susceptible mosquitoes is one-tenth the transmission probability corresponding to symptomatic humans, i.e. $\beta_A=0.048$.   \\  
	\item $f$ We adopt the convention utilized in \cite{chitnis2008determining}, where an estimate of $0.48$ will be used for high transmission areas and an estimate of $0.24$ will be used for low. Kanduna is a high transmission area, thus we let $\beta_Y=0.48$. An alternate choice would be to choose the average value of the parameters corresponding to the dominant species of parasite. The average value of the parameters appearing in rows one through five in column one of Table A.6 in \cite[p.~21]{chitnis2008determining} is approximately $0.36$. The origin of the parameters used in this calculation is \cite{thomson1957malarial,boyd1949epidemiology,draper1953observations}.    \\  
	\item $g$ The effective daily vector to human contact rate is obtained by dividing the value in row two column four of Table 3 in \cite[p.~541]{killeen2000simplified} by $\Omega_{_{\lm}\hspace{5pt}}$. Thus, $\beta_M=0.29 / \Omega_{_{\lm}\hspace{5pt}}=0.29 / 9 \approx 0.032$. \\
	\item $h$  The human latent period $\tilde{\gamma}$, measure in days, corresponding to \textit{P. falciparum} infection is taken from row three column one in Table A.7 in \cite[p.~21]{chitnis2008determining}. The range $9-10$ is given. The average value is then chosen, i.e. $\tilde{\gamma}\approx 9.5$. It follows that the average duration of the intermediate host latent period is $\gamma=1 / \tilde{\gamma}=1 / 9.5 \approx 0.11$. The parameter source is \cite{molineaux1980garki}.  \\
	\item $i$ Let $\tilde{\tau}$ denote the \textit{Plasmodium} incubation period, i.e. the number of days required for parasite development. Thus, by using the entry in row eight column three of Table 2 in \cite[p.~539]{killeen2000simplified}, it follows that the average duration of the definitive host latent period is $\tau=1 / \tilde{\tau}=1 / 11.6 \approx 0.09$. Technically, this parameter was calculated from the mean and median temperatures listed in the original source \cite{craig1999climate}.  \\
\end{tablenotes}	
 \end{threeparttable}
\end{table}}
%--------------------------------------------------------------------------------------------------
\newpage
%--------------------------------------------------------------------------------------------------
\begin{tablenotes}		
	\item $j$ The malaria death rates are taken from \cite{WLE} and are given in units of per $100,000$ people per year. As in the case of the human demographic data listed in Table \ref{table:POPULATIONhuman}, these rates correspond to the overall country that the region is contained in. Using the data provided in \cite{WLE}, it follows that $\delta=60.46/365.25/100,000 \approx 1.66 \times 10^{-6}$.	\\
	\item $k$ The average man biting rate is estimated by dividing the entry in row one column four of Table 3 in \cite[p.~541]{killeen2000simplified} by $\Omega_{_{\lm}\hspace{5pt}}$. More precisely, we have that $\sigma=3.8/9 \approx 0.42$.  \\
	\item $l$, $m$ Asymptomatic carriers transmit malaria to a lesser extent than symptomatic carriers and recover faster. Due to this we assume that the recovery rate $\lambda_{YR}$ for symptomatic individuals is one-tenth the rate $\lambda_{AR}$ of asymptomatic individuals. Let the quantity $1/\lambda$ denote the average duration of the human infectious period, provided that the individual has had no treatment. If $\Omega_I$ stands for the average duration of the \textit{P. falciparum} infectious period in humans.  We select the average of the entries appearing in row four column one of  Table A.9 in \cite[p.~21]{chitnis2008determining}, so that $\Omega_I=18$. Therefore,  $\lambda_{YR} \approx \lambda=1/ \Omega_I=1 /18 \approx 0.06$ and it follows that $\lambda_{AR}=0.6$. The original source of the parameter value is \cite{bloland2002roundtable}. 	\\
	\item $n$ The temporary immunity loss rate $\lambda_{RS}$ is such that $1 / \lambda_{RS}$ is equal to the average duration of the human immune period. As in \cite{chitnis2008determining}, we assume that the immune period lasts for an average of $5$ years in areas of high transmission and $1$ year in areas of low transmission. For Kaduna it follows that $\lambda_{RS}=1 / (5)365.25 \approx 5.48 \times 10^{-4}$.  \\
	\item $o$ This baseline assumption is due to the fact that the population under consideration is \textit{$A$-dominant}.    
    \end{tablenotes}

{\footnotesize
\begin{table}[H]
\label{table:Tanzania}
\begin{threeparttable}
\caption{Namawala}
\begin{tabularx}{.9\textwidth}{lXXr}
\textbf{Parameter} & \textbf{Dimension} & \textbf{Value} & \textbf{Source}  \\ [0.5ex] % inserts table %heading
\hline
\medskip
$\Omega_H$ & humans $\times$ $\text{time}^{-1}$ & $9.82 \times 10^{-5}$ & \cite{central2015world}\tnote{$a$}\\
$\Omega_M$ & mosquitoes $\times$ $\text{time}^{-1}$& $4928.13$ & \cite{killeen2000simplified}\tnote{$b$}\\
$\xi_H$ & $\text{time}^{-1}$ & $0.016$ & \cite{central2015world}\tnote{$c$}\\
$\xi_M$ & $\text{time}^{-1}$ & $0.09$ & \cite{molineaux1979assessment,chitnis2008determining}\tnote{$d$}\\
$\beta_A$ & $\text{time}^{-1}$ & $0.048$ & assumed\tnote{$e$}\\
$\beta_Y$ & $\text{time}^{-1}$ & $0.48$ & \cite{chitnis2008determining}\tnote{$f$} \\
$\beta_M$ & $\text{time}^{-1}$ & $0.002$ & \cite{killeen2000simplified}\tnote{$g$}\\
$\gamma$  & $\text{time}^{-1}$ & $0.11$ & \cite{chitnis2008determining}\tnote{$h$}\\
$\tau$ & $\text{time}^{-1}$ & $0.09$ &\cite{killeen2000simplified,craig1999climate}\tnote{$i$}\\
$\delta$ & $\text{time}^{-1}$ & $1.16 \times 10^{-6}$ &  \cite{WLE}\tnote{$j$} \\
$\sigma$ & $\text{time}^{-1}$ & $0.13$ & \cite{killeen2000simplified}\tnote{$k$}\\
$\lambda_{AR}$ & $\text{time}^{-1}$ & $0.6$ & assumed\tnote{$l$}\\
$\lambda_{YR}$ & $\text{time}^{-1}$ & $0.06$ & assumed\tnote{$m$}\\
$\lambda_{RS}$ & $\text{time}^{-1}$ & $5.48 \times 10^{-4}$ & \cite{chitnis2008determining}\tnote{$n$}\\
$u_{_{\low}}$ & n/a &  $0.5$ & assumed\tnote{$o$}\\
 \\ [1ex]
\hline
\end{tabularx}
\begin{tablenotes}
	\item $a$ From Table \ref{table:POPULATIONhuman}, it follows that $\Omega_H=(36.39-0.54)/365.25/1,000 \approx 9.82 \times 10^{-5}$. \\
	\item $b$ By utilizing the data provided in Table 3 in \cite[p.~541]{killeen2000simplified}, it follows that $\Omega_M=(1.8 \times 10^{6})/365.25 \approx 4928.13$. \\
	\item $c$ The average natural human mortality rate is calculated in a similar fashion, as in the case of Kaduna, by utilizing the data provided in Table \ref{table:POPULATIONhuman}. \\
	\item $d$ The dominant vector species of Namawala at the time the field measurements were taken was \textit{An. gambiae}, \cite{smith1993absence}. The data provided in \cite{gillies1965study} and appearing in row six column one of Table A.3 in \cite[p.~19]{chitnis2008determining} corresponds to \textit{An. gambiae} activity in Tanzania. Using this data, it follows that $\Omega_{_{\lm}\hspace{5pt}}=11.26$. Therefore, the average daily \textit{An. gambiae} mosquito mortality rate is $\xi_M=1/11.26 \approx 0.09$.   \\	
	\item $e$ As previously mentioned, the asymptomatic human to susceptible mosquito transmission probability is assumed to be $\beta_A=0.048$.   \\   
	\item $f$ Since Namawala is a high transmission area, we let $\beta_Y=0.48$.    \\     
    \item $g$ By dividing the value in row two column five of Table 3 in \cite[p.~541]{killeen2000simplified} by $\Omega_{_{\lm}\hspace{5pt}}$, it follows that $\beta_M=0.017 / 11.26 \approx 0.002$. \\
	\item $h$  The average duration of the intermediate host latent period is taken to be $\gamma=1 / \tilde{\gamma}=1 / 9.5 \approx 0.11$. The parameter value $\tilde{\gamma}$ is taken from row three column one in Table A.7 in \cite[p.~21]{chitnis2008determining}. The average value is then chosen, i.e. $\gamma \approx 9.5$. The original parameter source is \cite{molineaux1980garki}.      \\
	\item $i$ By using the entry in row eight column four of Table 2 in \cite[p.~539]{killeen2000simplified}, it follows that the average duration of the definitive host latent period is $\tau=1 / 11.6 \approx 0.09$. \\ 
	\item $j$  By the data provided in \cite{WLE}, it follows that $\delta=42.42/365.25/100,000 \approx 1.16 \times 10^{-6}$.       \\
	\item $k$ By using the data in row one column five of Table 3 in \cite[p.~541]{killeen2000simplified}, we have that $\sigma=1.5/11.26 \approx 0.13$. \\
	\item $l$ As in the previous table, the recovery rate pertaining to asymptomatic individuals is assumed to be $\lambda_{AR}=0.6$.      \\
	\item $m$ The average of the entries appearing in row four column one of  Table A.9 in \cite[p.~21]{chitnis2008determining} is selected, so that $\Omega_I=18$.  Therefore, $\lambda_{YR} \approx \lambda=1/ \Omega_I=1 /18 \approx 0.06$. The original source of the parameter value is \cite{bloland2002roundtable}.     \\       
	\item $n$ As Namawala is a high transmission area, the temporary immunity loss rate is taken to be $\lambda_{RS}=1 / (5)365.25 \approx 5.48 \times 10^{-4}$.  \\
	\item $o$ This baseline assumption is due to the fact that the population under consideration is \textit{$A$-dominant}.                  
    \end{tablenotes}
  \end{threeparttable}
\end{table}}

{\footnotesize
\begin{table}[H]
\label{table:PapuaNewGuinea}
\begin{threeparttable}
\caption{Butelgut}
\begin{tabularx}{.9\textwidth}{lXXr}
\textbf{Parameter} & \textbf{Dimension} & \textbf{Value} & \textbf{Source}  \\ [0.5ex] % inserts table %heading
\hline
\medskip
$\Omega_H$ & humans $\times$ $\text{time}^{-1}$ & $6.67 \times 10^{-5}$ & \cite{central2015world}\tnote{$a$}\\
$\Omega_M$ & mosquitoes $\times$ $\text{time}^{-1}$& $4106.78$ & \cite{killeen2000simplified}\tnote{$b$}\\
$\xi_H$ & $\text{time}^{-1}$ & $0.015$ & \cite{central2015world}\tnote{$c$}\\
$\xi_M$ & $\text{time}^{-1}$ & $0.14$ & \cite{molineaux1979assessment,chitnis2008determining}\tnote{$d$}\\
$\beta_A$ & $\text{time}^{-1}$ & $0.048$ & assumed\tnote{$e$}\\
$\beta_Y$ & $\text{time}^{-1}$ & $0.48$ & \cite{chitnis2008determining}\tnote{$f$} \\
$\beta_M$ & $\text{time}^{-1}$ & $0.006$ & \cite{killeen2000simplified}\tnote{$g$}\\
$\gamma$  &  $\text{time}^{-1}$ & $0.11$ & \cite{chitnis2008determining}\tnote{$h$}\\
$\tau$ & $\text{time}^{-1}$ & $0.1$ &\cite{killeen2000simplified,craig1999climate}\tnote{$i$}\\
$\delta$ & $\text{time}^{-1}$ & $1.03 \times 10^{-6}$ &  \cite{WLE}\tnote{$j$} \\
$\sigma$ & $\text{time}^{-1}$ & $0.14$ & \cite{killeen2000simplified}\tnote{$k$}\\
$\lambda_{AR}$ & $\text{time}^{-1}$ & $0.6$ & assumed\tnote{$l$}\\
$\lambda_{YR}$ & $\text{time}^{-1}$ & $0.06$ & assumed\tnote{$m$}\\
$\lambda_{RS}$ & $\text{time}^{-1}$ & $5.48 \times 10^{-4}$ & \cite{chitnis2008determining}\tnote{$n$}\\
$u_{_{\low}}$ & n/a &  $0.5$ & assumed\tnote{$o$}\\
 \\ [1ex]
\hline
\end{tabularx}
\begin{tablenotes}
	\item $a$ From Table \ref{table:POPULATIONhuman}, it follows that $\Omega_H=(24.38-0.00)/365.25/1,000 \approx 6.67 \times 10^{-5}$. \\
	\item $b$ By utilizing the data provided in Table 3 in \cite[p.~541]{killeen2000simplified}, it follows that $\Omega_M=(1.5 \times 10^{6})/365.25\approx 4106.78$.   \\ 
	\item $c$ Similarly, the average natural human mortality rate is calculated by utilizing the data provided in Table \ref{table:POPULATIONhuman}. \\
 	\item $d$ The dominant vector species of Butelgut at the time the field measurements were taken was \textit{An. punctulatus}, \cite{rates1988human}. The data provided in \cite{peters1960part} and appearing in row thirteen column one of Table A.3 in \cite[p.~19]{chitnis2008determining} corresponds to \textit{An. punctulatus} activity in Papua New Guinea. Using this data, it follows that $\Omega_{_{\lm}\hspace{5pt}}=7.1$. Therefore, the average daily \textit{An. punctulatus} mosquito mortality rate is $\xi_M=1/7.1 \approx 0.14$.    \\
	\item $e$ As previously demonstrated, the asymptomatic human to susceptible mosquito transmission probability is assumed to be $\beta_A=0.048$.   \\  
	\item $f$  Since Butelgut is a high transmission area, we let $\beta_Y=0.48$.   \\           
	\item $g$ By dividing the value in row two column six of Table 3 in \cite[p.~541]{killeen2000simplified} by $\Omega_{_{\lm}\hspace{5pt}}$, it follows that $\beta_M=0.042 / 7.1 \approx 0.006$. \\
	\item $h$ The average duration of the intermediate host latent period is taken to be $\gamma=1 / \tilde{\gamma}=1 / 9.5 \approx 0.11$. The parameter value $\tilde{\gamma}$ is taken from row three column one in Table A.7 in \cite[p.~21]{chitnis2008determining}. The average value is then chosen, i.e. $\gamma \approx 9.5$. The original parameter source is \cite{molineaux1980garki}. \\
	\item $i$ By using the entry in row eight column five of Table 2 in \cite[p.~539]{killeen2000simplified},  it follows that the average duration of the definitive host latent period is $\tau=1 / 9.6 \approx 0.1$. \\
	\item $j$   By the data provided in \cite{WLE}, it follows that $\delta=37.57/365.25/100,000 \approx 1.03 \times 10^{-6}$.     \\
	\item $k$ By using the data in row one column six of Table 3 in \cite[p.~541]{killeen2000simplified}, we have that $\sigma=0.99/\Omega_{_{\lm}\hspace{5pt}}=0.99/7.1 \approx 0.14$.   \\        
	\item $l$ Due to similar reasoning as above, the recovery rate pertaining to asymptomatic individuals is assumed to be $\lambda_{AR}=0.6$. \\
	\item $m$ The average of the entries appearing in row four column one of  Table A.9 in \cite[p.~21]{chitnis2008determining} is selected, so that $\Omega_I=18$.  Therefore, $\lambda_{YR} \approx \lambda=1/ \Omega_I=1 /18 \approx 0.06$. The original source of the parameter value is \cite{bloland2002roundtable}. \\
 	\item $n$ As Butelgut is a high transmission area, the temporary immunity loss rate is taken to be $\lambda_{RS}=1 / (5)365.25 \approx 5.48 \times 10^{-4}$.      \\
    \item $o$ This baseline assumption is due to the fact that the population under consideration is \textit{$A$-dominant}.      
    \end{tablenotes}
  \end{threeparttable}
\end{table}}
%----------------------------
%################
%################
\newpage
\bibliographystyle{plainnat}
\bibliography{SEYAR_ref}
%################
\end{document}